\documentclass{aa}
\usepackage{graphicx}
\usepackage[varg]{txfonts}
\usepackage{longtable}
\usepackage{lscape}
\usepackage{amsmath}

\begin{document}

\title{The effects of granulation and supergranulation on Earth-mass planet detectability in the habitable zone around F6-K4 stars.}

\titlerunning{Effect of granulation and supergranulation on Earth-mass planet detectability}

\author{N. Meunier \inst{1}, A.-M. Lagrange \inst{1} 
  }
\authorrunning{Meunier et al.}

\institute{
Univ. Grenoble Alpes, CNRS, IPAG, F-38000 Grenoble, France\\
     }

\offprints{N. Meunier}

\date{Received ; Accepted}

\abstract{The detectability of exoplanets and the determination of their projected mass in radial velocity are affected by stellar magnetic activity and photospheric dynamics. Among those processes, the effect of granulation, and even more so of supergranulation, has been shown to be significant in the solar case. The impact for other spectral types has not yet been characterised. }
{ Our study is aimed at quantifying the impact of these flows for other stars and estimating how such contributions affect their performance. }
{We analysed a broad array of extended synthetic time series that model these processes to characterise the impact of these flows on exoplanet detection for main sequence stars with spectral types from F6 to K4. We focussed on Earth-mass planets orbiting within the habitable zone around those stars. We estimated the expected detection rates and detection limits, tested the tools that are typically applied to such observations, and performed blind tests. }
{We find that both granulation and supergranulation on these stars significantly affect planet mass characterisation in radial velocity when performing a follow-up of a transit detection: 
the uncertainties on these masses are sometimes below 20\% for a 1 M$_{\rm Earth}$ (for granulation alone or for low-mass stars), but they are much larger in other  configurations (supergranulation, high-mass stars). For granulation and low levels of supergranulation, the detection rates are good for K and late G stars (if the number of points is large enough), but poor for more massive stars. The highest level of supergranulation  leads to a very poor performance, even for K stars; this is both due to low detection rates and to high levels of false positives, even for a very dense temporal sampling over ten years. False positive levels estimated from standard false alarm probabilities sometimes significantly overestimate or underestimate the true level,  depending on the number of points: it is, therefore, crucial to take this effect into account when analysing observations.  }
{We conclude that granulation and supergranulation significantly affect  the performance of exoplanet detectability. Future works will focus on improving the following three aspects: decreasing the number of false positives, increasing detection rates, and improving the false alarm probability estimations from observations. }


\maketitle

\section{Introduction}

A large number of exoplanets have been detected using indirect techniques for over 20 years. However, because these techniques are indirect, they are very sensitive to stellar variability. The radial velocity (RV) technique is particularly sensitive to activity that is due to both  magnetic and dynamical processes at different temporal scales. Many studies have focussed on stellar magnetic activity \cite[recognised early on by][]{saar97} based on simulations of simple spot configurations \cite[e.g.][]{desort07,boisse12,dumusque12} as well as more complex patterns \cite[e.g.][]{lagrange10b,meunier10,meunier10a,borgniet15,santos15,dumusque16,herrero16,dumusque17,meunier19,meunier19b,meunier19c}. Flows on different spatial and temporal scales also play an important role: in addition to large-scale flows such as meridional circulation \cite[][]{makarov10,meunier20c}, oscillations, granulation, and supergranulation also affect RV time series. 

The properties of these small-scale flows  and the mitigating techniques used to remove them (mostly averaging techniques) have been studied in several works \cite[e.g.][]{dumusque11b,cegla13,meunier15,cegla15,sulis16,sulis17,cegla18,meunier19e,cegla19,chaplin19} for the Sun and other stars. More details can be found in the review by \cite{cegla19b}.
The impact of granulation on the use of standard  statistical tools has been pointed out by \cite{sulis17b}, who proposed a new method (based on periodogram standardisation) to improve these tools, so far for a solar type star. The RV jitter associated to granulation has also been studied for  chromospherically quiet stars covering a large range in spectral types and evolutionary stages by \cite{bastien14}. 

Granulation and supergranulation are challenging because of the shape of their power spectrum, which is flat (instead of decreasing, as in the case of oscillations) at low frequencies \cite[][]{harvey84}, and because it is not related to usual activity indicators.
Furthermore, in \cite{meunier19e}, hereafter referred to as Paper I, we showed that for the  Sun, the effect of supergranulation was unexpectedly strong and more problematic than the granulation signal. Here, we perform a similar analysis (with the addition of more complete blind tests) for main sequence stars extending over a large range of spectral types, that is, from F6 to K4 as in our magnetic activity simulations \cite[][hereafter referred to as Paper II]{meunier19}, where this contribution was added to the activity signal to build more realistic long-term time series of realistic  activity patterns.  In the present paper, we aim to study granulation and supergranulation contributions to RVs for stars with various spectral types and to perform a detailed analysis of the false positive levels from different points of view (theoretical and observational) and their effect on exoplanet detection rates. 

We adopted a systematic approach to study and quantify these effects for different conditions, including different spectral types, numbers of observations, and samplings. We consider exoplanet detectability using RV techniques, but also the mass characterisation which can be made using RV in transit follow-ups: when the planet has been detected and validated using transits, its radius is known (relative to the stellar radius) along with other parameters (orbital period, phase), but only the RV techniques can currently provide a mass estimate, which, in turn, allows us to estimate its density, thus  giving us a hint of its composition. We focus on Earth-like planets in the habitable zone of their host star. Such a systematic approach is also very important because there are few stars observed that have a very large (in the 500-1000 regime or above) number of observations currently available; thus, tests on observations are currently limited, in addition to the fact that they could have undetected planets, for stars other than the Sun \cite[][]{collier19}.

The outline of the paper is as follows. In Sect.~2, we present the synthetic time series and the approaches we implemented to analyse them, as well as, in particular, how we define theoretical levels of false positives. In Sect.~3, we analyse these time series using true false positive levels (i.e. assuming a perfect knowledge of the properties of the signal) to derive detection rates and mass detection limits. In Sect.~4, we focus on the observational point of view by comparing usual false alarm probability levels with the true false positive levels and characterising the detection limits proposed in \cite{meunier12} for this type of signal. Then we estimate the uncertainty on the mass estimation in transit follow-ups. We implement blind tests to fully characterise the performance in terms of detectability and false positive levels when a classical tool is used to evaluate detections. Finally, we test complementary samplings in Sect.~5 and present our conclusions in Sect.~6.

\section{Model and analysis}

In this section, we describe the time series and how we extrapolate data from  solar parameters \cite[][]{meunier15,meunier19e} to build stellar time series. Then we present the different approaches to analyse these synthetic time series and, in particular, we discuss how we determine false positive levels. 

\subsection{Time series of oscillations, granulation, and supergranulation}

Our reference times series are solar ones: we first provide the amplitudes we consider for the Sun and apply those to G2 stars. Then we describe our assumptions for other stars.

\begin{figure}
\includegraphics{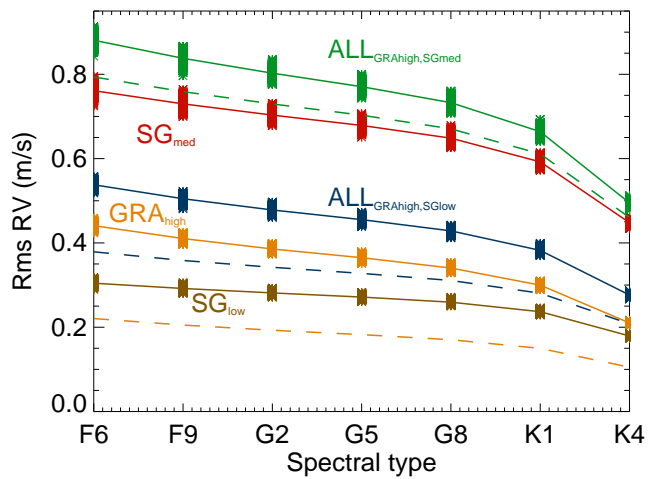}
\caption{
Rms RV vs. spectral type for GRAhigh (orange), SGmed (red), SGlow (brown), ALL$_{\rm GRAhigh,SGmed}$ (green), and ALL$_{\rm GRAhigh,SGlow}$ (blue), for the best sampling (3650 points, no gaps). The dashed lines correspond to the configurations including GRAlow (same colour code). Individual values are shown as stars. 
}
\label{rms}
\end{figure}

\subsubsection{Solar amplitudes }

We first define the solar values we consider in this study. The time series are derived from power spectra following \cite{harvey84} for granulation and supergranulation and following the shape of the envelope of the oscillations from \cite{kallinger14}, as in Papers I and II. This method has the advantage of allowing us to produce a large amount of very long time series. We  showed in \cite{meunier15} that the shape proposed by  \cite{harvey84} was well adapted, even at low frequencies: therefore, we use the parameters found in \cite{meunier15}. The choice of a one-hour binning is similar to what we chose in Paper I and corresponds to the timescales where the RV jitter due to granulation reaches an inflexion point \cite[][]{meunier15}: binning over a longer duration is not efficient enough to reduce this jitter further and so, this binning time is used to filter granulation out best. 

For granulation, in the majority of our study, we use an rms (root-mean-square) of 0.83 m/s before averaging (i.e. 0.39 m/s after averaging over one hour), hereafter GRAhigh, which stands as our reference value, provided by our simulations of about 1 million granules on the disk at any given time in \cite{meunier15}. As discussed in Paper I, such simulations were based on realistic properties of granules \cite[derived from hydrodynamical simulations of ][]{rieutord02}, which are known to reproduce realistic line profiles \cite[][]{asplund00}. However, lower values were derived from the observation  of two specific spectral lines: about 0.32 m/s by \cite{elsworth94} from the Potassium line at 770 nm and 0.46 m/s from the Sodium doublet at 589 nm by \cite{palle99}. More recently, the residuals on timescales lower than one day obtained by \cite{collier19} on solar integrated RV times series and covering the whole spectrum obtained by HARPS-N are also of the order of 0.40 m/s (when averaging over typically five minutes).  Similar amplitudes have been obtained by \cite{sulis20} using MHD simulations.  The difference between these estimates and the results of \cite{meunier15}  may be due to some subtle effects in the centre-to-limb dependence which are not taken into account in \cite{meunier15}, but also to the fact that observations were made in a single lines which may not be representative of the whole spectrum. For that reason, a twice lower level (hereafter GRAlow) with respect to our reference level will also be considered in mass characterisations and blind tests in Sect.~4.3 and 4.4. 
We note that \cite{cegla19} obtained very low rms RV for granulation using a reconstruction based on MHD simulations of the solar surface, around 0.1 m/s. The reason for this discrepancy is not clear at this stage, although it may be due the fact that strong vertical magnetic fields were used.

Concerning supergranulation, \cite{meunier15} provide a large range of possible values based on our current knowledge of these flows. Here, we consider two values, their median level (0.7 m/s, hereafter SGmed), and their lower level (0.27 m/s, hereafter SGlow), as in \cite{meunier19e}: these are in agreement with typical amplitudes obtained for a few stars by \cite{dumusque11b}. The median level is also close to the rms found by \cite{palle99} for the Sun, with 0.78 m/s for the Sodium doublet lines. Because of the longer timescales of supergranulation, the rms RV is almost the same after the 1 hour averaging. The amplitude of the oscillations is derived from \cite{davies14}, as in Paper I. The time scale is the same one obtained in \cite{meunier15} as in Paper I, that is, 1.1 10$^6$ s.

We mainly use five types of time series throughout the paper: high  level of granulation alone (GRAhigh), supergranulation alone (SGmed, median level, and SGlow, low level), all contributions for oscillations, a high level of granulation, and median supergranulation (ALL$_{\rm GRAhigh,SGmed}$) or low  supergranulation (ALL$_{\rm GRAhigh,SGlow}$). In the following, ALL always represents the superposition of oscillations, granulation, and supergranulation. The other three configurations (GRAlow alone, ALL$_{\rm GRAlow,SGmed}$, ALL$_{\rm GRAlow,SGlow}$) are mostly be considered for the mass characterisation and blind tests to provide a complete view. 
The configuration ALL$_{\rm GRAhigh,SGmed}$ was used in combination with magnetic activity in Paper II. 
The contribution attributed to any  of these combinations is referred to as the OGS (for oscillations, granulation, supergranulation) signal in the following. The oscillations are not studied alone here because we consider one-hour averages and they are well averaged out \cite[][]{chaplin19} at such timescales: they did not prevent us from obtaining  excellent detection rates when considered independently (Paper I). 

\subsubsection{Stellar time series}

We considered seven spectral types covering the F6-K4 range, that is, F6, F9, G2, G5, G8, K1, and K4. The amplitudes of the different components were scaled to G2 stars (i.e. solar values from the previous section) as in Paper II (previous section). We recall them here in brief. Granulation parameters are scaled from G2 stars to other spectral types using results from \cite{beeck13a}. Oscillation parameters are scaled using laws from \cite{kjeldsen95}, \cite{samadi07}, 
\cite{bedding03}, \cite{kippenhahn90}, and \cite{belkacem13}\footnote{There are other scaling laws, for example by \cite{yu18}, but they are not very different. Because the oscillations are strongly averaged in this work, this choice is not critical.}. Supergranulation is scaled following the granulation scaling, assuming supergranulation is strongly related to granulation properties \cite[][]{rieutord00,roudier16}, including the time scale, which can differ by up to about 20\%, so that the impact should be small.

All time series were produced for a ten-year period of duration with a 30-second time step and  are then binned over one hour. We then selected one such point per night. Examples of time series (subsets over short periods) are shown in Appendix A for F6, G2, and K4 stars, as well as examples of the power functions versus frequency. In addition to this full sample of 3650 nights, we consider several other configurations with a gap of four months per year  to simulate the fact that a star can usually not be observed all year long. Then $N_{\rm obs}$ nights were randomly selected out of the remaining nights over the ten-year duration. Each realisation of this selection for a given value of $N_{\rm obs}$ corresponds to a different sampling. We use  $N_{\rm obs}$ = 180, 542, 904, 1266, 1628, 1990, and 2352 nights with the four-month gap each year), and 3650 nights (no gap), leading to a total of eight configurations. In Paper I, we found that using a random selection or considering packs of adjacent nights did not lead to significant differences. In addition, \cite{burt18} tested different ways of building the sampling  for magnetic activity time series and found that the random sampling was optimal (the uniform sampling was not extremely different however).  Testing of additional sampling configurations is presented in Sect.~5. 

Figure~\ref{rms} summarises  the rms RV versus spectral type for the eight configurations of OGS time series used in this paper. We note   a general decrease towards lower mass stars. 
When considering all components and spectral types, the RV jitter varies between 0.28 and 0.9 m/s typically when considering GRAhigh. The dashed lines show the levels when the granulation level is divided by 2 (GRAlow). In this case the granulation rms varies between 0.22 and 0.1 m/s, and when combined with the low level of supergranulation it varies between 0.2 and 0.37 m/s. We note that even for such a large number of points, there is little  dispersion in RV jitter from one realisation to the next.    

\begin{figure}
\includegraphics{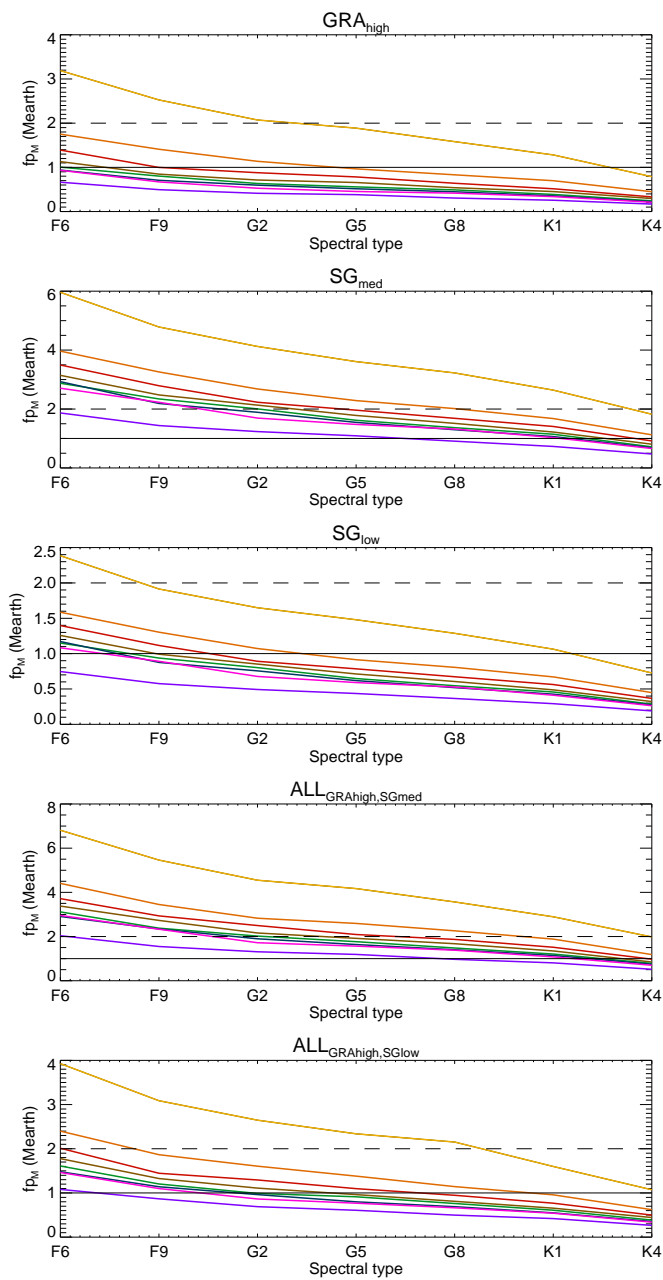}
\caption{
False positive level in mass fp$_M$ vs. spectral type for different numbers of points $N_{\rm obs}$: 180 (yellow), 542 (orange), 904 (red), 1266 (brown), 1628 (green), 1990 (blue), 2352 (pink), and 3650 (purple), and for different OGS configurations (from top to bottom). Values of fp$_M$ correspond to 1\% of false positives and PHZ$_{\rm med}$. 
The two horizontal lines indicate the 1 M$_{\rm Earth}$ (solid line) and 2 M$_{\rm Earth}$ (dashed line) levels for comparison. }
\label{fpmass}
\end{figure}

\begin{figure}
\includegraphics{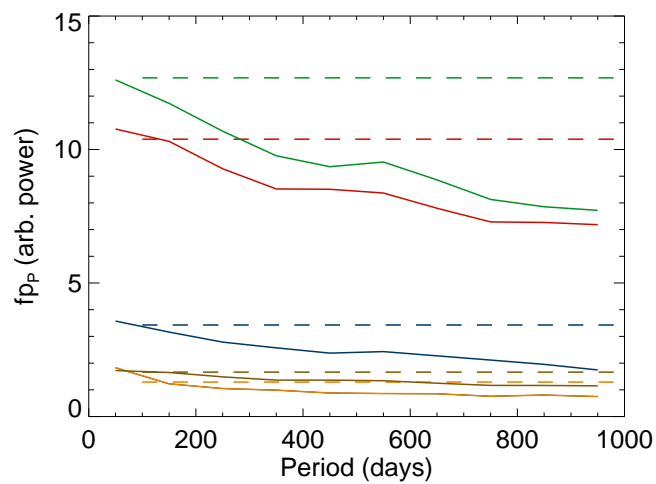}
\caption{
False positive level in power fp$_P$  vs. period for highest number of points, G2 stars, and five OGS configurations (GRAhigh in orange, SGmed in red, SGlow in brown, ALL$_{\rm GRAhigh,SGmed}$  in green, and ALL$_{\rm GRAhigh,SGlow}$ in blue). The solid lines represent fp$_P$ computed in 100d ranges, while the dashed horizontal lines correspond to the single value of fp$_P$ computed over 100-1000d. 
}
\label{fppuiss}
\end{figure}

\subsection{Principle of the analysis}

Here, we describe the planet properties considered in this paper and then discuss issues related to detectability as well as mass characterisation in transit follow-ups.

\subsubsection{Planets}

We focus our analysis on low-mass planets orbiting in the habitable zone of their host stars. We define the limits of the habitable zone as a function of spectral type as in \cite{meunier19b}, following \cite{kasting93}, \cite{jones06}, and \cite{zaninetti08}. We consider three typical orbital periods, corresponding to the inner side (PHZ$_{\rm in}$), the middle (PHZ$_{\rm med}$), and the outer side (PHZ$_{\rm out}$) of the habitable zone: the resulting orbital periods vary between 409-1174 days for F6 stars to 179-501 days for K4 stars. 
Furthermore, we consider only circular orbits, for simplicity. 

Most of the computations are carried out with projected masses of 1 and 2 M$_{\rm Earth}$. 
For inclinations higher than  40-50$^\circ$, the performance obtained with these masses is representative of this whole range of inclinations, while for lower inclinations the performance should be significantly worse than the one presented in this paper.  Therefore, additional blind tests, presented in Sect.~4.4, are also performed, considering a distribution of inclinations between 0$^\circ$ and 90$^\circ$ when building the data set and with the assumption that the orbital plane is the same as the stellar equatorial plane. 
In the case of a transit follow-up using RV to characterise the mass, however, the projected mass can be considered to be the true mass.

\subsubsection{Detectability}

In subsequent sections, the analysis of the time series is made using two complementary approaches (i.e. two test statistics), which are then compared. The steps are as follows: (i) we analyse the periodograms\footnote{We use the Lomb Scargle periodogram with no normalisation to be able to compare powers between different types of contributions. They are computed between 2 and 2000 days.} of the time series, and compute the maximum amplitude around the considered PHZ (frequential analysis, computed in 0.9--1.1 PHZ range); (ii) we fit the planetary signal, considering a period guess corresponding to the period of this  peak with maximum amplitude (temporal analysis) or of interest (PHZ) depending on the case. This fit is made using a $\chi^2$ minimization. 

We first consider the detectability of such exoplanets in the presence of the stellar contribution defined in the previous section. 
Because we consider synthetic time series, we  can study them  with the certainty that there is no planet present in the signal. As a consequence, we can estimate a true level of false positive (FP) for a given test statistics (frequential and temporal analysis)  and for a given probability (e.g. 1\%), and it is then possible to compute detection rates for a given planet (on the time series where the planet have been added), considering this level of false positives. 
The method we apply to determine the FP is described in Sect.~2.3. Once we have determined a true FP level corresponding to a certain percentage of false positives, and a detection rate for a given mass, we can also determine which mass corresponds to a good detection rate (e.g. 95\%), which provides a detection limit. This approach is explored in Sect. 3.

From the point of view of the observer, however, the determination of the true level of false positive due to a given stellar contribution is not possible because it is not possible to know 
if it includes other, additional signals (of a planet for instance) and because we have only one realisation of the signal. This is why the analysis of observed time series always relies on other methods, such as the use of false alarm probability levels using bootstrap analysis, although this approach makes assumptions on the signal which may not be correct, as pointed out by \cite{sulis17}. 
In a second step, we therefore test this type of approach and compare it with the one based on the true false positive level. We also compare the detection limits based on the periodogram analysis proposed by \cite{meunier12}, the  local power analysis (LPA) method with the true detection limits. A blind test is implemented to estimate the detection rates and false positive levels and compare them with the true ones.  This approach is explored in Sect. 4.

\subsubsection{Mass characterisation}

The latter issue, also studied in Sect.~4, concerns  the performance with regard to mass characterisations of planets  detected by transit in photometric light curves. In this case, we consider that  the planet presence is confirmed,meaning that the transits do not require any validation using RV observations.  There is, therefore, no issue  with false positives in this case and we also know its orbital period and phase with very good precision from the transit. We can then fit the RV amplitude due to the planet (temporal analysis) at this orbital period to determine the precision for the mass characterisation.

\subsection{False positives from synthetic time series}

Here, we describe how we estimate the false positive (FP) level at the 1\% level, both in mass (temporal analysis) and power (frequential analysis)
This level corresponds to the behavior in the frequency of the OGS signal alone and for a given test of statistics (here, the power at the period we are interested in or the fitted mass; see previous section) since it is computed based on a large number of time series of the OGS signal alone. This is done with no correction of the signal (apart from the 1 hour binning).

To estimate the FP from our time series, we produce 1000 realisations of the OGS signal and sampling (for a given spectral type and number of points $N_{\rm obs}$) as described in Sect.~2.1. For each of the three orbital periods corresponding to the habitable zone (Sect.~2.2), we fit a planetary signal at this period, which provides 1000 values of the mass. The period used as a guess before minimisation is the period of the peak with maximum power in the periodogram around the period we are interested in (namely in the 0.9--1.1 PHZ range as above). The 1\% false positive level fp$_M$ is defined as the mass such that 1\% of the 1000 values are higher. This level is therefore estimated for each spectral type, $N_{\rm obs}$ and PHZ. 

To ascertain that a planet has been detected, we compare the fitted mass (temporal analysis) to fp$_M$: if it is higher than fp$_M$, we consider the planet as  detected. Figure~\ref{fpmass} shows fp$_M$ versus spectral type and $N_{\rm obs}$, for PHZ$_{\rm med}$. The values of fp$_M$ decrease towards lower mass stars and with higher values of $N_{\rm obs}$. There are many configurations where fp$_M$ is higher than 1 M$_{\rm Earth}$ with values as high as several M$_{\rm Earth}$ for F6 stars, but below 1 M$_{\rm Earth}$ for K4 stars. For a given spectral type and OGS configuration, fp$_M$ decreases as $N_{\rm obs}$ increases, but the variation is not linear in $\sqrt{N_{\rm obs}}$, and after a sharp decrease at low $N_{\rm obs}$, the level does not change much, as shown in Fig.~\ref{fpmass}. More details about the dependence on $N_{\rm obs}$ is shown in Sect.~5.

For each of these 1000 realisations, we also compute the periodogram and the maximum peak amplitude in two ways: between 100 and 1000 days (which includes most of our PZH values) and in 10 ranges of 100 days between 0 and 1000 days, to check whether the FP depends on the period. As before, the 1\% level, fp$_P$, is computed out of each 1000 series of values. The results are shown in Fig.~\ref{fppuiss}. There is a clear trend with period and the whole range of values corresponds roughly to the lowest period. In the following, we consider fp$_P$ computed for the different period ranges to take this trend  into account.

\section{Simulated detection rates of Earth-mass planets in the habitable zone}

In this section, we consider the synthetic time series produced in the previous section and add planets with different masses at different orbital periods to estimate the effect of the OGS signal on exoplanet detectability. We use the level of false positives (corresponding to 1\% in the following) defined in Sect.~2.3, both in mass for the temporal analysis (fp$_M$) and in power for the frequential analysis (fp$_P$). We then compute detection rates for various masses and detection limits corresponding to these well identified detection rates. We use only GRAhigh in this section (with five OGS configurations).

\begin{figure}
\includegraphics{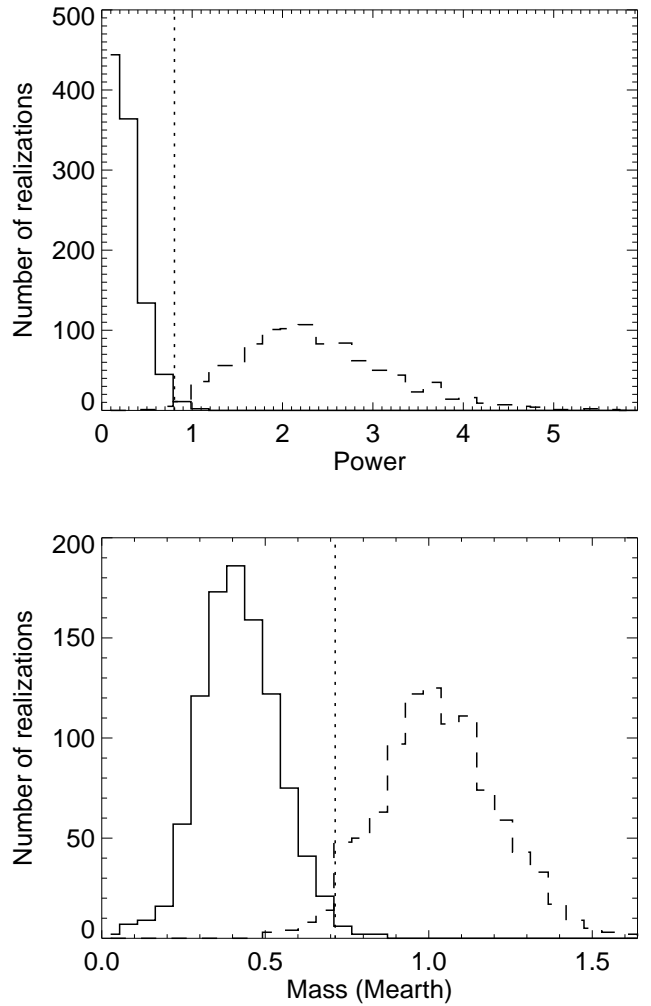}
\caption{
Example of distributions of power (upper panel) and mass (lower panel) in the presence of 1 M$_{\rm Earth}$ planet (dashed line) and with no planet (false positive values,  solid line). The distributions are for a G2 star, 1266 points and PHZ$_{\rm med}$ and GRAhigh. The vertical line indicates the position of the 1\% false positive level deduced from the solid line distribution. 
}
\label{exdist}
\end{figure}

\begin{figure}
\includegraphics{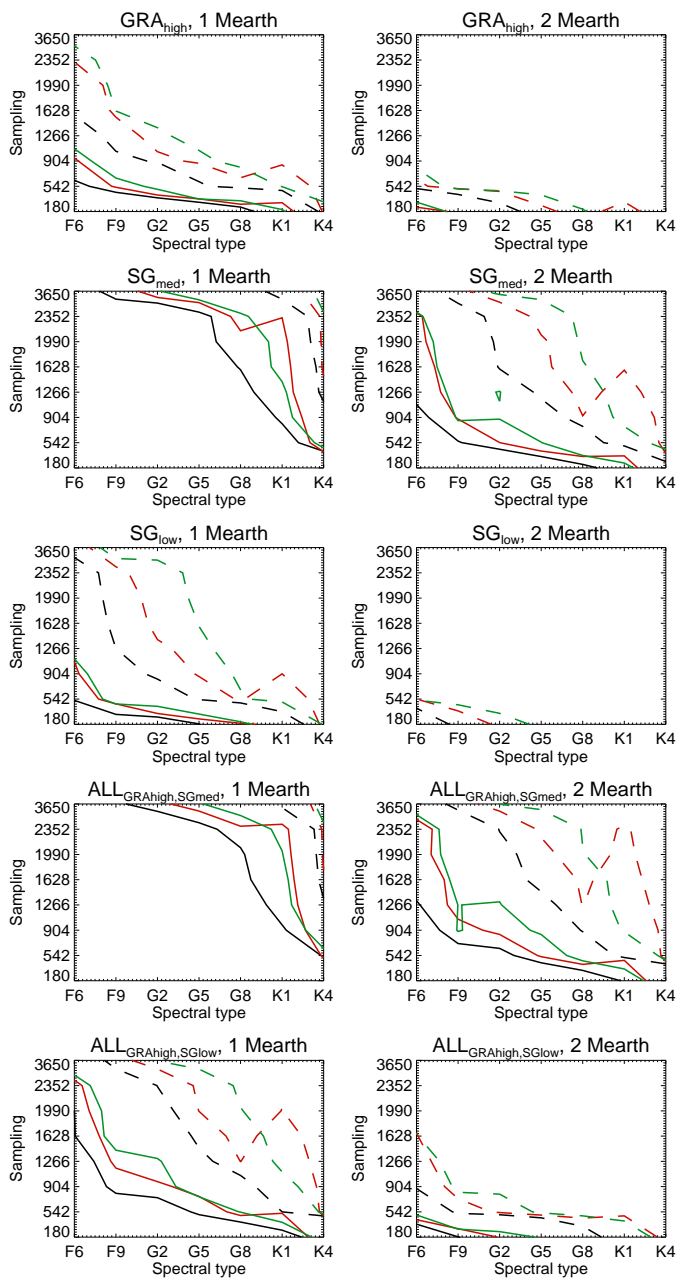}
\caption{
Detection rates of 50\% (solid lines) and 95\% (dashed lines) based on frequential analysis vs. spectral type and $N_{\rm obs}$, for different OGS configurations (from top to bottom) and for different orbital periods: PHZ$_{\rm in}$ (black), PHZ$_{\rm med}$ (red), and PHZ$_{\rm out}$ (green). }
\label{coupe_taux_p}
\end{figure}

\begin{figure}
\includegraphics{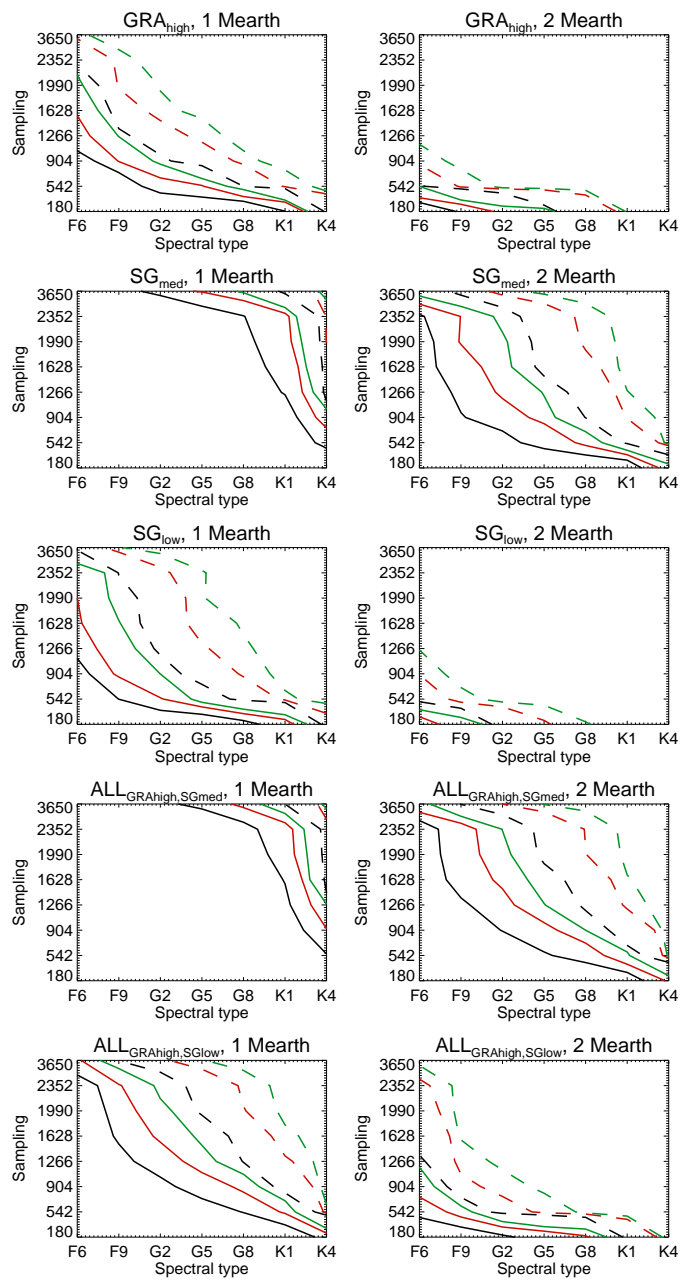}
\caption{
Same as Fig.~\ref{coupe_taux_p} but based on temporal analysis. 
}
\label{coupe_taux_m}
\end{figure}

\subsection{Detection rates for Earth-mass planets}

We consider planets with projected masses of 1 M$_{\rm Earth}$ and 2 M$_{\rm Earth}$ (see Sect.~2.2.1 for a discussion) on circular orbits and at three positions in the habitable zone of each spectral type as described in Sect.~2. The signal due to such planets (with a random phase) is added to each of the 1000 realisations of the OGS signals and sampling for each spectral type and $N_{\rm obs}$. 
For the frequential analysis, we use the amplitude of the peak at the orbital period we are interested in. 
For the temporal analysis, the fit is made with an initial guess for the period corresponding to this orbital period, which leads to a certain mass. 
The 1000 values of the peak amplitude and of the mass are then compared to the corresponding values of fp$_M$ and fp$_P$ at the considered period (see Sect.~2.3).  The false positives are computed around the targeted periodicities, which biases them since there may be other false positives at other periods as well: This will be studied in Sect.~4. 
The percentage of those 1000 values above the true false positive level is the detection rate, associated to the chosen level of false positive (here 1~\%). An example of the distributions of the 1000 values is shown in Fig.~\ref{exdist} to illustrate the procedure. Values beyond  the vertical lines correspond to detections. For the same configuration, the detection rate seems slightly larger for the frequential analysis compared to the temporal analysis, meaning that the frequential analysis is more robust for obtaining good detection rates.

Figure~\ref{coupe_taux_p} shows the resulting detection rates obtained with the frequential analysis depending on N$_{\rm obs}$. For each spectral type, the curves indicate the necessary number of points $N_{\rm obs}$ to reach a 50\% detection rate (solid lines) or a 95\% detection rate (dashed lines). Curves at a low level mean that it is very easy to detect planets (small values of $N_{\rm obs}$ are sufficient), while curves at the top correspond to configurations for which a detection is difficult to obtain (high values of $N_{\rm obs}$).
Higher values of $N_{\rm obs}$ are necessary for longer orbital periods, as expected (since the planetary signal is dropping). For 2 M$_{\rm Earth}$, the detection rates are very good for granulation and low supergranulation levels (or ALL$_{\rm GRAhigh,SGlow}$), as excellent rates can be reached with a low number of observations. Very good detection rates require a very large number of observations (a few hundreds to a few thousands depending on spectral type) when considering SGmed. Adding granulation to SGmed does not change much the performance. For 1 M$_{\rm Earth}$, the performance is not as good, and higher numbers of points are required to get good detection rates. The low level of supergranulation leads to good detection rates, but only with a high number of points, except for F stars for which even our maximum $N_{\rm obs}$ of 3650 nights does not allow us to reach detection rates of 95\%. The situation is significantly worse for the median level of supergranulation, with conclusions similar to what was found in Paper I for G2 stars. 
We also observe a bump for K1 stars and PHZ$_{\rm med}$: this is due to the fact that for this particular configuration, the orbital period is equal to 366 days\footnote{This period of 366 days corresponds to the middle of the habitable zone for a K1 star. The Earth has a one-year period but lies closer to the inner side of the habitable zone of a G2 star.}, and given the gap introduced every one year in the sampling, planets at such periods would naturally be more difficult to detect.  As expected, the frequential analysis is therefore quite sensitive to the temporal window. We conclude that the performance is good for a 2 M$_{\rm Earth}$ planet, while for a 1 M$_{\rm Earth}$ planet good results can be achieved only with a very high frequency of observations, mostly due to supergranulation.

Figure~\ref{coupe_taux_m} shows similar curves for the temporal analysis, that is, with the fitted mass used as a criterion for estimating the detection rates. The global trends are similar to the frequential analysis, with two main differences. All curves correspond to higher numbers of points, that is, more observations are requested to obtain the same detection rate. This is due to the difference in false positive levels already noted in Sect.~2.3: the frequential analysis criterion allows us to get better detection rates. On the other hand, there is no more bump for K1 stars and PHZ$_{\rm med}$ with this approach, as the temporal analysis is less sensitive to the temporal window than the frequential analysis.

\subsection{Detection limits}

\begin{figure}
\includegraphics{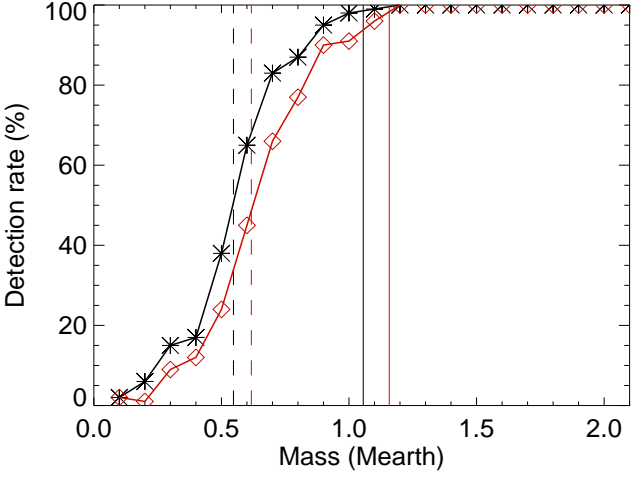}
\caption{
Example of detection rate vs. planet mass, for G2 stars, 1266 points, and GRAhigh, in two cases: Based on frequential analysis (black curve) and on temporal analysis (red curve). The vertical solid lines indicate the corresponding 95\% level, and the dashed lines the 50\% level. 
}
\label{limdet_ex}
\end{figure}

\begin{figure}
\includegraphics{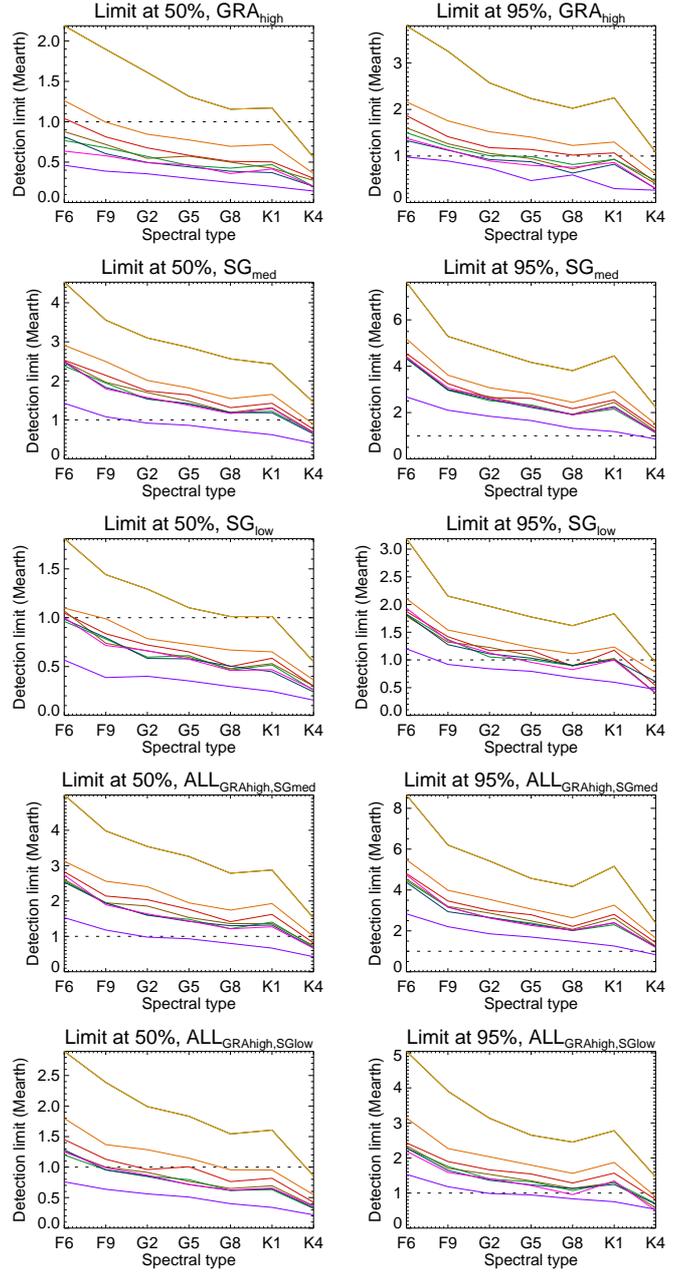}
\caption{
        Detection limits vs. spectral type for 50\% detection rate (left panels) and 95\% detection rate (right panels) for frequential analysis, PHZ$_{\rm med}$ and for different values of $N_{\rm obs}$ (from low $N_{\rm obs}$ to high $N_{\rm obs}$, see Sect.~2.1): Yellow (180), orange (542), red (904), brown (1266), green (1628), blue (1990), pink (2352), purple (3650). The horizontal dotted line corresponds to a 1 M$_{\rm Earth}$ planet.  
}
\label{limdet_cut}
\end{figure}

\begin{figure}
\includegraphics{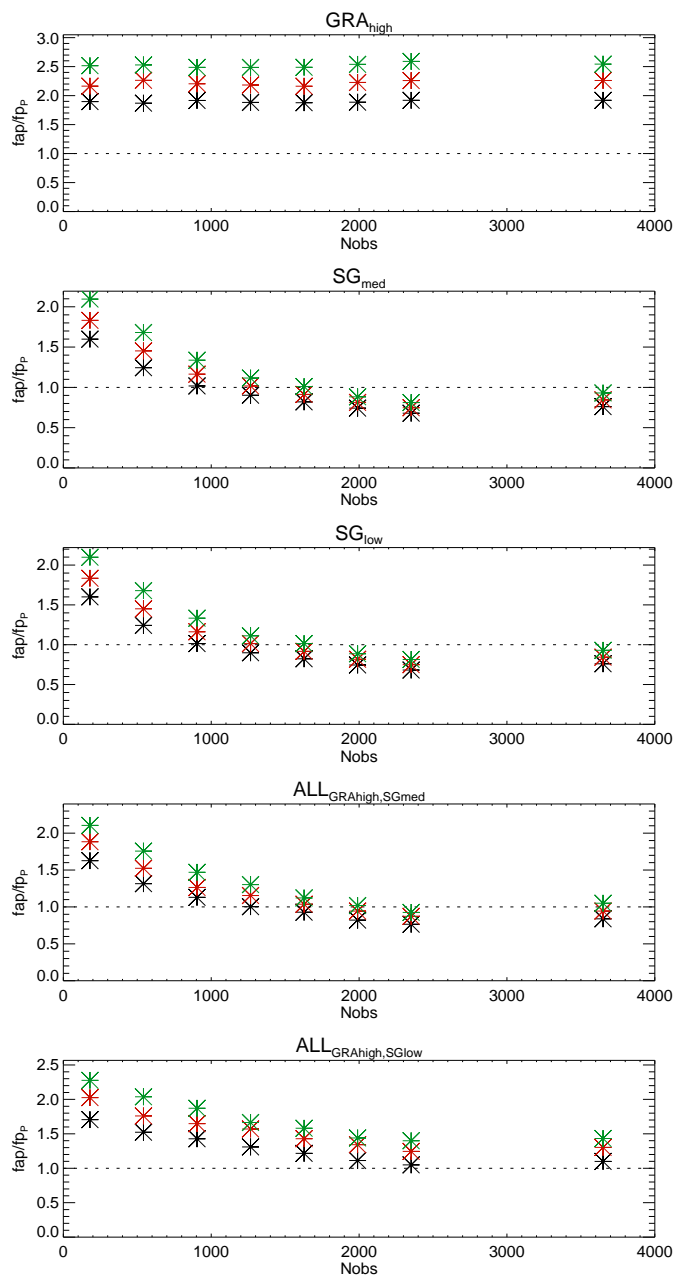}
\caption{
Average ratio fap/fp$_P$ for PHZ$_{\rm med}$  vs. $N_{\rm obs}$.  The average is computed over all realisations and spectral types.
The colour code represents the period: Inner side (black), middle (red), and outer side (green) of the habitable zone. 
}
\label{fap1}
\end{figure}

\begin{figure}
\includegraphics{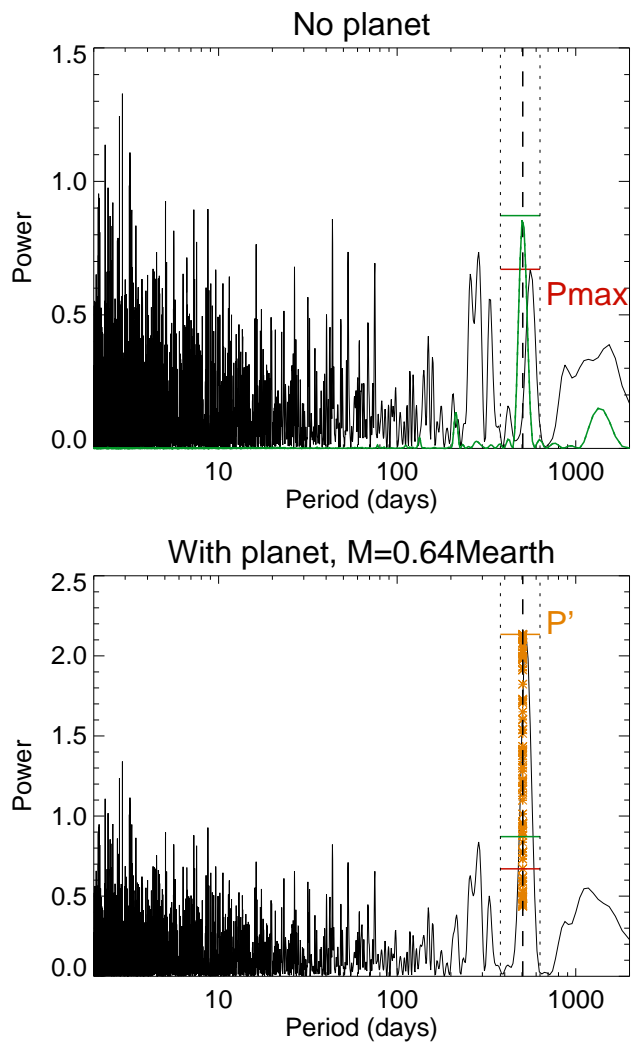}
\caption{
Example of the periodogram for OGS alone (upper panel, GRAhigh alone) and with a planet at LPA mass (0.53 M$_{\rm Earth}$, lower panel) to illustrate the LPA computations. The red and green horizontal lines corresponds to the maximum of the OGS periodogram in the window delimited by the dotted lines and multiplied by 1.3 respectively. The green solid line periodogram in the upper panel is for the planet alone. The example of periodogram with OGS+planet is for an arbitrary phase of the planet, and the horizontal orange line corresponds to the maximum power. Orange stars are for 100 realisations of the planet phase. 
}
\label{lpa_protocole}
\end{figure}

Detections rates are  computed as in the previous section but for a large range of planet masses and a 0.1 M$_{\rm Earth}$ step. This  allows us to determine at which mass, for a given spectral type, $N_{\rm obs}$, and OGS configuration, the detection rate is equal to 95\% for example (given a false positive level of 1\%). Only 100 realisations of the signal OGS+planet are performed because such computations are time consuming. For the same reason, computations are made only for the middle of the habitable zone PHZ$_{\rm med}$. An example of detection rate versus planet mass is shown in Fig.~\ref{limdet_ex} to illustrate the procedure. As already noted, there is a shift between the frequential analysis and the temporal analysis, of the order of 0.1 M$_{\rm Earth}$ in this example.

Figure~\ref{limdet_cut} shows the detection limits versus spectral types for different OGS contributions and different values of $N_{\rm obs}$, for various numbers of nights (between 180 and 3650) covering 10 years. At the 50\% level, they are often below 1 M$_{\rm Earth}$  (especially for low mass stars) if $N_{\rm obs}$ is sufficiently high: this is the case  for GRAhigh, SGlow and ALL$_{\rm GRAhigh,SGlow}$. 
They are mostly above 1 M$_{\rm Earth}$  for SGmed and ALL$_{\rm GRAhigh,SGmed}$ with values up to 2.5 M$_{\rm Earth}$  for F6 stars however. At the 95\% level, only the highest values of $N_{\rm obs}$ allow to reach 1 M$_{\rm Earth}$, and this is true for K4 stars only when considering the median level of supergranulation. 

We conclude that in most configurations, the detection limits are higher than 1 M$_{\rm Earth}$. This is the case especially for the most massive stars and  when a limited number of nights is available (typically a few hundreds for granulation, but a few thousands for supergranulation).

\begin{figure}
\includegraphics{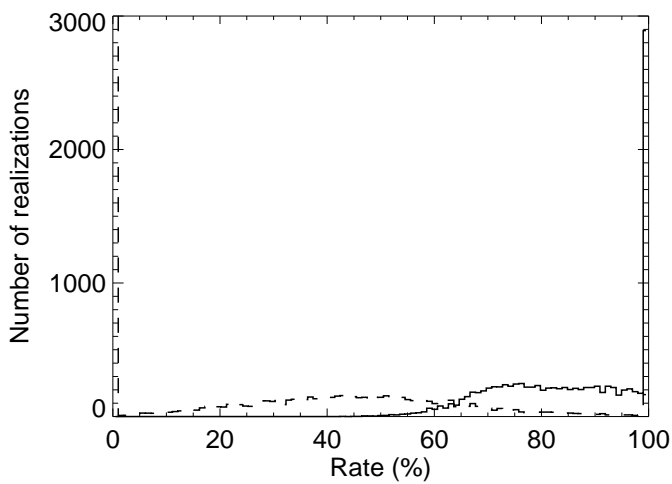}
\caption{
Distribution of exclusion rates (solid line) and detection rates (dashed line) for all LPA tests, i.e. covering all spectral types, three values of $N_{\rm obs}$, different OGS configurations, and PHZ$_{\rm med}$.
}
\label{lpadist}
\end{figure}

\section{Observational approach}

The results presented in the previous section are based on a perfect knowledge of the OGS signal. This allowed us to compute true false positive levels and to deduce detection rates corresponding to a given level (1\%) of a false positive: given the true false positive levels, this approach provided the best detection rates possible, with a controlled false positive level.  We now consider the point of a view of an observer, who is interested in a time series which may contain other contributions and for which we have only one realisation: different tools must then be used, and actual detection rates may be lower, or the resulting detection rates may correspond to a higher false positive level. 
It is, therefore, important to compare what tools. We first compare the false alarm probability (FAP) obtained using a bootstrap analysis with the true false positive level. Then we compute the detection limits using the LPA method \cite[][]{meunier12} and determine which true detection rates and exclusion rates these detection limits correspond to. Finally, we characterise the mass uncertainty in transit follow-ups and we implement several blind tests to estimate the detection rates and false positives obtained when a usual FAP analysis of the data is performed.

\subsection{Classical bootstrap false alarm probability}

In this section, we focus on the comparison between the FAP level and the true false positive level, fp$_P$, with no injected planet, both at the 1\% level. The effect on detection rates will be studied in the blind tests in Sect.~4.4. Only GRAhigh is used in this section. 
For each time series (with no planet), we compute the 1\% FAP level using a bootstrap analysis. Because it is time consuming, only ten realisations of the OGS signal are considered for each spectral type and value of $N_{\rm obs}$. The maximum of the periodogram to compute the FAP is computed over the whole periodogram, that is, between 2 and 2000 days. For each configuration (spectral type, $N_{\rm obs}$) and a given orbital period (one of the three PHZ values), we compute the following values: the percentage of simulations with a FAP higher than the true false positive level fp$_P$  at 1\% obtained in the previous sections (this is necessarily noisy since there are only ten realisations); the ratio of the FAP\ and FP, namely, fap/fp$_P$ (averaged over the ten realisations); the number of peaks above the FAP (averaged over the ten realisations). 

The results are summarised in Fig.~\ref{fap1}. 
The fap/fp$_P$  and the percentage of simulations with FAP larger than the FP are strongly correlated, therefore only the ratio is shown. Although the results show some dispersion because of the low number of realisations (a larger number of realisations performed on a few typical configurations gives similar results, however), some trends can be observed. 
The ratio covers a wide range, with values between 0.6 and 3 (after averaging on the ten realisations). 
For GRAhigh, the percentage is always 100\%, and it is almost always the case for ALL$_{\rm GRAhigh,SGlow}$: the FAP is then always  overestimating the false positive level, on average by a factor  of two (corresponding to a factor of four on the mass). 
In the other configurations, there is a high proportion of simulations where the FAP is larger than the true false positive level when $N_{\rm obs}$ is small, and it tends to be the opposite for a large number of points, with a transition for $N_{\rm obs}$ in the 1000-2000 range. 
The limit between the two regimes occurs  at  higher $N_{\rm obs}$ for a given orbital period (alternatively, for a given $N_{\rm obs}$, the ratio is larger at longer periods). 
Finally, the average number of peaks over all configurations is low (0.24) but there are several peaks above the FAP in some  configurations, mostly for supergranulation alone and ALL$_{\rm GRAhigh,SGmed}$, especially when $N_{\rm obs}$ is large, in agreement with the ratio. 

The true false positive level corresponds to the true frequency behaviour of the OGS signal, while the FAP assumes a white noise with a similar rms RV and a similar distribution of RV values. The shape of the power spectrum of the OGS signal is such that the usual FAP computation is not always adapted: it appears to overestimate the false positive level when the number of point is low (or for GRAhigh and ALL$_{\rm GRAhigh,SGlow}$ in all configurations) when, rather, it should underestimate the detection rate, corresponding to a conservative approach of the detection. When the number of points is high however, for SGlow SGmed, and ALL$_{\rm GRAhigh,SGmed}$, the FAP underestimates the false positive level, which should lead to potentially good detection rates but corresponding to much higher false positive levels in reality. These results are compatible with those of \cite{sulis17}, who proposed a new method (periodogram standardisation) to be able to use standard tools such as the FAP. 

Finally, we note that the FAP is computed over the whole range over which we compute the periodogram (2-2000 days), while in the previous section, we consider the FP dependent on the period (see Sect.~2.3 and Fig.~\ref{fppuiss}): given its shape, the FP level we are interested when searching for planets in the habitable zone is lower than at short periods. We expect the FAP to agree better  with the FP at low periods. 

\begin{figure}
\includegraphics{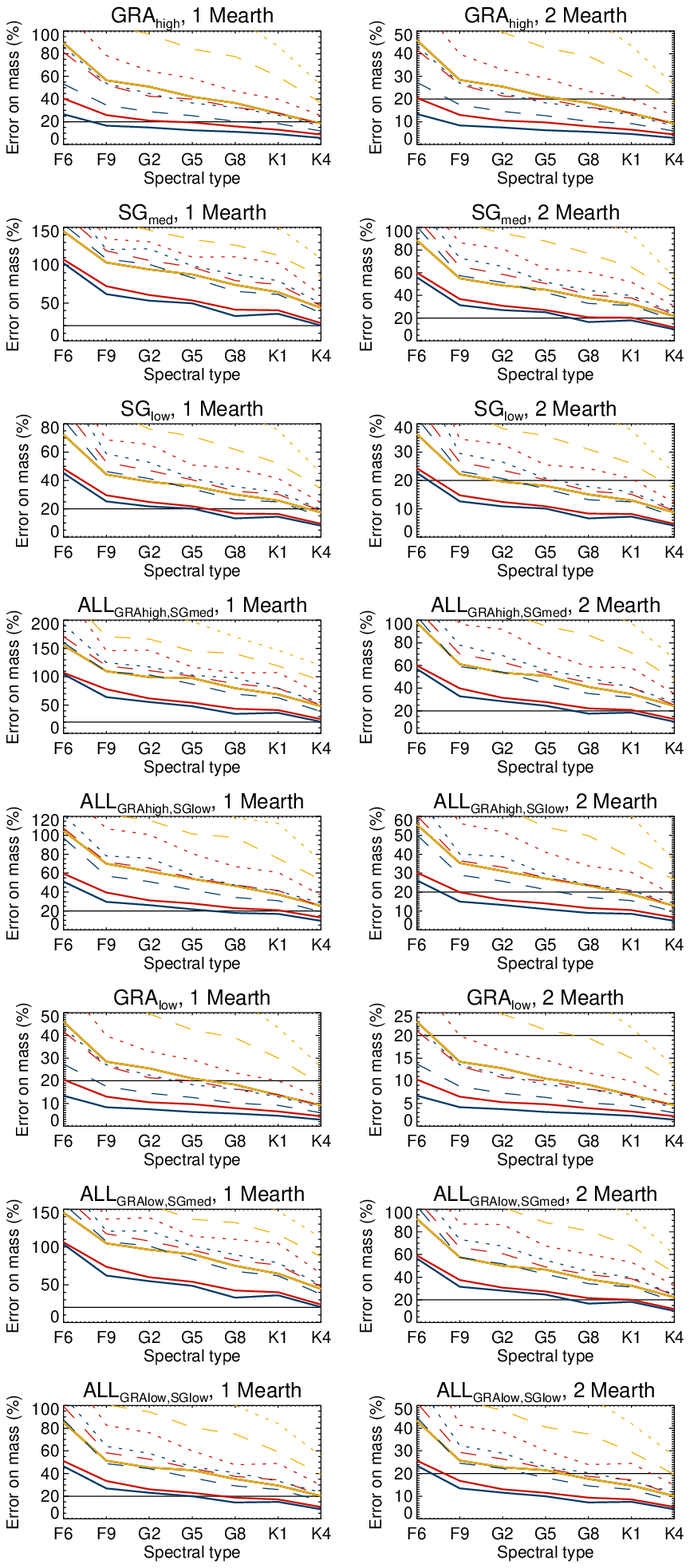}
\caption{
Uncertainty on mass in transit follow-up vs. spectral type for two planet masses (1 M$_{\rm Earth}$ on left hand-side and 2 M$_{\rm Earth}$ on  right hand-side), for PHZ$_{\rm med}$, and for different values of $N_{\rm obs}$: 180 points (yellow), 1266 points (brown), and 2352 points (pink).  The 1$\sigma$ levels are shown as solid lines, the 2$\sigma$ levels as dashed lines, and the 3$\sigma$ levels as dotted lines. 
The black horizontal line shows the 20\% level for reference. The 2-$\sigma$ and 3-$\sigma$ uncertainties are in some cases out of scale  for clarity.}
\label{errm}
\end{figure}

\begin{figure}
\includegraphics{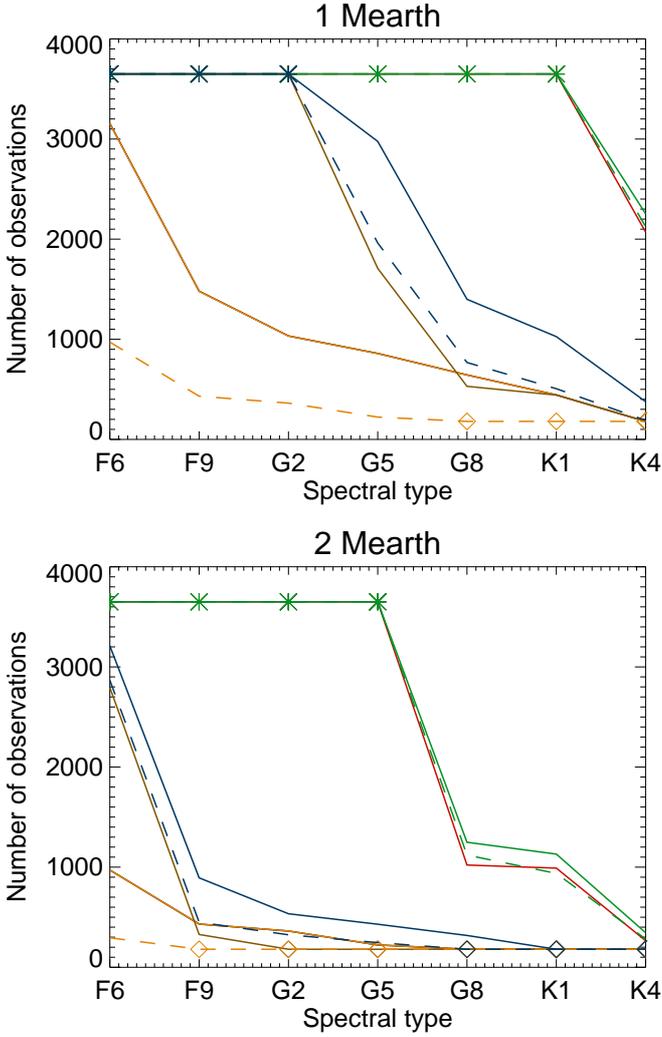}
\caption{
Number of points necessary for 20\% uncertainties on mass characterisation, for 1 M$_{\rm Earth}$ (upper panel) and 2 M$_{\rm Earth}$  (lower panel) and different OGS contributions: 
GRAhigh (orange), SGmed (red), SGlow (brown), ALL$_{\rm GRAhigh,SGmed}$ (green), and ALL$_{\rm GRAhigh,SGlow}$ (blue). The dashed lines correspond to the configurations including GRAlow (same colour code as in Fig.~\ref{rms}). Stars indicate that even with our largest number of points the uncertainties are in fact higher than 20\% (lower limit for N$_{\rm obs}$). Diamonds indicate that even with 180 points the uncertainties are in fact lower than 20\% (upper limit for N$_{\rm obs}$). 
}
\label{carac20}
\end{figure}

\begin{figure}
\includegraphics{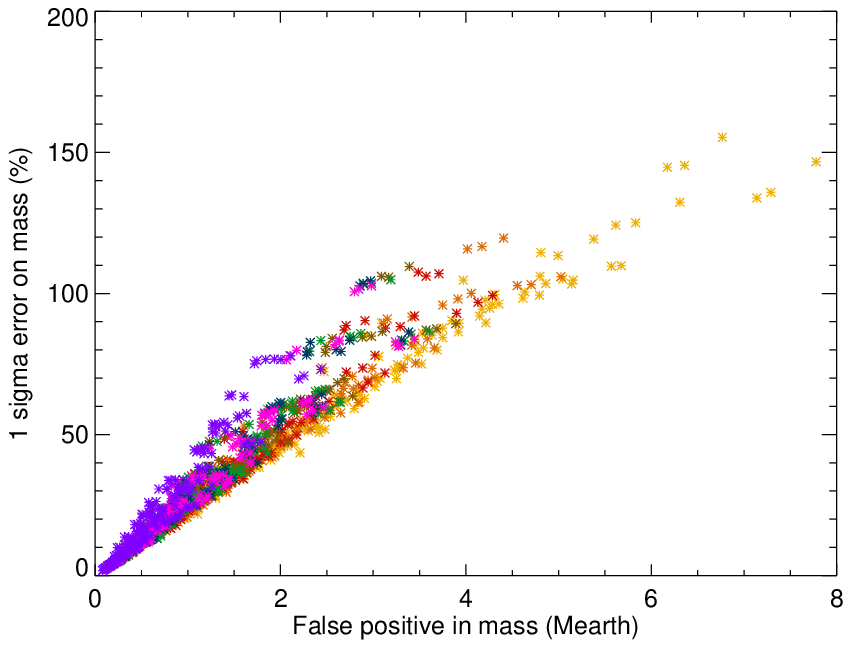}
\caption{
Uncertainty on mass characterisation at 1$\sigma$ level vs. false positive level in mass fp$_M$ for all spectral types, $N_{\rm obs}$ and different OGS configurations. The colour code corresponds to the different $N_{\rm obs}$ values:  180 (yellow), 542 (orange), 904 (red), 1266 (brown), 1628 (green), 1990 (blue), 2352 (pink), and 3650 (purple).  
}
\label{errm_fp}
\end{figure}

\subsection{LPA detection limits: exclusion rates and detection rates}

The LPA method proposed in \cite{meunier12} computes detection limits as a function of orbital period from a given RV time series, taking the power around the considered orbital period due to stellar contribution into account since stellar activity produces signal at some specific periods.
This fast computing method  has been used in several works \cite[for example][]{lagrange13,borgniet17,lannier17,lagrange18,borgniet19,grandjean19}. Here, we recall the method in brief, which is also illustrated in the upper panel of Fig.~\ref{lpa_protocole}. For a given orbital period P$_{\rm orb}$, we compute the maximum power P$_{\rm max}$ in the periodogram in a window around P$_{\rm orb}$. The detection limit is defined as the mass which would give a peak amplitude equal to 1.3$\times$P$_{\rm max}$ \cite[][]{lannier17}: we exclude the presence of planets with masses above the LPA detection limit because otherwise they would have produced a larger amplitude than observed (around that period), meaning that it is an exclusion limit. There is, however, a simplification in this computation. This is because when the planetary signal is superposed on a stellar signal, depending on its phase, the amplitude of the resulting peak can vary a great deal, as shown, for example, in Paper I. This effect is not taken into account in the LPA computation, although the 1.3 factor gives a good margin. It is  useful to estimate, for different OGS configurations (only GRAhigh is used here, i.e. five configurations of OGS), which exclusion rates such a definition corresponds to: the objective is that this rate is close to 100\% for good exclusion performance derived from this limit and as robust as possible for all configurations. 

For that purpose, we implement the following procedure, illustrated in the lower panel of Fig.~\ref{lpa_protocole}. For each spectral type and each $N_{\rm obs}$ value chosen among a subsample (180, 1266, 2353 points), we consider 100 (N1) realisations of the OGS signal and sampling. One of these realisations is shown in Fig.~\ref{lpa_protocole}. For each of these N1 realisations, the LPA detection limit $M_{\rm lpa}$ is computed for orbital periods equal to PHZ$_{\rm med}$ and we perform 100 (N2) realisations of the planetary signal of this mass $M_{\rm lpa}$  and period (i.e. N2 random phases), which is added to the corresponding OGS signal. The maximum peak in the periodogram, P', computed in the same window as above, is compared to P$_{\rm max}$(maximum power in the periodogram around the considered period) for each of these N2 realisations (the N2 values of P' are shown in the lower panel of Fig.~\ref{lpa_protocole}: the percentage of realisations (out of the N2 values) where this maximum is higher than P$_{\rm max}$ (i.e. unobserved) is the exclusion rate. In addition, the maximum peak in the periodogram can also be compared to the true false positive fp$_P$ (from Sect.~2.3), leading to a detection rate computed from the N2 realisations. For each configuration, we therefore derive 100 exclusion rates and 100 detection rates.

We find that the exclusion rate is quite constant for most spectral types (and slightly lower for F6 stars), with a median of 87\%.  Therefore, when computing the LPA limits with the above threshold, there would therefore still be a 13\% chance to miss a planet at the detection limit. The detection rates, on the other hand are rather low, typically in the 20-40\% range.
The average detection limits  are below 1 M$_{\rm Earth}$, except when the median supergranulation level is considered, in which case it is above 1 M$_{\rm Earth}$ for the most massive stars. The LPA detection limit naturally depends on the spectral type, but also  depends strongly on the number of points $N_{\rm obs}$. As a summary, Fig.~\ref{lpadist} shows the distribution of the different rates for all realisations. The exclusion rate shows a higher peak at 100\% (about a quarter of all simulations) and all are above 50\%. The detection rates are much lower, with a high peak at 0.

\begin{table}
\caption{Effect of the amplitude factor on LPA exclusion rates}
\label{tab_lpa}
\begin{center}
\renewcommand{\footnoterule}{}  
\begin{tabular}{lllll}
\hline
Factor & Mean & Median & \%  of real. & Minimum \\
  &  (\%) & (\%) &  for 100\% & (\%) \\ 
  &        &          &   excl. rate           &      \\\hline
1.3 & 83 & 81 & 21 & 56 \\
1.5 & 89 & 91 & 34 & 58 \\
1.7 & 91 & 95 & 42 & 60 \\
1.9 & 93 & 99 & 50 & 64 \\
2.1 & 94 & 100  &  58  & 64    \\
\hline
\end{tabular}
\end{center}
\tablefoot{The amplitude factor is  the ratio applied to the maximum power in the periodogram in the period range we are interested in to be compared to the planet amplitude (see text). 
}
\end{table}

Finally, for G2 stars and 1266 points,  we investigated  the effect of the chosen factor (1.3) to compute the LPA limit on the exclusion rates. The results are shown in Table~\ref{tab_lpa}. As expected, the exclusion rates are improved by a larger factor. A median exclusion rate of 99\% is reached for a factor of 1.9, for which half of the cases correspond to a 100\% exclusion rate: this would correspond to a LPA mass that is higher by 21\% (compared to the mass obtained with the 1.3 factor). We note, however, that the minimum exclusion rate increases very slowly. 

We conclude that the LPA corresponds to a good exclusion rate, although it is not 100\%. The LPA masses are also lower than the detection limits computed in the previous section.

\subsection{Mass characterisation for Earth-mass planets in the habitable zone}

Before considering the detectability issue  from the point of view of an observer, we  consider the performance in terms of mass characterisation during a transit follow-up in RV. The transit provides an excellent estimate of the orbital period and of the phase of the planetary signal (the length of the transit is extremely small compared to the orbital periods considered here). The mass of the injected planet is extremely close to the true mass (orbit seen edge-on). We consider 1000 realisations of the OGS signal (8 configurations) as defined in Sect.~2.1  and samplings for each spectral type, values of $N_{\rm obs}$ and PHZ, and add a 1 M$_{\rm Earth}$ or a 2 M$_{\rm Earth}$ planet with an arbitrary phase to each of them. Results for additional masses are shown in Appendix B. The planetary signal is then fitted (amplitude only as the period and phase are known) and from this we deduce the planet's mass. For each configuration, the 1000 values of the mass can then be compared to the input value.

For K4 stars, the mass distributions are quite narrow and are well separated between the two input masses we consider. The distributions are very good for GRAhigh and GRAlow, but when added to supergranulation (in particular, SGmed) the distributions are dominated by supergranulation. Distributions are close to Gaussian. For G5 stars, the distributions widen and for the input of 1 M$_{\rm Earth}$ and SGmed (or ALL$_{\rm GRAhigh,SGmed}$), the distributions are wide enough to include no planet, hence, there are large uncertainties on the mass. Finally, for F6 stars, the distributions are much wider and the median level of supergranulation leads to very large dispersion, (much larger than the mass). Thus, they correspond to very poor mass characterisations.

The average fitted mass is always in excellent agreement with the input mass, with no significant bias. The dispersion decreases with increasing $N_{\rm obs}$ and decreasing stellar mass. 
For example, for G2 stars and ALL$_{\rm GRAhigh,SGlow}$, at the 3$\sigma$ level, masses are between 0 and 3 M$_{\rm Earth}$ (for an input of 2 M$_{\rm Earth}$) and between 0 and 3 M$_{\rm Earth}$ (for an input of 1 M$_{\rm Earth}$) for 180 points. The ranges are reduced to 1.1-3 and 0.1-1.7 M$_{\rm Earth}$ , respectively for 1266 points, and to 1.2-2.6 and 0.2-1.6 for 2352 points. 
For K4 stars in the same conditions, the 3$\sigma$ uncertainties are already very good for 180 points (0.2-1.8 and 1.2-2.8 M$_{\rm Earth}$) and falls to 0.7-1.2 and 1.8-2.2 M$_{\rm Earth}$ for the higher number of points. 

The uncertainties on the mass are summarised in Fig.~\ref{errm}. For 1 M$_{\rm Earth}$ and GRAhigh, the uncertainties at the 1$\sigma$ level are below 35\% and except for the most massive stars, they are around 20\% or below, which are good mass estimates. With SGlow, the uncertainties are larger, but remain below a few 10\% (40\% for F6 stars with a very good sampling). They are, however, significantly higher when considering SGmed (alone or added to granulation and oscillations), and can be as high as  100\% for F6 stars and are always above 20\%.  The low level of granulation alone provides very good uncertainties: for F6 they are below 20\% for $N_{\rm obs}$ above 1266 for 1 M$_{\rm Earth}$, and for K4 they are much below 20\% even for a small $N_{\rm obs}$. Performance is still good when the low level of supergranulation is added (except for stars with spectral types earlier than G2, even for very high $N_{\rm obs}$), providing a large $N_{\rm obs}$, but again are mostly above 20\% for the median level of supergranulation is added and can reach values up to 50\% for F6 stars. 
In absolute values, the uncertainties are not very different between 1 M$_{\rm Earth}$ and 2 M$_{\rm Earth}$, so that the relative uncertainty for 2 M$_{\rm Earth}$ is about twice smaller than for 1 M$_{\rm Earth}$.
Overall, there is a significant gain in performance between 180 (very poor in general) and 1266 points, but not between 1266 and 2352 points, which does not improve the situation significantly. The dependence on $N_{\rm obs}$ is discussed in detail in Sect.~5.1. 
For practical purposes, a representation of the values of $N_{\rm obs}$ to reach a precision of 20\% on the mass is shown in Fig.~\ref{carac20}. Values are lower or upper limits in a few cases: upper limits mean that even with 180 points, uncertainties are below 20\%, so that a lower number of points are sufficient. Lower limits shown by the diamond symbols mean that even with 3650 points over ten years it is impossible to reach a 20\% uncertainties. Apart for K4, the only OGS contributions allowing to reach 20\% with $N_{\rm obs}$ within the range that we considered are granulation alone (high or low), SGlow, and combination of both.

The uncertainty on the mass estimation is strongly correlated with the true false positive level (in mass) computed in Sect.~2.3, as illustrated in Fig.~\ref{errm_fp}. When considering all spectral types, $N_{\rm obs}$ values, orbital periods, and different OGS configurations together, the correlation between the two variables is 0.96. The correlation slightly depends on the OGS configurations, with values between 0.93 and 0.99, but remains very high. There is a tendency for high values of $N_{\rm obs}$ to lead to higher uncertainties at a given false positive level (however, they naturally correspond on average to lower false positive levels). 
For example, the false positive level at 2 M$_{\rm Earth}$ corresponds to a 1$\sigma$ uncertainty between 40\% and 60\%. For 1 M$_{\rm Earth}$, it is between  20\% and 35 \%. To guarantee uncertainties below 20\%, the theoretical false positive level should be below $\sim$0.5 M$_{\rm Earth}$.


\subsection{Blind tests}

\begin{table*}
\caption{Blind tests: Configuration and results}
\label{tab_det}
\begin{center}
\renewcommand{\footnoterule}{}  
\begin{tabular}{lllllllll}
\hline
Planet & Peak & Pfit=Ptrue &  Colour & Detection of & Detection of & Detection  & Bias on \\
 configuration    & significance &  & code & true planet & false planet & status & statistics\\
\hline
No injection & & & & & & & \\
 & good & -  & red & - & yes & False positive & nb./prop.\\
 &      &     &  &  & & Detection of &  \\
 &      &    &  &  & & non-existent planet&  \\
 & bad  & -  & black & - & no& Good recovery & no bias \\
\hline
Injection  & & & & & & & \\
 & good & yes & green & yes & no& Good recovery & no bias \\
 & good & no &  brown & no & yes & Wrong planet  & prop.\\
  & bad & yes&  orange & no & no  & Rejection of & nb./prop.\\
 &      &    &  &  & & true planet &  \\
 & bad & no  & blue & no & no & Missed planet & nb./prop.\\
\hline
\end{tabular}
\end{center}
\tablefoot{The peak significance is based on the FAP (bootstrap of the signal, 1\% level). "Pfit=Ptrue" means that the difference between the two is lower than a certain threshold (see text). The colour code corresponds to Fig.~\ref{blind_pc}. The bias on planet statistics can either be on the number of detected planets or on their properties (in particular the orbital period). 
}
\end{table*}

\begin{table*}
\caption{Examples of detections and false positive rates from blind tests 
}
\label{tab_pc}
\begin{center}
\renewcommand{\footnoterule}{}  
\begin{tabular}{c|c|c|c|c|c|c|c|c}
\hline
Spectral type & \multicolumn{4}{c|}{G2} & \multicolumn{4}{c}{K4} \\ \hline
OGS & GRA & ALL  & ALL & ALL & GRA & ALL  & ALL & ALL \\ 
    &    (high)     & $_{\rm GRAhigh}$ & $_{\rm GRAhigh}$ & $_{\rm GRAlow}$  &   (high)      & $_{\rm GRAhigh}$ & $_{\rm GRAhigh}$ & $_{\rm GRAlow}$  \\
    &         & $_{\rm SGmed}$ & $_{\rm SGlow}$ & $_{\rm SGmed}$  &        & $_{\rm SGmed}$ & $_{\rm SGlow}$ & $_{\rm SGmed}$  \\
\hline
\% detection (good planet) (green) & 60& 8& 35& 15 &100& 66& 99 & 74\\
\% no detection (no injection) (black) &95 & 66& 97&43& 100 & 92& 95 & 87\\
\% bad planet detected (all) (red+brown) & 2.5 & 29.5&4 & 48 &0 &10 &3 & 12 \\
\% false positive (red) &4.6 &34 &3 & 57 &0 &8 &5 &  13 \\
\% wrong planet detected (brown) &0 &25 &5& 39  &0 &12 &1 & 10\\
\% missed+rejected planet (orange+blue) & 40& 67& 60&47 &  0& 21& 0& 16\\
\hline
\end{tabular}
\end{center}
        \tablefoot{Blind tests are performed with 1266 points for 1 M$_{\rm Earth}$. The colour code is indicated to relate the configurations to Table.~\ref{tab_det}.
        }
\end{table*}

\begin{figure}
\includegraphics{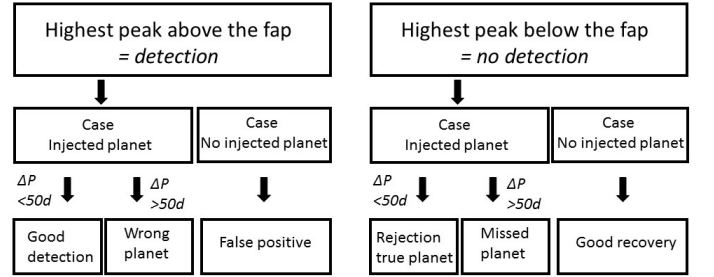}
\caption{
Decision algorithm to attribute a category (according to Table~\ref{tab_det}) to each realisation of the blind test. $\Delta$P is equal to $|P_{\rm peak}-P_{\rm true}|$. 
}
\label{blind_decision}
\end{figure}

\begin{figure}
\includegraphics{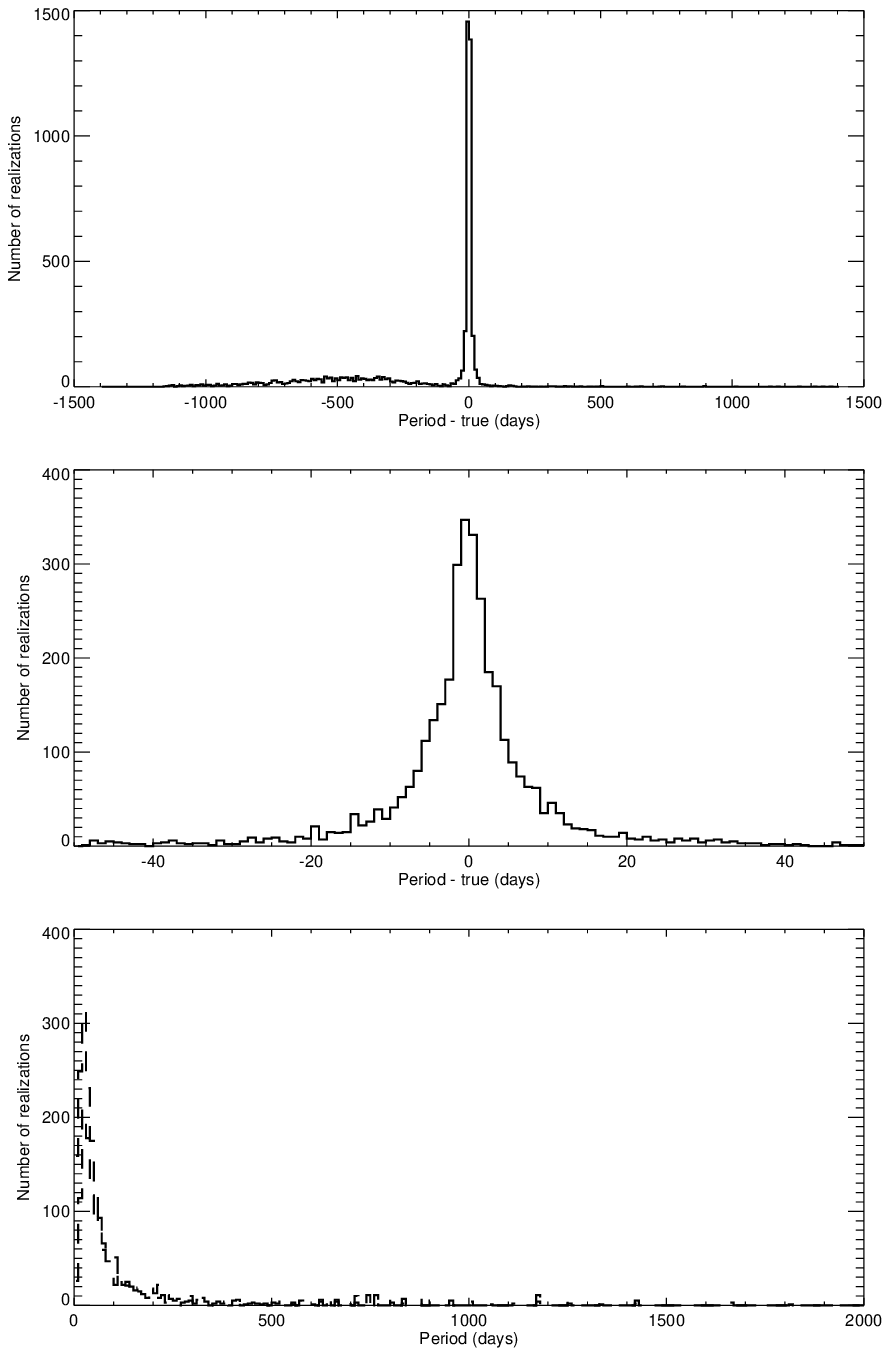}
\caption{
Distribution of the difference between the highest peak  period and true period for all OGS configurations and all realisations with injected planet in the blind test (1266 points, 1 M$_{\rm Earth}$), corresponding to a total of 5546 realisations. The middle panel is a zoom in the [-50d,50d] range. The lower panel shows the distribution of the  periods found outside the [-50d,50d] range when a planet is injected (solid line), and when no planet is injected (dashed line).}
\label{blind_dist}
\end{figure}

In this last section, we implement blind tests to estimate the level of false positives and the detection rates when applying the FAP criterion to the OGS time series in two cases: when a planet is injected or when there is no planet. We describe the principle of the blind tests, how the data sets are built and analysed, and, finally, our results. 

\subsubsection{Principle}

For each OGS signal and spectral type, statistically half of the realisations of the time series remain unchanged while a planet is added to the other half. The analysis of each time series allows to determine whether a planet is detected on not. In a second independent step, we  determine the level of false positives and the detection rate for each set of simulations, by comparing the outputs with what was actually injected or not. 
We focus our analysis on one of the $N_{\rm obs}$ values (1266 points), which corresponds to good conditions, but with still a reasonable rate of observations in future dense monitorings.

The fitting challenge implemented in \cite{dumusque17} which focusses on stellar magnetic activity defined several detectability criteria. We use similar criteria and terminology  with a few modifications: 1) We decide whether there is a detection or not using a binary choice, but since there is no further comparison with activity indicator for example, there is no intermediate case; 2) False positives are counted separately for realisations with an injected planet and with no planet; 3) The identification of the planet in \cite{dumusque17} was based on whether the amplitude K and the period P corresponded to the injected planet, while here we use only the period as a criterion, because given the dispersion in mass this criteria would be quite subjective and can be used in a second step if necessary. The different categories are summarised in Table~\ref{tab_det}. False positives and missed or wrong planets can bias statistics on exoplanets: their effect is also indicated in Table~\ref{tab_det}. We also note that the detection criteria in \cite{dumusque17} was not the same in all methods as it depended on the fitting method, and may be different from ours. 

\subsubsection{Building of the data sets}

The first step of the procedure consists of building the data sets. For each configuration (one spectral type, and one of the eight OGS configurations), we consider 200 realisations of the OGS signal and sampling. Computations are made for 1266 points and 1 M$_{\rm Earth}$ unless otherwise  noted (figures for other  values and approaches such as including a distribution in inclinations are shown  in Appendix C).  Based on a random variable, on average, half of the realisations remain unchanged, while a planet is added to the other half. The planet has the following properties: the orbital period P$_{\rm orb}$ is chosen randomly in the PHZ$_{\rm in}$-PHZ$_{\rm out}$ range (i.e. we consider the habitable zone globally, using a uniform distribution), and the phase is chosen randomly in the [0-2$\pi$] interval. The projected mass is equal to 1 M$_{\rm Earth}$ (projected mass, see discussion in Sect.~2.2.1) unless noted otherwise: these blind tests serve as  our reference. Figures corresponding to other masses are shown in Appendix C, along with  blind tests that include inclination distribution.

\subsubsection{Analysis of the time series and detectability criteria}

For each configuration (one spectral type, and one of the 8 OGS configurations), each of the 200 realisations of the time series are analysed as follows. 
The FAP at the 1\% level is computed (using 200 realisations of the bootstrap, which we checked does not give very different results from a larger number of realisations). The periodogram of the time series is  computed and the highest peak is identified (in the range 2-2000 d). If the amplitude of the peak is lower than the FAP, then we establish that there is no detection, whereas if the amplitude of the peak is higher, we consider this to be a detection. In this latter case, a sinusoid (we recall that we consider only circular orbits here) is fitted, with the period fixed to the peak period, to obtain the mass.

We note that the conditions are different from the theoretical results presented in Sect.~3. In Sect.~3, each computation was focusing on the behaviour at a given period (for example the middle of the habitable zone) and, therefore, on the power at this particular period or the mass corresponding to a fit at this period. Here, we address a different question, since we do not focus on a given period: we place ourselves at the point of view of an observer and we do not know where the planet is injected, that is, we consider the whole 2-2000 day range and not only the habitable zone. The analysis can even lead to a (wrong) detection outside of the period range where the planet is injected. This can then induce a higher rate of false positives (unless the criteria to make the detection is much higher than the true false positive level). 

In a second independent step, we compare the results  with the input parameters: this allows us to associate one of the categories of Table~\ref{tab_det} to each realisation. The decision algorithm is shown in Fig.~\ref{blind_decision}. To define whether a peak is attributable to the true planet or not,  we use the criterion  $|P_{\rm peak}-P_{\rm true}|<0.1 P_{\rm true}$  to determine if the planet is the correct one (see next section).

\subsubsection{Results of the blind tests: Planet properties and detection rates}

\begin{figure}
\includegraphics{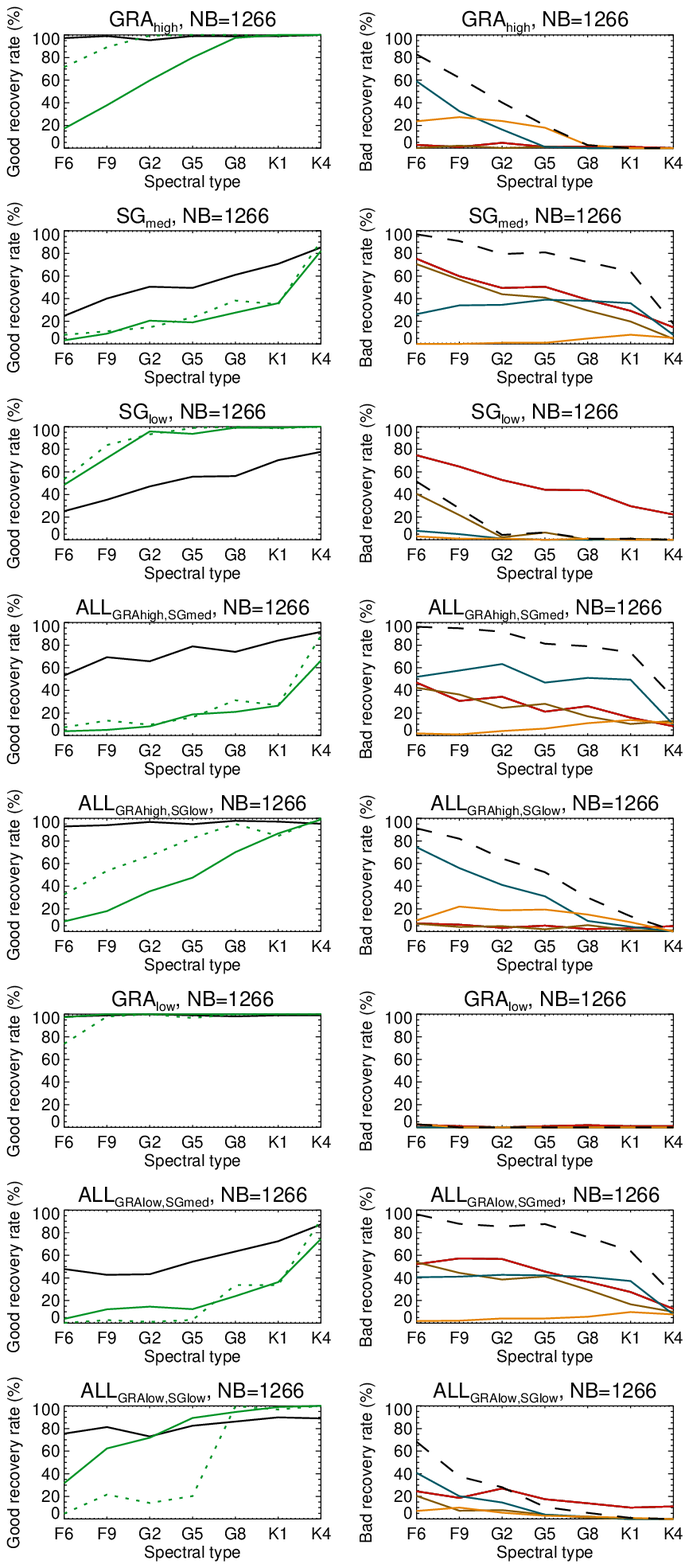}
\caption{
Good recovery percentages (left-hand side panels) and bad recovery percentages (right-hand side panels) vs. spectral type in the main blind test (1266 points, 1 M$_{\rm Earth}$). Good recoveries include no detection when no injected planet (black) and good planet recovered when injected (green). The green dotted line corresponds to the detection rate obtained in Sect.~3 with the theoretical false positive levels for the middle of the habitable zone for comparison. Bad recoveries include the false positive rate when no planet is injected (red), wrong planet detected (brown), rejection of true planet (orange), and missed planet (blue). The dashed black line is the sum of all bad recovery rates when a planet is injected (brown+orange+blue).  
}
\label{blind_pc}
\end{figure}

\begin{figure}
\includegraphics{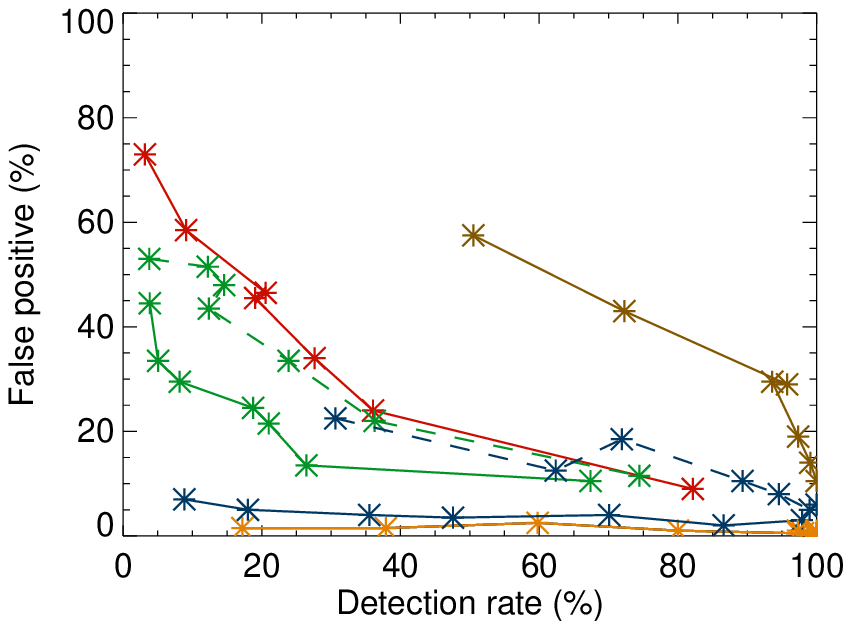}
\caption{
False positive rate vs. detection rate for each OGS configuration (GRAhigh in orange, SGmed in res, SGlow in brown, ALL$_{\rm GRAhigh,SGmed}$ in green, and ALL$_{\rm GRAhigh,SGlow}$ in blue) in main blind tests. The dashed lines correspond to configurations including GRAlow. The orange dashed line is not visible (all points in the lower right corner). }
\label{blind_roc}
\end{figure}

The outputs of each blind test are mainly the properties of the fitted planet parameters when detected and the percentages corresponding to the different categories defined in Table~\ref{tab_det}. 
We first focus on the period since a criterion on the period obtained during the analysis must be defined to assign each realisation to one of the categories. 
Figure~\ref{blind_dist} shows the distribution of the difference between the  periods provided by the analysis and the true periods over all realisations (i.e. all OGS configurations and all realisations with an injected planet), independently of the significance of the peak. 
The realisations outside this range correspond mostly to peaks found at low periods, with a maximum of the distribution in the 20-30 days range, as shown in the lower panel of Fig.~\ref{blind_dist}: 95\% of those peaks are at periods below the true orbital period and many of them are, in effect, below the FAP. In practice, the width of the peak varies with the period, and a threshold of 10\% of the period allows us to separate the peak from outliers.

Table~\ref{tab_pc} shows a few examples of percentages, for G2 and K4 stars and a subset of OGS (GRA, ALL$_{\rm GRAhigh,SGmed}$, ALL$_{\rm GRAhigh,SGlow}$). Ideally, we would like to obtain 100\% on the first two lines and 0\% on the other lines.  The categories correspond to Table~\ref{tab_det}, some of them being regrouped. For example, the percentage of bad planet detections (i.e. the global false positive rate) corresponds to planets detected when none was injected, added to the planet detected with a wrong period. For G2 stars and granulation, the recovery rate is very good when no planet is injected but lower when there is an injected planet: most of the lost planets correspond to peaks below the FAP. The performance is much poorer for ALL$_{\rm GRAhigh,SGmed}$, with very low detection rates when a planet is injected and high false positive level. Even for ALL$_{\rm GRAhigh,SGlow}$, the recovery rate when a planet is injected is only 35\%. For K4 stars, performance is perfect of granulation and very good for ALL$_{\rm GRAhigh,SGlow}$. For ALL$_{\rm GRAhigh,SGmed}$, the detection rate is only 67\%, however.

Figure~\ref{blind_pc} summarises the percentages for all configurations (1266 points, 1 M$_{\rm Earth}$). The good recovery rates are shown on the left-hand side panels. When no planet is injected (black curves), they are very good for GRAhigh and GRAlow, and above 80-90\% when added to SGlow.  They are strongly degraded in other configurations, for all spectral types (and more so for high mass stars). The detection rates when a planet is injected (green curves) are good for all stars for GRAlow only, and for K stars and sometimes G stars for GRAhigh, SGlow, ALL$_{\rm GRAhigh,SGlow}$, and ALL$_{\rm GRAlow,SGlow}$ (the threshold depends on the configuration) but strongly degraded for all other cases. It could seem surprising that the performance is better when considering ALL$_{\rm GRAhigh,SGlow}$ compared to SGlow alone (no injected planet): this is likely due to the fact that when adding the GRAhigh signal, even though the rms is increased, the power spectrum is then more similar to the GRAhigh power spectrum corresponding to good performance in the habitable zone.

The green dotted lines correspond to the detection rate obtained for the  theoretical false positive level of 1\% (Sect.~3), which is to be compared to the green solid line observed in the blind test.  The two estimations are sometimes similar, corresponding to the FP that is  very close to the FAP (Sect.~3), while in other cases, the blind test detection rates are lower than expected from the theoretical false positive level due to the difference between the FAP and the true FP. There is, therefore, a complex relationship between the theoretical results and the detection rates derived from the blind tests. We conclude that the FAP provides a detection rate which corresponds to a different false positive level from the one expected (i.e. in our case, diverging by 1\%).

\subsubsection{Results of the blind tests: False positives}

The right-hand panels in Fig.~\ref{blind_pc}  show the bad recovery rates. When a planet is injected, the bad recoveries (dashed black line) naturally serve as the complement of the green curve from the left panels. It represents a wide variety of situations: it is sometimes dominated by the missed planet (bad period and below the FAP, in blue), sometimes by the rejection of true planet (in orange); globally, that is, because peaks are below the FAP and sometimes because the highest peak is above the FAP but does not correspond to the planet (in brown, i.e. a false positive). We note that the false positive rate when a planet is injected (brown) is different from the false positive when no planet is injected (in red, completary to the black curves on the left-hand side panels) for supergranulation (especially SGlow)  alone, but it is similar when granulation and supergranulation are superposed.

For GRA, ALL$_{\rm GRAhigh,SGmed}$, and ALL$_{\rm GRAhigh,SGlow}$, the red and brown curves are similar, that is, the percentage of false positives is the same whether a planet (of 1 M$_{\rm Earth}$) is injected or not. However, the situation is different for SGmed and SGlow because when no planet is injected, the percentage of false positives is the same, even though they have very different rms RV. 
This is due to the fact that here the comparison of the power is made with the FAP. Because the shape of the power spectrum is the same between SGmed and SGlow, and because the FAP values are scaled with the rms, both power and FAP increases from SGlow to SGmed in a similar manner and the percentage of false positives is similar. In the case of ALL$_{\rm GRAhigh,SGlow}$, the signal is dominated by GRAhigh, hence, a level that is similar betwen GRAhigh and ALL$_{\rm GRAhigh,SGlow}$, while the situation is intermediate for ALL$_{\rm GRAhigh,SGmed}$.
For GRAlow, the rate of false positives is very small in all cases. However, when added to supergranulation (either SGmed or SGlow), the latter dominates, and rates are very similar to those obtained when combining with GRAhigh, only slightly lower.

The level of false positives here may be large because our analysis is too simplistic. When a peak is detected above the FAP, we should test the robustness of the detection to determine whether the peak is stable or not for example.
More sophisticated methods will have to applied in this area in the future (see Sect.~6). 

Another representation of these results is shown in Fig.~\ref{blind_roc}, showing the percentage of false positives (sum of the two contributions described above) versus the detection rate (computed on the cases with an injected planet), which is similar to a ROC curve (but where each point corresponds to a spectral type). Each curve corresponds to one of the OGS configurations. Ideally, we would like points to be in the lower right corner. Points at the top have a high false positive level and points on the left correspond to poor detection rates. If we compare the global level of false positive here and the rms for each type of OGS configuration, we see that there is not a direct correspondence, because a granulation-like signal provides better performance due to their more suitable power spectrum (for a given rms). High-mass stars are to the left of each curve and lead to high rates of false positives  and low detection rates except for granulation alone (for GRA$_{\rm low}$ all points are in the lower right corner), and to a lesser extent ALL$_{\rm GRAhigh,SGlow}$. 
We also note that the highest level of false positives is obtained for SG alone. However, when granulation is added to supergranulation, the rms increases, but the level of false positive decreases because the shape of the power spectrum is closer to the granulation shape, leading to better performance: This  explains why the level of false positive is higher when SGmed and GRAlow are superimposed (dashed green curve) compared to SGmed and GRAhigh (solid green curve), that is, closer to the SG behavior (large false positive rates) even though the rms is lower.

\subsubsection{Additional configurations}

\begin{figure}
\includegraphics{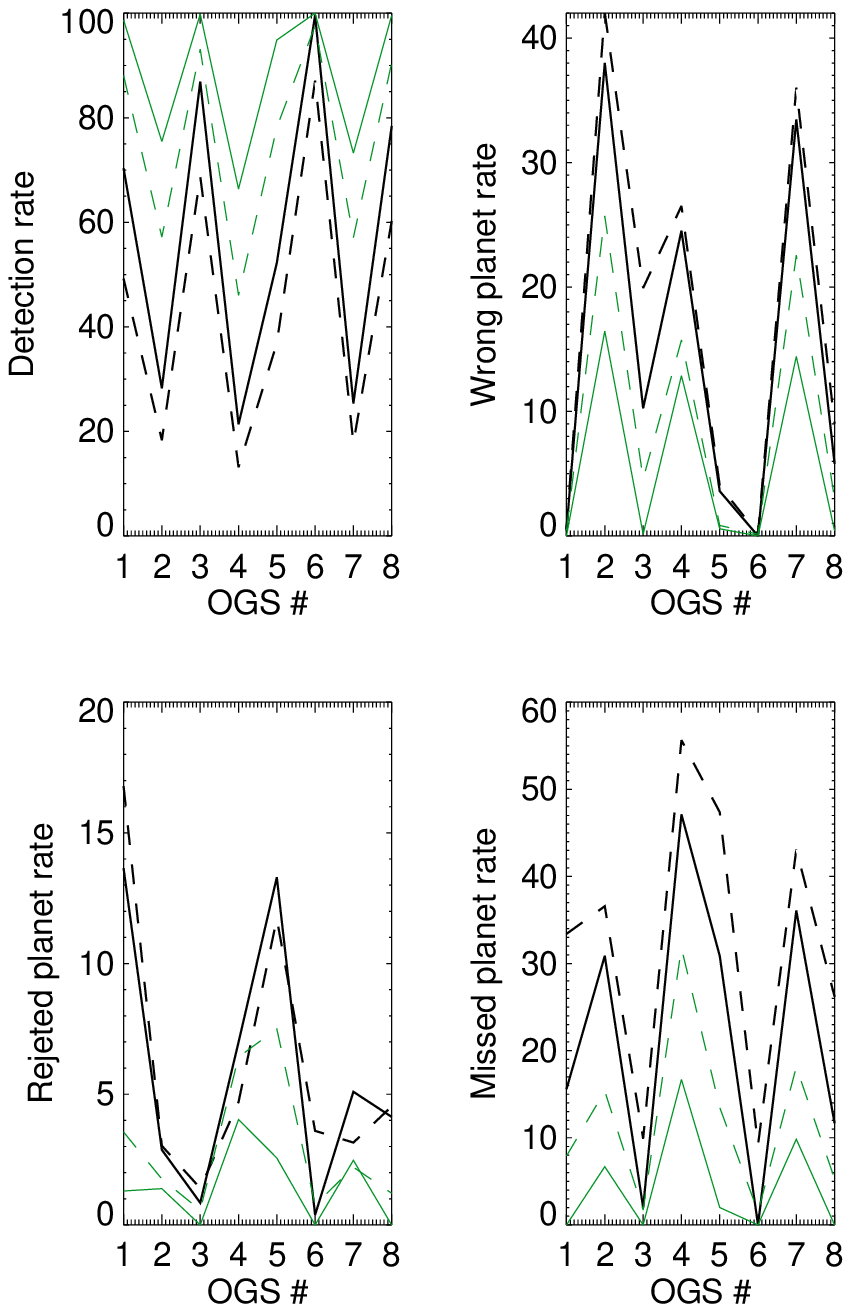}
\caption{
Comparison of average rates for 1 M$_{\rm Earth}$ (black) and 2 M$_{\rm Earth}$ (green), and without taking projection into account (solid lines, the mass is the apparent mass) and taking inclination into account (dashed lines, the mass is the true mass). The number associated to each OGS configuration corresponds to the order of the plots in Fig.~\ref{blind_pc} (from top to bottom, 
 i.e. GRA$_{\rm high}$ is number 1, SG$_{\rm med}$ is number 2 and so on). The detection rate plot corresponds to the green curves in the left panels in Fig.~\ref{blind_pc}, the wrong planet rate plot to the brown curves, the rejected planet rate plot to the orange curves, and the missed planet rate plot to the blue curves in the right panels in Fig.~\ref{blind_pc}. 
}
\label{blind_incl}
\end{figure}

Additional configurations are tested in Appendix C.1 (180 points only) and C.2 (2 M$_{\rm Earth}$). The performance for 180 points is very poor. The level of false positives is quite low, which can be explained by the results shown in Sect.~4.1: here, the FAP overestimates the true false positive level and, therefore, there are few peaks above the FAP. The detection rates are very low, however. On the other hand, the performance is much better for a 2 M$_{\rm Earth}$ planet compared to a 1 M$_{\rm Earth}$, although it is not perfect in all cases: for F and early G stars, the detection rates reach values below 50\% when supergranulation is high. 

We also implemented a similar blind test, but in which 1~M$_{\rm Earth}$  or 2 M$_{\rm Earth}$ are the true planet mass. We assume that the orbital plane is similar to the equatorial plane and take the distribution of stellar inclination into account.  We expect slightly lower detection rates than before (for cases with injected planets), which is indeed observed as shown in Appendix C.3 and C.4. Figure~\ref{blind_incl} shows the average of the rates over all spectral types for each OGS configuration, without taking inclination into account (previous results) and, conversely, taking it into account. The detection rates are slightly lower when considering inclination (i.e. the true mass), typically by a different of about 12-13 points on the percentage. The difference is mostly due to the larger amount of missed planet when the mass is the projected mass only.

\subsubsection{Corresponding LPA limits}

Finally, we compute the LPA detection limits  (see Sect.~4.2 for the definition): with an injected planet with a mass of 1 M$_{\rm Earth}$, we want the LPA detection limit (M$_{\rm lpa}$) to be higher than 1 M$_{\rm Earth}$. We compute ten values of M$_{\rm lpa}$ over the habitable zone, which are then averaged together for each spectral type. 
The average M$_{\rm lpa}$ and the percentage of realisations where M$_{\rm lpa}$  is higher than 1 M$_{\rm Earth}$. In all cases, M$_{\rm lpa}$ is indeed above 1 M$_{\rm Earth}$, and the percentage above 70\%, which is in agreement with expectation. 
When no planet is injected on the other hand, we want M$_{\rm lpa}$ to be as low as possible. For SGmed and ALL$_{\rm GRAhigh,SGmed}$, they are above 1 M$_{\rm Earth}$ for F6-G8 stars, so that in those cases, the exclusion of the presence of low mass planet (below 1 M$_{\rm Earth}$) is not possible. This is strongly related to the performance in terms of detection rates described above. For all other configurations (OGS, spectral types), they are always below 1 M$_{\rm Earth}$. 
We conclude that the LPA provides results  that are consistent with the presence of the injected planet.

\subsubsection{Comparison of the detection rates with the K/N criterion}

\begin{figure}
\includegraphics{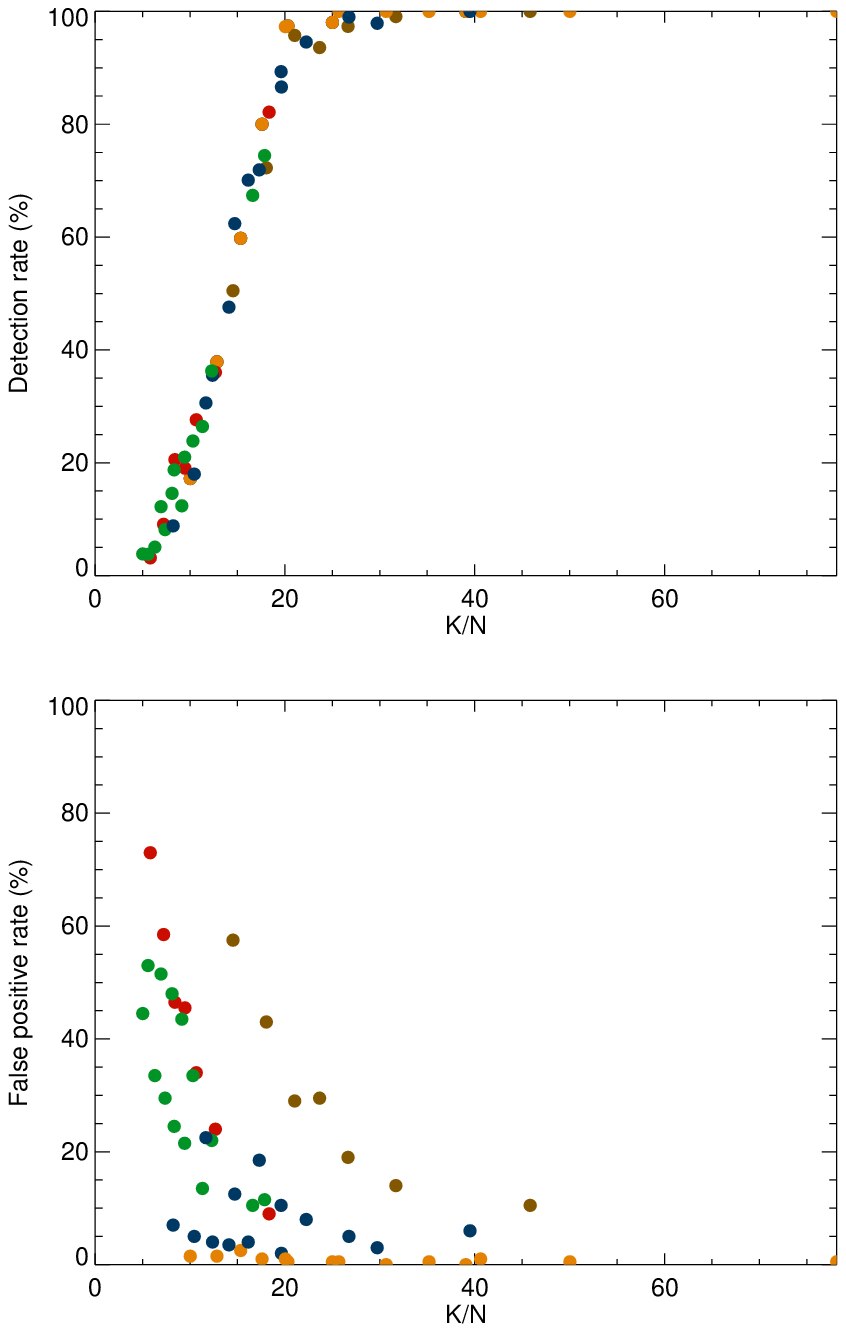}
\caption{
Detection rate (planet injected, upper panel) and  false positive rate  (lower panel) from blind test vs. K/N criterion for each OGS configuration (GRAhigh and GRAlow in orange, SGmed in res, SGlow in brown, ALL$_{\rm GRAhigh,SGmed}$ and ALL$_{\rm GRAlow,SGmed}$  in green, and ALL$_{\rm GRAhigh,SGlow}$ and ALL$_{\rm GRAlow,SGlow}$  in blue), for 1266 points, 1 M$_{\rm Earth}$. 
}
\label{kn}
\end{figure}

In this section, we compute the K/N criterion proposed in \cite{dumusque17} and defined as $K_{\rm pl}\sqrt{N_{\rm obs}}$/RV$_{\rm rms}$, where $K_{\rm pl}$  is the amplitude of a planetary signal in RV (for a given mass, period, host star), $N_{\rm obs}$ is the number of observation, and RV$_{\rm rms}$ is the RV jitter\footnote{In the original definition in \cite{dumusque17}, the RV jitter is computed after correction with a linear correlation with chromospheric emission index and trend in time, which is not done here because it is irrelevant to the type of signal we consider.}. K/N is used by \cite{dumusque17} as a  criterion for estimate the quality of recovery rates. Therefore, we compute this practical criterion for a 1 M$_{\rm Earth}$ planet and compare it to the detection rates obtained previously for the same planet mass (cases with injected planet). The results are shown in Fig.~\ref{kn}. We find a very clear relationship between the two: all OGS configurations and spectral types lie along the same curve with very little dispersion, so the criterion is adequate to describe the detection rate in these conditions. Detection rates better than 50\% correspond to K/N above $\sim$7, and K/N must be above $\sim$9 to reach detection rates better than 95\%. This is very similar  to  the rough threshold between bad recoveries and good recoveries of $\sim$7.5 in \cite{dumusque17}, who focused on magnetic activity. On the other hand, there is not a one-to-one relationship between this criterion and the false positives, as the different OGS configurations correspond to different levels, as shown on the lower panel of Fig.~\ref{kn}.

Although the curve for a given number of points, $N_{\rm obs}$ , and the mass are well-defined, it is, in fact, very dependent on the configurations. For example, for a lower number of points (see Fig.~\ref{kn_N180} in Appendix C.1 for 180 points)  the curve is very different: the curve is also well-defined, but for a similar K/N, the detection rates are lower than for 1266 points. The same is observed for 2 M$_{\rm Earth}$, with the 50\% level reached at lower K/N values compared to a 1 M$_{\rm Earth}$ planet. Thus, the criterion is not universal. We  then consider the performance as a function of the number of points in Sect. 5.

\section{Effect of the sampling}

In this last section, we focus on the effects of the sampling. We first summarise the dependence of the performance obtained in Sects. 3 and 4 on N$_{\rm obs}$. Then we test the effect of the sampling in a limited amount of cases: regular sampling instead of random, with a duration limited to three years instead of ten years, and including data binning.

\subsection{Summary of the effect of the number of points}

\begin{figure}
\includegraphics{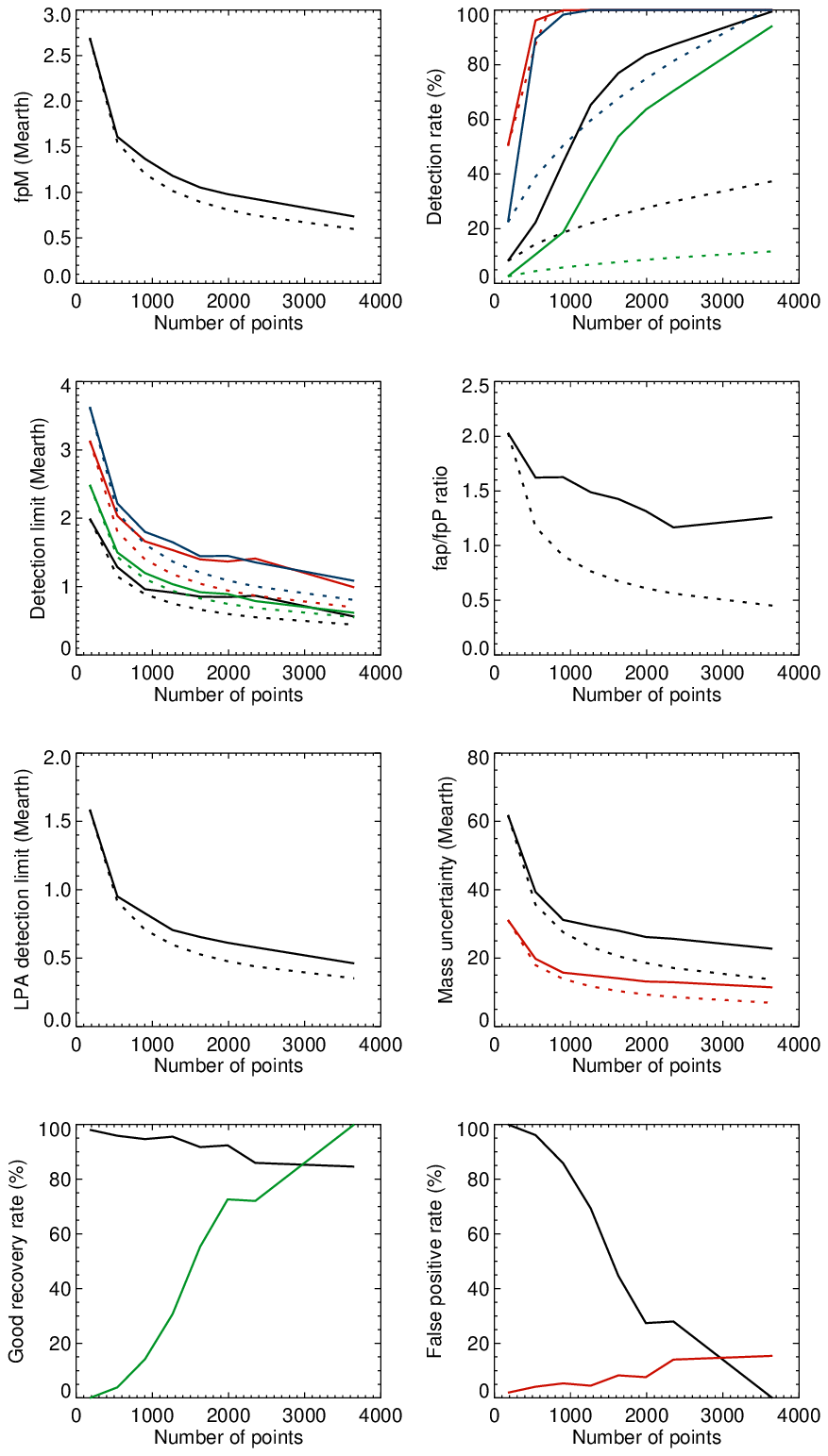}
\caption{
Effect of $N_{\rm obs}$ on performance studied in Sect.~2, 3, and 4 for G2 stars, PHZ$_{\rm med}$ and ALL$_{\rm GRAhigh,SGlow}$. The different panels represent: fp$_M$ from Sect.~2.3; detection rates using the true false positive level in power (black line) and mass (red line) from Sect.~3.1;  true detection limits in power (black line for 50\% detection rate, red line for 95\% detection rate) and in mass (green line for 50\% detection rate, blue line for 95\% detection rate) from Sect.~3.2; fap/fp$_P$ from Sect.~4.1; Average LPA detection limit from Sect.~4.2;  1$\sigma$ uncertainty on the mass characterisation from Sect.~4.3 (black for 1 M$_{\rm Earth}$ and red for 2 M$_{\rm Earth}$); Detection rate from the blind test in Sect.~4.4 with planet injected (green) and good recovery when no planet is injected (black); False poitives when a planet is injected (dashed black line) and no planet is injected (red) from the same blind tests. 
The dotted lines correspond to what would be obtained if the variability was following a $N_{\rm obs}^{-0.5}$ law ($N_{\rm obs}^{0.5}$ in the case of the detection rate), scaled to the values at 180 days.
}
\label{nbp}
\end{figure}

Figure~\ref{nbp} summarises the performance obtained in the previous sections for G2 stars, PHZ$_{\rm med}$ , and ALL$_{\rm GRAhigh,SGlow}$ versus the number of points. 
Below 500 points, curves obtained with the  theoretical false positive threshold in mass, detection limits, and mass characterisation are not very different from a 1/$\sqrt{N_{\rm obs}}$ dependence. However, above 500 points (and for all values for the fap/fp ratio), they decrease more slowly than the 1/$\sqrt{N_{\rm obs}}$ law. This is, therefore, important to optimise the observing time. The uncertainty on the mass appears, for example, to be saturating at high N$_{\rm obs}$. On the other hand, detection limits (upper right panel) vary strongly  with $N_{\rm obs}$ and do not follow a /$\sqrt{N_{\rm obs}}$ law. The same is true for the detection rates in the blind tests. Increasing the number of points may also increase the level of false positives however (when no planet is injected).

\subsection{Regular vs. random sampling}

In previous sections, we consider a random sampling during the period of observations. We now consider  the effect of this choice by testing the performance of a regular sampling in a few cases (G2 and K4 stars) for the blind test and over all spectral types for the mass characterisation. This test is done as in Sect.~4.4, that is, with 1266 points over ten years, and GRAhigh. We find that the mass uncertainties are extremely similar to what is obtained with the random sampling. The blind tests show that the detection rates when a planet is injected are also very similar, the random sampling providing slightly better detection rates. However, when no planet is injected, the regular sampling provides better false positive rates for certain OGS signals (SG alone and ALL$_{\rm GRAhigh,SGmed}$ ) while they are very similar for GRAhigh alone and ALL$_{\rm GRAhigh,SGlow}$. We conclude that in the future, depending on the observational constrains and type of signals, the two types of sampling must be tested to decide which one provides the best performance.
 
\subsection{Temporal coverage}

\begin{figure}
\includegraphics{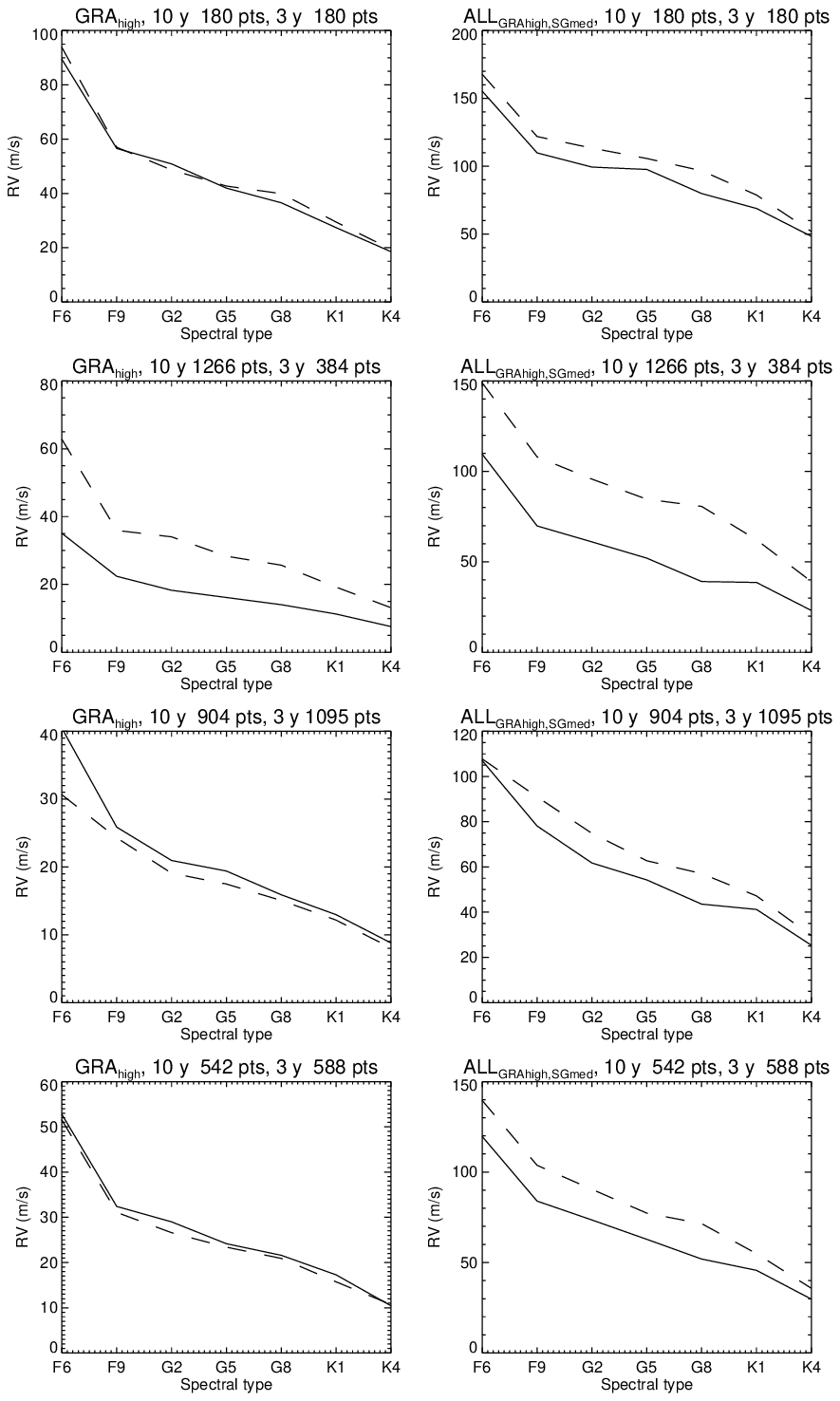}
\caption{
Uncertainty on mass for GRAhigh (left-hand side panels) and ALL$_{\rm GRAhigh,SGmed}$ (righ-hand side panels) vs. spectral type comparing ten-year coverage (solid lines) and three-year coverage (dashed line) for different values of $N_{\rm obs}$ (from top to bottom), and for PHZ$_{\rm med}$. 
}
\label{cov3y}
\end{figure}

In this work, we observed that high values of $N_{\rm obs}$ were necessary for obtaining good performance and we tested only across a long duration (ten years). In this section, we estimate the performance in a few cases if only three years of data are available, both on the blind tests (detectability) and mass characterisation. 
We keep the four-month gap every year (except for the highest value of $N_{\rm obs}$, 1095) and consider the following number of points with this gap: 180 (to be compared with the same number of points spread over 10 years), 284, 384 (to be compared with a $N_{\rm obs}$  of 1266 in the previous simulations because it corresponds to the same density of points), 486, 588, and 690. We consider all spectral types. The figures are shown in Appendix B.1 and C.5.  
Figure~\ref{cov3y} shows a comparison in mass uncertainty between a few ten-year and three-year coverage configurations for GRAhigh and ALL$_{\rm GRAhigh,SGlow}$. For 180 points for both coverages, the performance is similar for GRAhigh but worse when supergranulation is added for the three-year coverage compared to the ten-year period. When N$_{\rm obs}$ increases the differences remain when supergranulation is added. The same behavior is observed for 542 points over ten years and 588 over three years (with a similar number of points). It is, for example, more efficient to obtain 904 observations over ten years than 1095 over three years in this case. We conclude that for granulation alone, the temporal  coverage is not a critical choice, but longer time series provides better performance when considering supergranulation. Figure~\ref{carac20_3y} also shows the number of points necessary to reach a 20\% uncertainty on the mass: In most cases, when supergranulation is included, 1095 is a lower limit, that is, it is not possible to reach such a level for a 1 M$_{\rm Earth}$ planet; saturation is present only for SGmed for 2 M$_{\rm Earth}$. With granulation alone, it is possible to reach 20\% for a 1 M$_{\rm Earth}$ planet in most cases. 
The blind tests were carried out for 384 observations over the three years. Compared to the 1266 points over 10 years, the detection rates are significantly lower, although the false positive rates are not much affected. The relationship between K/N and the detection rate is also shifted compared to Sect.~4.4.

\subsection{Temporal binning}

We compare the performance after binning the time series using 30-day bins with the preceding  results. The objective is mostly to test whether binning the signal over several days to average out supergranulation is efficient. Since we are interested in long orbital periods, such a binning should not a priori affect the planetary signal very much. The protocol is otherwise similar to the one described in Sect. 4.3 for the mass characterisation in transit follow-up and in Sect.~4.4 for the blind tests (1 M$_{\rm Earth}$, 1266 points). The figures are shown in Appendix B.2 and C.6. The mass characterisation is not improved by the binning: depending on the configuration it is similar to the no-binning results or worse. The number of observations necessary to reach a precision of 20\% on the mass is higher than without binning. The blind test shows that 
when no planet is injected, performance in terms of good recovery is slightly better than with no binning. However, when a planet is injected, the performance is worse. The level of false positives is very low. We conclude that such a binning does not significantly help to improve the detectability performance.

\section{Conclusion}

In this paper, we study in detail the effect of granulation and supergranulation on Earth-mass planet mass characterisation and detectability for stars between F6 and K4 stars for different numbers of points. The two strong advantages of our approach include: the application of a large set of time series due to these flows and a systematic analysis of their impact and performance in terms of false positive, detection rates, detection limits, and mass characterisation. This work is based on several assumptions, which we recall here: 1) The shape of the power spectrum is similar to what we found in \cite{meunier15}, although we test different granulation and supergranulation levels (the power at long orbital period depending on the rms of the signal and the timescale, which is fixed here), and the supergranulation amplitude versus spectral type follows the granulation dependence on spectral type; 2) We do not add any other signal (magnetic activity, instrumental, photon noise...) except for planets; 3) We focus on a long orbital period in the habitable zone around these stars; 4) No correction technique is applied except for the one-hour binning and the test involving a 30-day binning. 

Our main conclusions, noted here and detailed below, are: 1) Both granulation and supergranulation affect the detection rates and the false positive levels, but supergranulation plays the main role; 2) Different tools give different results because they are based on different assumptions (mainly on the false positive definition) and should be used with caution (e.g. FAP computed from a bootstrap analysis).

Our results can be summarised as follows. The presence of granulation and supergranulation affects mass characterisation in RV when performing a follow-up of a transit detection. The uncertainties on these masses are sometimes below 20\% for a 1 M$_{\rm Earth}$ (mostly for granulation alone or for low mass stars), but they are much larger in certain configurations (supergranulation, high-mass stars). This contribution is, therefore, important to consider when performing mass characterisations. 

We estimated detection rates and detection limits corresponding to a good detection rate using theoretical levels of false positive (i.e. assuming a perfect knowledge of the signal). Aside from when the temporal window is not very good (for example period close to the one-year period), the frequential analysis (periodogram analysis) leads to better detection rates than the temporal analysis (fit of the planetary signal). The performance is poor for a large fraction of our configurations, and always requires a large amount of points. Granulation alone or added to low levels of supergranulation leads to good detection rates (although a very high number of points is required for F stars), but the performance is very poor for the median level of supergranulation.

When adopting the point of view of an observer (i.e. without knowing whether any other contribution than the stellar signal is present), we found that the FAP (obtained with a standard bootstrap anaysis of the observed time series) does not provide the true false positive level: apart from GRA and SGlow (always an overestimation of the true level), they   overestimate the true level for a low number of points (meaning a conservative detection) and underestimate it when the number of points was large (with the risk of false positives). Current surveys are in the regime of a low number of points (the FAP estimate is, therefore, conservative), but future observations using a large N$_{\rm obs}$ to improve the detection rates are likely to be more sensitive to an underestimation of the FAP. Here, we characterise the exclusion rates associated to the LPA detection limits \cite[][]{meunier12} when applied to this type of signal, showing that the threshold used in previous works corresponds to a median exclusion rate of 83 \% (masses  should be increased by about 20\% to correspond to 99\%). This should be kept in mind when using them to compute occurrence rates.

Finally, we performed several blind tests corresponding to different conditions in terms of planet mass, number of points, and different sampling issues (binning, duration...). As for the theoretical approach, the performance both in terms of detection rates and false positives is poor for F and G stars, whereas it is good for K stars. These rates strongly depend on the number of points as well and we find that the detection rate as a function of the K/N criterion \cite[][]{dumusque17} follows a single curve for all OGS configuration for a given number of points, but not when considering different number of points: the performance fortunately increases faster than $\sqrt{N_{\rm obs}}$. 

An important result from the blind tests comes from the comparison between the detection rates and false positives in our various configurations: 
 We find that for most stars, the detection rates are well below 100\% and always associated to a high level of false positives. 
The blind tests we implemented  used a simple analysis method, that is, based only on the FAP, given that we lack 'activity' indicators for this type of signal, which is in contrast to the case dealing with magnetic activity (see below). As a consequence, to improve this performance, future works will need to concentrate on both aspects.
The scope of the present paper is focussed on estimating the performance across a wide variety of configurations but without using mitigating techniques, which have yet to be developed.  

Some approaches in the literature may help to decrease the number of false positives. Periodogram standardisation may help to better define the false positive level, as discussed, for example, by  \cite{sulis16,sulis17}. Stacked periodograms, as proposed by \cite{mortier17}, may  also aid in this purpose. However, it remains to be seen whether these methods allow us to increase the detection rate, that is, to recover missed planets (although the second one may help to a certain extent with regard to planet peaks that are not too far below  the FAP).  Improving the detection rates will, however, require the development of new methods. Gaussian processes, which may be fitting to describe this type of signal due to their flexibility, may also absorb planets at long orbital periods: this will have to be checked with similar simulations.  One difficulty arises from the fact that usual activity indicators cannot be used (e.g. the $\log R'_{HK}$). We do not expect a correlation with photometry (which is not often simultaneous with the RV, anyway) from the simulations of \cite{meunier15} due to the high stochasticity of the granulation signal and it is not present for supergranulation \cite[][]{meunier07c}. There may be a small correlation with the bisector shape variation \cite[][]{cegla19} for granulation (but its use when superposed on the bisector variations due to other processes may be limited), however, we do not expect any for supergranulation because it involves relatively large scale flows (little dependence on line depth expected) which is relatively symmetric across the disk (no strong effect as there would be e.g. for a spot crossing the disk). However, this aspect has not yet been measured nor simulated so it remains to be checked in future studies.

\begin{acknowledgements}

We thank L. Bigot and S. Sulis for useful discussions. 
This work has been funded by the ANR GIPSE ANR-14-CE33-0018.
This work was supported by the "Programme National de Physique Stellaire" (PNPS) of CNRS/INSU co-funded by CEA and CNES.
This work was supported by the Programme National de Plan\'etologie (PNP) of CNRS/INSU, co-funded by CNES.

\end{acknowledgements}

\bibliographystyle{aa}
\bibliography{biblio}

\begin{thebibliography}{61}
\expandafter\ifx\csname natexlab\endcsname\relax\def\natexlab#1{#1}\fi

\bibitem[{{Asplund} {et~al.}(2000){Asplund}, {Nordlund}, {Trampedach}, {Allende
  Prieto}, \& {Stein}}]{asplund00}
{Asplund}, M., {Nordlund}, {\AA}., {Trampedach}, R., {Allende Prieto}, C., \&
  {Stein}, R.~F. 2000, \aap, 359, 729

\bibitem[{{Bastien} {et~al.}(2014){Bastien}, {Stassun}, {Pepper}, {Wright},
  {Aigrain}, {Basri}, {Johnson}, {Howard}, \& {Walkowicz}}]{bastien14}
{Bastien}, F.~A., {Stassun}, K.~G., {Pepper}, J., {et~al.} 2014, \aj, 147, 29

\bibitem[{{Bedding} \& {Kjeldsen}(2003)}]{bedding03}
{Bedding}, T.~R. \& {Kjeldsen}, H. 2003, \pasa, 20, 203

\bibitem[{{Beeck} {et~al.}(2013){Beeck}, {Cameron}, {Reiners}, \&
  {Sch{\"u}ssler}}]{beeck13a}
{Beeck}, B., {Cameron}, R.~H., {Reiners}, A., \& {Sch{\"u}ssler}, M. 2013,
  \aap, 558, A48

\bibitem[{{Belkacem} {et~al.}(2013){Belkacem}, {Samadi}, {Mosser}, {Goupil}, \&
  {Ludwig}}]{belkacem13}
{Belkacem}, K., {Samadi}, R., {Mosser}, B., {Goupil}, M.-J., \& {Ludwig}, H.-G.
  2013, in Astronomical Society of the Pacific Conference Series, Vol. 479,
  Progress in Physics of the Sun and Stars: A New Era in Helio- and
  Asteroseismology, ed. H.~{Shibahashi} \& A.~E. {Lynas-Gray}, 61

\bibitem[{{Boisse} {et~al.}(2012){Boisse}, {Bonfils}, \& {Santos}}]{boisse12}
{Boisse}, I., {Bonfils}, X., \& {Santos}, N.~C. 2012, \aap, 545, A109

\bibitem[{{Borgniet} {et~al.}(2019){Borgniet}, {Lagrange}, {Meunier}, {Galland
  }, {Arnold}, {Astudillo-Defru}, {Beuzit}, {Boisse}, {Bonfils}, {Bouchy},
  {Debondt}, {Deleuil}, {Delfosse}, {Desort}, {D{\'\i}az}, {Eggenberger},
  {Ehrenreich}, {Forveille}, {H{\'e}brard}, {Loeillet}, {Lovis}, {Montagnier},
  {Moutou}, {Pepe}, {Perrier}, {Pont}, {Queloz}, {Santerne}, {Santos},
  {S{\'e}gransan}, {da Silva}, {Sivan}, {Udry}, \& {Vidal-Madjar}}]{borgniet19}
{Borgniet}, S., {Lagrange}, A.~M., {Meunier}, N., {et~al.} 2019, \aap, 621, A87

\bibitem[{{Borgniet} {et~al.}(2017){Borgniet}, {Lagrange}, {Meunier}, \&
  {Galland}}]{borgniet17}
{Borgniet}, S., {Lagrange}, A.~M., {Meunier}, N., \& {Galland}, F. 2017, \aap,
  599, A57

\bibitem[{{Borgniet} {et~al.}(2015){Borgniet}, {Meunier}, \&
  {Lagrange}}]{borgniet15}
{Borgniet}, S., {Meunier}, N., \& {Lagrange}, A.-M. 2015, \aap, 581, A133

\bibitem[{{Burt} {et~al.}(2018){Burt}, {Holden}, {Wolfgang}, \&
  {Bouma}}]{burt18}
{Burt}, J., {Holden}, B., {Wolfgang}, A., \& {Bouma}, L.~G. 2018, \aj, 156, 255

\bibitem[{{Cegla}(2019)}]{cegla19b}
{Cegla}, H. 2019, Geosciences, 9, 114

\bibitem[{{Cegla} {et~al.}(2013){Cegla}, {Shelyag}, {Watson}, \&
  {Mathioudakis}}]{cegla13}
{Cegla}, H.~M., {Shelyag}, S., {Watson}, C.~A., \& {Mathioudakis}, M. 2013,
  \apj, 763, 95

\bibitem[{{Cegla} {et~al.}(2018){Cegla}, {Watson}, {Shelyag}, {Chaplin},
  {Davies}, {Mathioudakis}, {Palumbo}, {Saar}, \& {Haywood}}]{cegla18}
{Cegla}, H.~M., {Watson}, C.~A., {Shelyag}, S., {et~al.} 2018, \apj, 866, 55

\bibitem[{{Cegla} {et~al.}(2015){Cegla}, {Watson}, {Shelyag}, \&
  {Mathioudakis}}]{cegla15}
{Cegla}, H.~M., {Watson}, C.~A., {Shelyag}, S., \& {Mathioudakis}, M. 2015, in
  Cambridge Workshop on Cool Stars, Stellar Systems, and the Sun, Vol.~18, 18th
  Cambridge Workshop on Cool Stars, Stellar Systems, and the Sun, ed. G.~T.
  {van Belle} \& H.~C. {Harris}, 567--574

\bibitem[{{Cegla} {et~al.}(2019){Cegla}, {Watson}, {Shelyag}, {Mathioudakis},
  \& {Moutari}}]{cegla19}
{Cegla}, H.~M., {Watson}, C.~A., {Shelyag}, S., {Mathioudakis}, M., \&
  {Moutari}, S. 2019, \apj, 879, 55

\bibitem[{{Chaplin} {et~al.}(2019){Chaplin}, {Cegla}, {Watson}, {Davies}, \&
  {Ball}}]{chaplin19}
{Chaplin}, W.~J., {Cegla}, H.~M., {Watson}, C.~A., {Davies}, G.~R., \& {Ball},
  W.~H. 2019, \aj, 157, 163

\bibitem[{{Collier Cameron} {et~al.}(2019){Collier Cameron}, {Mortier},
  {Phillips}, {Dumusque}, {Haywood}, {Langellier}, {Watson}, {Cegla}, {Costes},
  {Charbonneau}, {Coffinet}, {Latham}, {Lopez-Morales}, {Malavolta},
  {Maldonado}, {Micela}, {Milbourne}, {Molinari}, {Saar}, {Thompson},
  {Buchschacher}, {Cecconi}, {Cosentino}, {Ghedina}, {Glenday}, {Gonzalez},
  {Li}, {Lodi}, {Lovis}, {Pepe}, {Poretti}, {Rice}, {Sasselov}, {Sozzetti},
  {Szentgyorgyi}, {Udry}, \& {Walsworth}}]{collier19}
{Collier Cameron}, A., {Mortier}, A., {Phillips}, D., {et~al.} 2019, \mnras,
  487, 1082

\bibitem[{{Davies} {et~al.}(2014){Davies}, {Chaplin}, {Elsworth}, \&
  {Hale}}]{davies14}
{Davies}, G.~R., {Chaplin}, W.~J., {Elsworth}, Y., \& {Hale}, S.~J. 2014,
  \mnras, 441, 3009

\bibitem[{{Desort} {et~al.}(2007){Desort}, {Lagrange}, {Galland}, {Udry}, \&
  {Mayor}}]{desort07}
{Desort}, M., {Lagrange}, A.-M., {Galland}, F., {Udry}, S., \& {Mayor}, M.
  2007, \aap, 473, 983

\bibitem[{{Dumusque}(2016)}]{dumusque16}
{Dumusque}, X. 2016, \aap, 593, A5

\bibitem[{{Dumusque} {et~al.}(2017){Dumusque}, {Borsa}, {Damasso},
  {D{\'{\i}}az}, {Gregory}, {Hara}, {Hatzes}, {Rajpaul}, {Tuomi}, {Aigrain},
  {Anglada-Escud{\'e}}, {Bonomo}, {Bou{\'e}}, {Dauvergne}, {Frustagli},
  {Giacobbe}, {Haywood}, {Jones}, {Laskar}, {Pinamonti}, {Poretti}, {Rainer},
  {S{\'e}gransan}, {Sozzetti}, \& {Udry}}]{dumusque17}
{Dumusque}, X., {Borsa}, F., {Damasso}, M., {et~al.} 2017, \aap, 598, A133

\bibitem[{{Dumusque} {et~al.}(2012){Dumusque}, {Pepe}, {Lovis},
  {S{\'e}gransan}, {Sahlmann}, {Benz}, {Bouchy}, {Mayor}, {Queloz}, {Santos},
  \& {Udry}}]{dumusque12}
{Dumusque}, X., {Pepe}, F., {Lovis}, C., {et~al.} 2012, \nat, 491, 207

\bibitem[{{Dumusque} {et~al.}(2011){Dumusque}, {Udry}, {Lovis}, {Santos}, \&
  {Monteiro}}]{dumusque11b}
{Dumusque}, X., {Udry}, S., {Lovis}, C., {Santos}, N.~C., \& {Monteiro},
  M.~J.~P.~F.~G. 2011, \aap, 525, A140

\bibitem[{{Elsworth} {et~al.}(1994){Elsworth}, {Howe}, {Isaak}, {McLeod},
  {Miller}, {New}, {Speake}, \& {Wheeler}}]{elsworth94}
{Elsworth}, Y., {Howe}, R., {Isaak}, G.~R., {et~al.} 1994, \mnras, 269, 529

\bibitem[{{Grandjean} {et~al.}(2019){Grandjean}, {Lagrange}, {Keppler},
  {Meunier}, {Borgniet}, {Chauvin}, {Desidera}, {messina}, {Sterzik},
  {Pantoja}, {Rodet}, \& {Zircher}}]{grandjean19}
{Grandjean}, A., {Lagrange}, A.-M., {Keppler}, M., {et~al.} 2019, accepted in
  A\&A, 0

\bibitem[{{Harvey}(1984)}]{harvey84}
{Harvey}, J.~W. 1984, in Probing the depths of a Star: the study of Solar
  oscillation from space, ed. R. W. Noyes, \& E. J. Rhodes Jr., JPL, 400, 327

\bibitem[{{Herrero} {et~al.}(2016){Herrero}, {Ribas}, {Jordi}, {Morales},
  {Perger}, \& {Rosich}}]{herrero16}
{Herrero}, E., {Ribas}, I., {Jordi}, C., {et~al.} 2016, \aap, 586, A131

\bibitem[{{Jones} {et~al.}(2006){Jones}, {Sleep}, \& {Underwood}}]{jones06}
{Jones}, B.~W., {Sleep}, P.~N., \& {Underwood}, D.~R. 2006, \apj, 649, 1010

\bibitem[{{Kallinger} {et~al.}(2014){Kallinger}, {De Ridder}, {Hekker},
  {Mathur}, {Mosser}, {Gruberbauer}, {Garc{\'{\i}}a}, {Karoff}, \&
  {Ballot}}]{kallinger14}
{Kallinger}, T., {De Ridder}, J., {Hekker}, S., {et~al.} 2014, \aap, 570, A41

\bibitem[{{Kasting} {et~al.}(1993){Kasting}, {Whitmire}, \&
  {Reynolds}}]{kasting93}
{Kasting}, J.~F., {Whitmire}, D.~P., \& {Reynolds}, R.~T. 1993, \icarus, 101,
  108

\bibitem[{{Kippenhahn} \& {Weigert}(1990)}]{kippenhahn90}
{Kippenhahn}, R. \& {Weigert}, A. 1990, {Stellar Structure and Evolution}, 192

\bibitem[{{Kjeldsen} \& {Bedding}(1995)}]{kjeldsen95}
{Kjeldsen}, H. \& {Bedding}, T.~R. 1995, \aap, 293, 87

\bibitem[{{Lagrange} {et~al.}(2010){Lagrange}, {Desort}, \&
  {Meunier}}]{lagrange10b}
{Lagrange}, A.-M., {Desort}, M., \& {Meunier}, N. 2010, \aap, 512, A38

\bibitem[{{Lagrange} {et~al.}(2018){Lagrange}, {Keppler}, {Meunier}, {Lannier},
  {Beust}, {Milli}, {Bonnavita}, {Bonnefoy}, {Borgniet}, {Chauvin}, {Delorme},
  {Galland}, {Iglesias}, {Kiefer}, {Messina}, {Vidal-Madjar}, \&
  {Wilson}}]{lagrange18}
{Lagrange}, A.~M., {Keppler}, M., {Meunier}, N., {et~al.} 2018, \aap, 612, A108

\bibitem[{{Lagrange} {et~al.}(2013){Lagrange}, {Meunier}, {Chauvin}, {Sterzik},
  {Galland}, {Lo Curto}, {Rameau}, \& {Sosnowska}}]{lagrange13}
{Lagrange}, A.-M., {Meunier}, N., {Chauvin}, G., {et~al.} 2013, \aap, 559, A83

\bibitem[{{Lannier} {et~al.}(2017){Lannier}, {Lagrange}, {Bonavita},
  {Borgniet}, {Delorme}, {Meunier}, {Desidera}, {Messina}, {Chauvin}, \&
  {Keppler}}]{lannier17}
{Lannier}, J., {Lagrange}, A.~M., {Bonavita}, M., {et~al.} 2017, \aap, 603, A54

\bibitem[{{Makarov} {et~al.}(2010){Makarov}, {Parker}, \& {Ulrich}}]{makarov10}
{Makarov}, V.~V., {Parker}, D., \& {Ulrich}, R.~K. 2010, \apj, 717, 1202

\bibitem[{{Meunier} {et~al.}(2010{\natexlab{a}}){Meunier}, {Desort}, \&
  {Lagrange}}]{meunier10a}
{Meunier}, N., {Desort}, M., \& {Lagrange}, A.-M. 2010{\natexlab{a}}, \aap,
  512, A39

\bibitem[{{Meunier} \& {Lagrange}(2019{\natexlab{a}})}]{meunier19b}
{Meunier}, N. \& {Lagrange}, A.~M. 2019{\natexlab{a}}, \aap, 628, A125

\bibitem[{{Meunier} \& {Lagrange}(2019{\natexlab{b}})}]{meunier19e}
{Meunier}, N. \& {Lagrange}, A.~M. 2019{\natexlab{b}}, \aap, 625, L6

\bibitem[{{Meunier} \& {Lagrange}(2020)}]{meunier20c}
{Meunier}, N. \& {Lagrange}, A.-M. 2020, arXiv e-prints, arXiv:2004.10611

\bibitem[{{Meunier} {et~al.}(2015){Meunier}, {Lagrange}, {Borgniet}, \&
  {Rieutord}}]{meunier15}
{Meunier}, N., {Lagrange}, A.-M., {Borgniet}, S., \& {Rieutord}, M. 2015, \aap,
  583, A118

\bibitem[{{Meunier} {et~al.}(2019){Meunier}, {Lagrange}, {Boulet}, \&
  {Borgniet}}]{meunier19}
{Meunier}, N., {Lagrange}, A.~M., {Boulet}, T., \& {Borgniet}, S. 2019, \aap,
  627, A56

\bibitem[{Meunier {et~al.}(2019)Meunier, Lagrange, \& Cuzacq}]{meunier19c}
Meunier, N., Lagrange, A.-M., \& Cuzacq, S. 2019, \aap, 632, A81

\bibitem[{{Meunier} {et~al.}(2012){Meunier}, {Lagrange}, \& {De
  Bondt}}]{meunier12}
{Meunier}, N., {Lagrange}, A.-M., \& {De Bondt}, K. 2012, \aap, 545, A87

\bibitem[{{Meunier} {et~al.}(2010{\natexlab{b}}){Meunier}, {Lagrange}, \&
  {Desort}}]{meunier10}
{Meunier}, N., {Lagrange}, A.-M., \& {Desort}, M. 2010{\natexlab{b}}, \aap,
  519, A66

\bibitem[{{Meunier} {et~al.}(2007){Meunier}, {Tkaczuk}, \&
  {Roudier}}]{meunier07c}
{Meunier}, N., {Tkaczuk}, R., \& {Roudier}, T. 2007, \aap, 463, 745

\bibitem[{{Mortier} \& {Collier Cameron}(2017)}]{mortier17}
{Mortier}, A. \& {Collier Cameron}, A. 2017, \aap, 601, A110

\bibitem[{{Pall{\'e}} {et~al.}(1999){Pall{\'e}}, {Roca Cort{\'e}s},
  {Jim{\'e}nez}, {GOLF Team}, \& {Virgo Team}}]{palle99}
{Pall{\'e}}, P.~L., {Roca Cort{\'e}s}, T., {Jim{\'e}nez}, A., {GOLF Team}, \&
  {Virgo Team}. 1999, in Astronomical Society of the Pacific Conference Series,
  Vol. 173, Stellar Structure: Theory and Test of Connective Energy Transport,
  ed. A.~{Gimenez}, E.~F. {Guinan}, \& B.~{Montesinos}, 297

\bibitem[{{Rieutord} {et~al.}(2002){Rieutord}, {Ludwig}, {Roudier}, {Nordlund},
  \& {Stein}}]{rieutord02}
{Rieutord}, M., {Ludwig}, H.-G., {Roudier}, T., {Nordlund}, A., \& {Stein}, R.
  2002, Nuovo Cimento C Geophysics Space Physics C, 25, 523

\bibitem[{{Rieutord} {et~al.}(2000){Rieutord}, {Roudier}, {Malherbe}, \&
  {Rincon}}]{rieutord00}
{Rieutord}, M., {Roudier}, T., {Malherbe}, J.~M., \& {Rincon}, F. 2000, \aap,
  357, 1063

\bibitem[{{Roudier} {et~al.}(2016){Roudier}, {Malherbe}, {Rieutord}, \&
  {Frank}}]{roudier16}
{Roudier}, T., {Malherbe}, J.~M., {Rieutord}, M., \& {Frank}, Z. 2016, \aap,
  590, A121

\bibitem[{{Saar} \& {Donahue}(1997)}]{saar97}
{Saar}, S.~H. \& {Donahue}, R.~A. 1997, \apj, 485, 319

\bibitem[{{Samadi} {et~al.}(2007){Samadi}, {Georgobiani}, {Trampedach},
  {Goupil}, {Stein}, \& {Nordlund}}]{samadi07}
{Samadi}, R., {Georgobiani}, D., {Trampedach}, R., {et~al.} 2007, \aap, 463,
  297

\bibitem[{{Santos} {et~al.}(2015){Santos}, {Cunha}, {Avelino}, \&
  {Campante}}]{santos15}
{Santos}, A.~R.~G., {Cunha}, M.~S., {Avelino}, P.~P., \& {Campante}, T.~L.
  2015, \aap, 580, A62

\bibitem[{Sulis {et~al.}(2016)Sulis, Mary, \& Bigot}]{sulis16}
Sulis, S., Mary, D., \& Bigot, L. 2016, in 2016 IEEE International Conference
  on Acoustics, Speech and Signal Processing (ICASSP), 4428--4432

\bibitem[{{Sulis} {et~al.}(2017{\natexlab{a}}){Sulis}, {Mary}, \&
  {Bigot}}]{sulis17}
{Sulis}, S., {Mary}, D., \& {Bigot}, L. 2017{\natexlab{a}}, IEEE Transactions
  on Signal Processing, 65, 2136

\bibitem[{{Sulis} {et~al.}(2017{\natexlab{b}}){Sulis}, {Mary}, \&
  {Bigot}}]{sulis17b}
{Sulis}, S., {Mary}, D., \& {Bigot}, L. 2017{\natexlab{b}}, in Proc. 25th
  European Signal Processing Conference, 1095--1099

\bibitem[{{Sulis} {et~al.}(2020){Sulis}, {Mary}, \& {Bigot}}]{sulis20}
{Sulis}, S., {Mary}, D., \& {Bigot}, L. 2020, \aap, 635, A146

\bibitem[{{Yu} {et~al.}(2018){Yu}, {Huber}, {Bedding}, \& {Stello}}]{yu18}
{Yu}, J., {Huber}, D., {Bedding}, T.~R., \& {Stello}, D. 2018, \mnras, 480, L48

\bibitem[{{Zaninetti}(2008)}]{zaninetti08}
{Zaninetti}, L. 2008, Serbian Astronomical Journal, 177, 73

\end{thebibliography}

\begin{appendix}

\section{Typical power and examples of time series}

\begin{figure}
\includegraphics{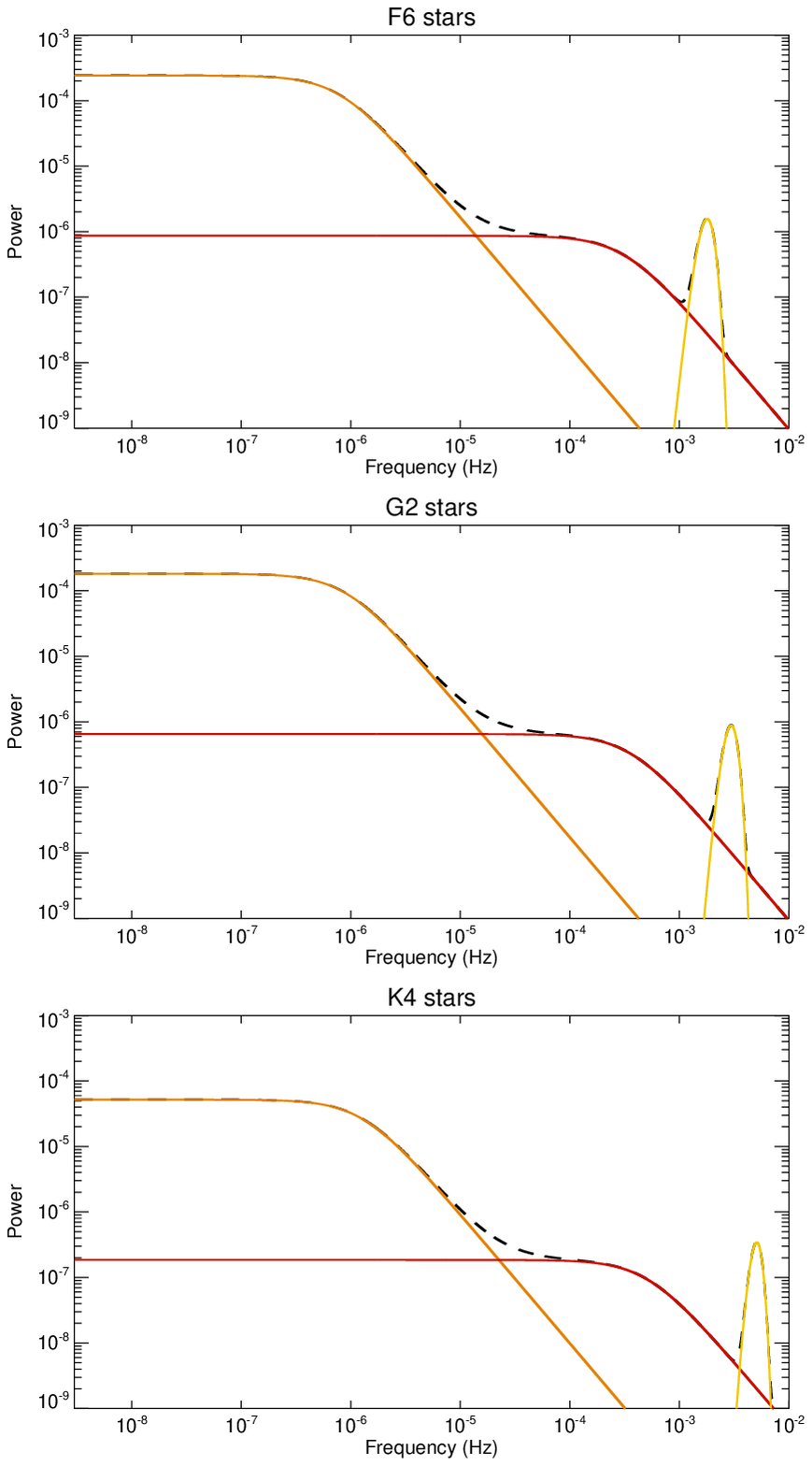}
\caption{
Example of power function used in Sect.2.1, for F6, G2, and K4 stars (from top to  bottom), for SGmed (red), for GRAhigh (orange), and oscillations (yellow). The dashed black line represents the sum of these three curves. 
}
\label{ex_power}
\end{figure}

Figure~\ref{ex_power} shows the theoretical power functions used to produce the time series analysed in this paper for GRAhigh and SGmed. The other power functions (GRAlow and/or SGlow) only differ in amplitudes. Figures~\ref{ex_G2} and \ref{ex_K4} show examples of subsets of time series (covering 8 hours, 5 days, and 50 days) for the different contributions and for G2 and K4 stars, respectively. 

\begin{figure*}
\includegraphics{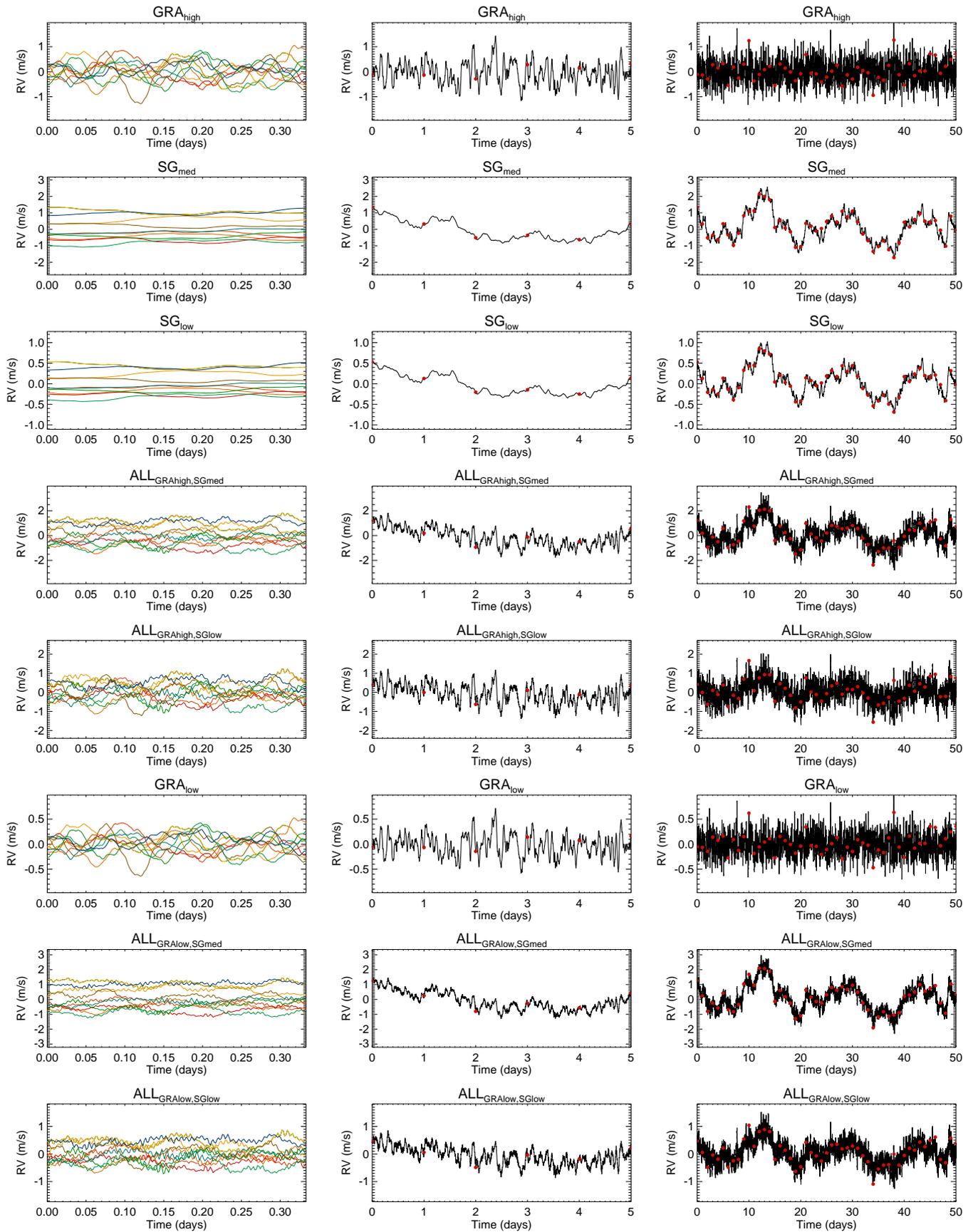}
\caption{
Extract from time series vs. time for F6 stars, different OGS configurations (from top to bottom) and for three temporal coverages: eight hours (left-hand side panels), five days (middle panels) and 50 days (right-hand side panels), smoothed over one hour. For the eight-hour coverage, ten examples corresponding to adjacent nights are superposed (each section of the RV time series is represented by a different color). For the two other coverage sets, the black line represents the full resolution time series (smoothed over one hour) and the red circles are the selected points used in the analysis (one point per day).}
\label{ex_F6}
\end{figure*}

\begin{figure*}
\includegraphics{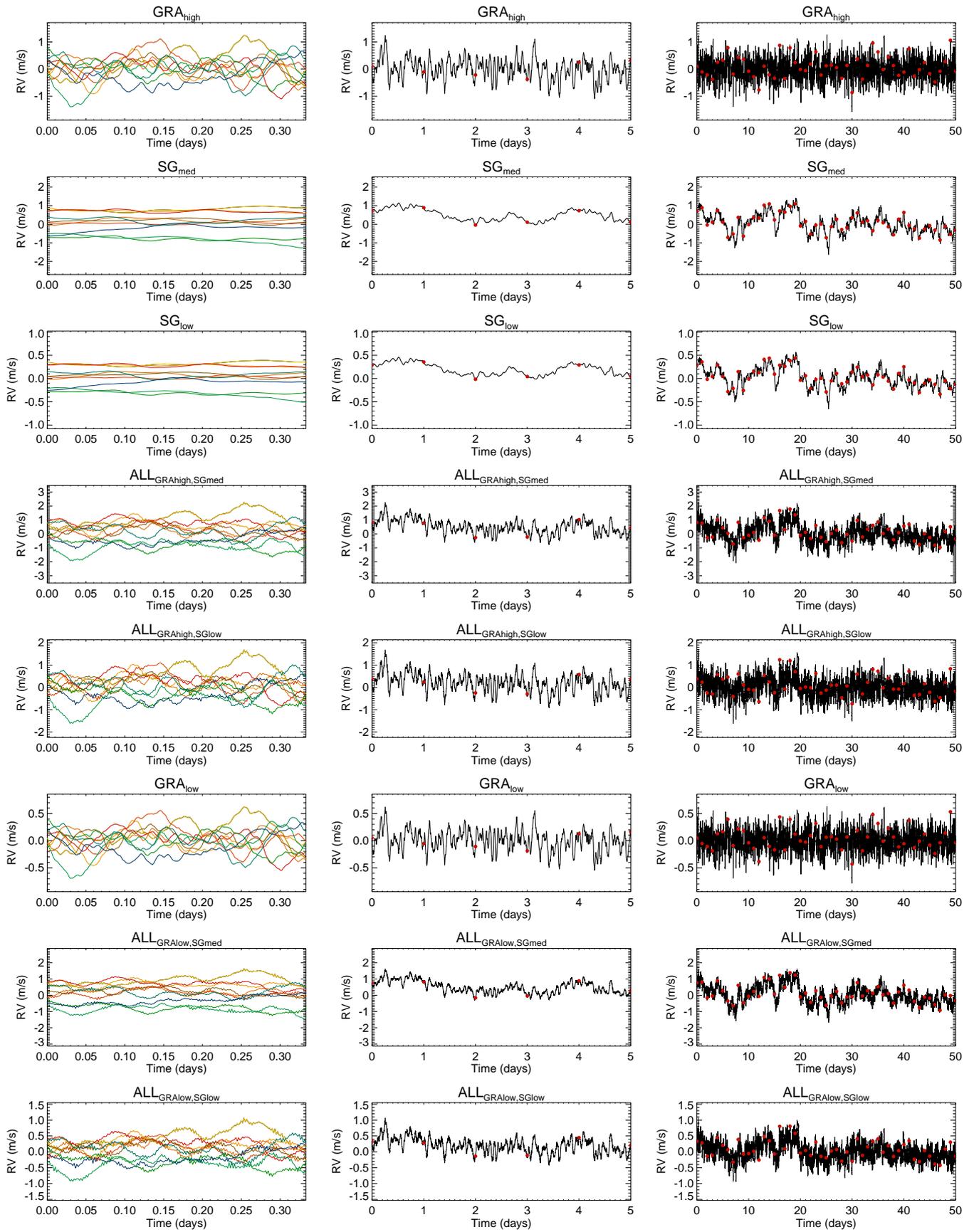}
\caption{
Same as Fig.~\ref{ex_F6} for G2 stars.  }
\label{ex_G2}
\end{figure*}

\begin{figure*}
\includegraphics{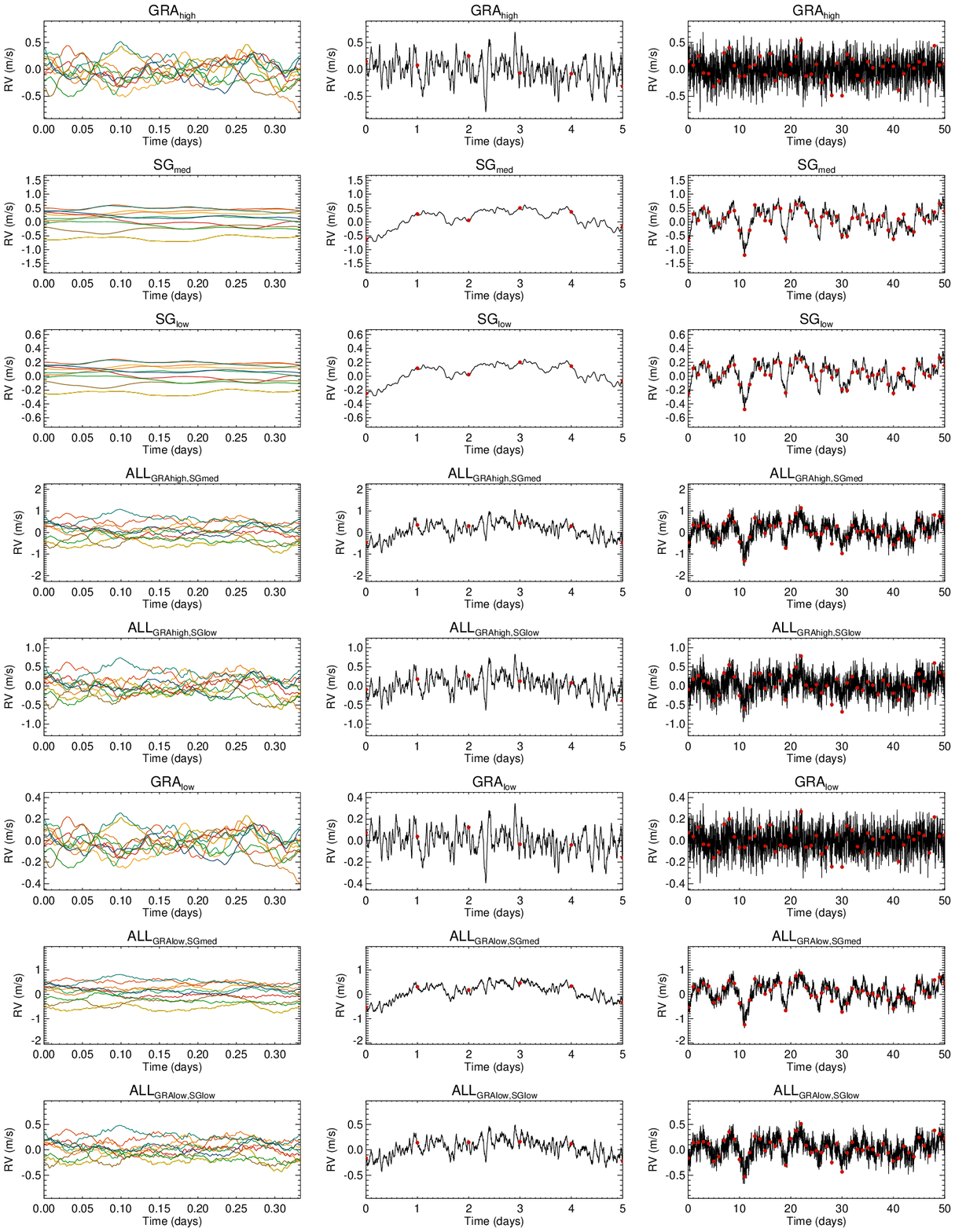}
\caption{
Same as Fig.~\ref{ex_F6} for K4 stars. }
\label{ex_K4}
\end{figure*}




\section{Additional mass characterisation results}

Additional mass characterisation results are presented here for the purposes of comparing different conditions. We explore the effect of the  sampling. 

\subsection{Case PHZ$_{\rm med}$, 1 and 2 M$_{\rm Earth}$ planets, 3 year coverage}

\begin{figure}
\includegraphics{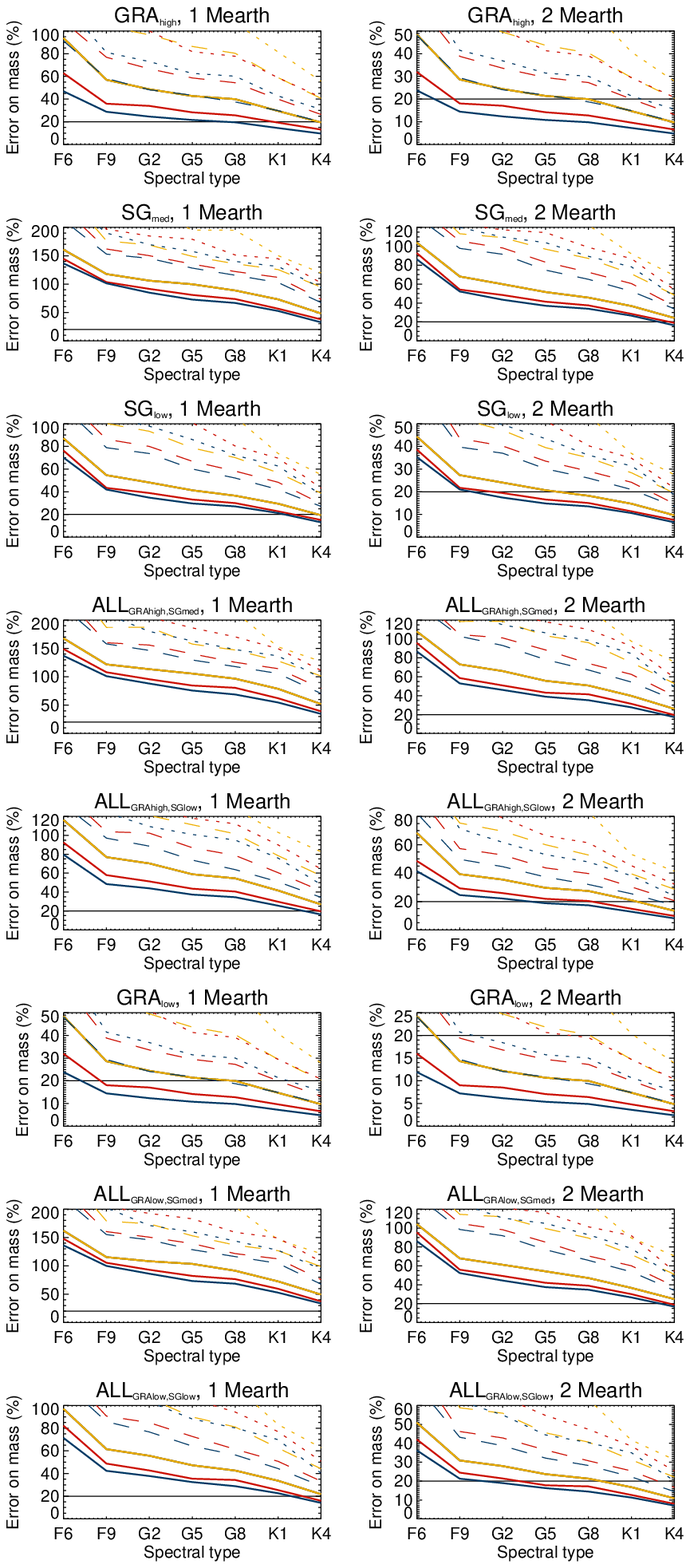}
\caption{
Same as Fig.~\ref{errm} but for a three-year coverage, and N$_{\rm obs}$ of 180, 384, and 690. 
}
\label{carac_3y}
\end{figure}

\begin{figure}
\includegraphics{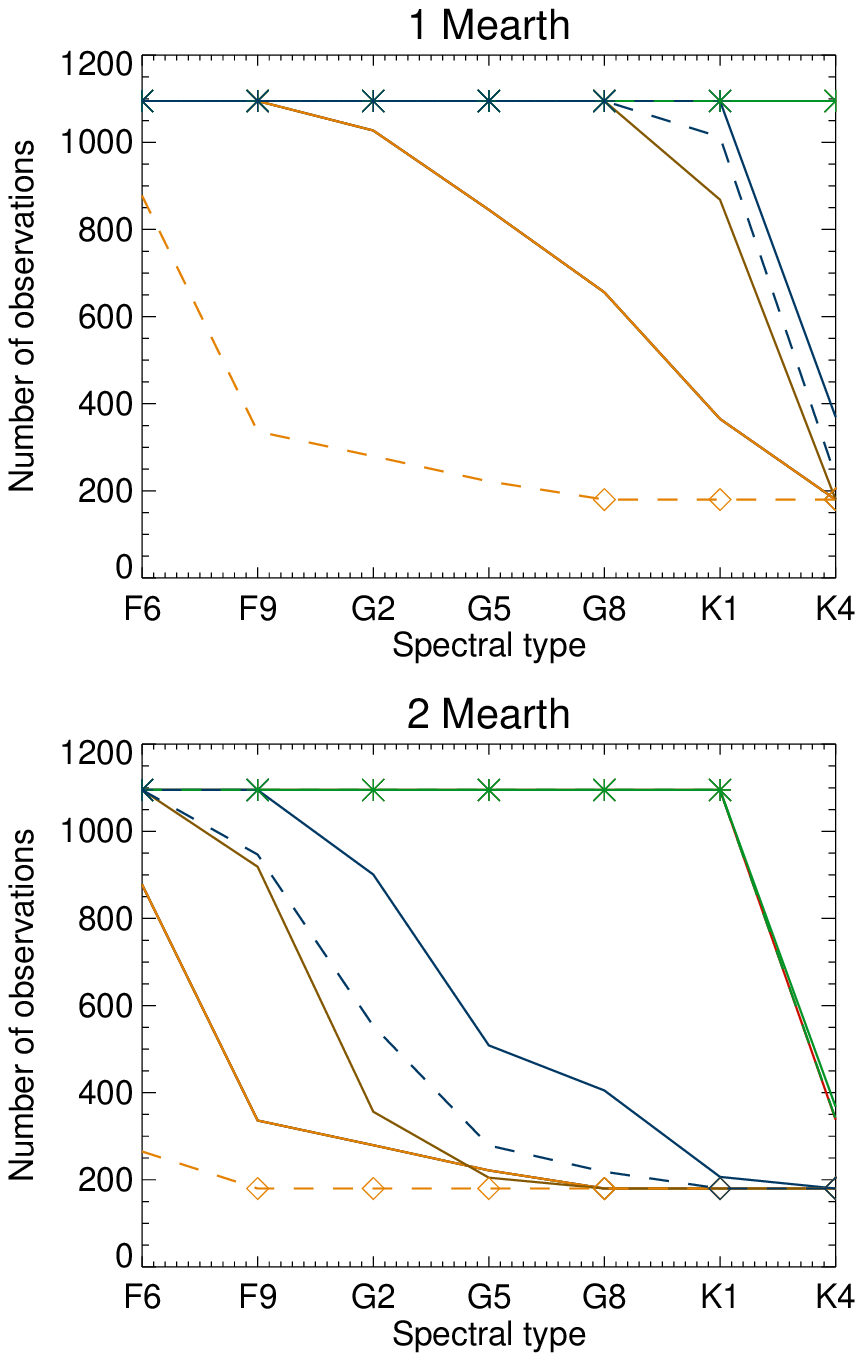}
\caption{
Same as Fig.~\ref{carac20} but for a three-year coverage. 
}
\label{carac20_3y}
\end{figure}

Figure~\ref{carac_3y} shows the uncertainty on the mass for a three-year coverage (see Sect.~5.4), and Fig.~\ref{carac20_3y} the number of observations necessary to reach an uncertainty on the mass of 20\%. The results are discussed in Sect.~5.3.

\subsection{Case PHZ$_{\rm med}$, 30 day binning, 1 and 2 M$_{\rm Earth}$ planets}

\begin{figure}
\includegraphics{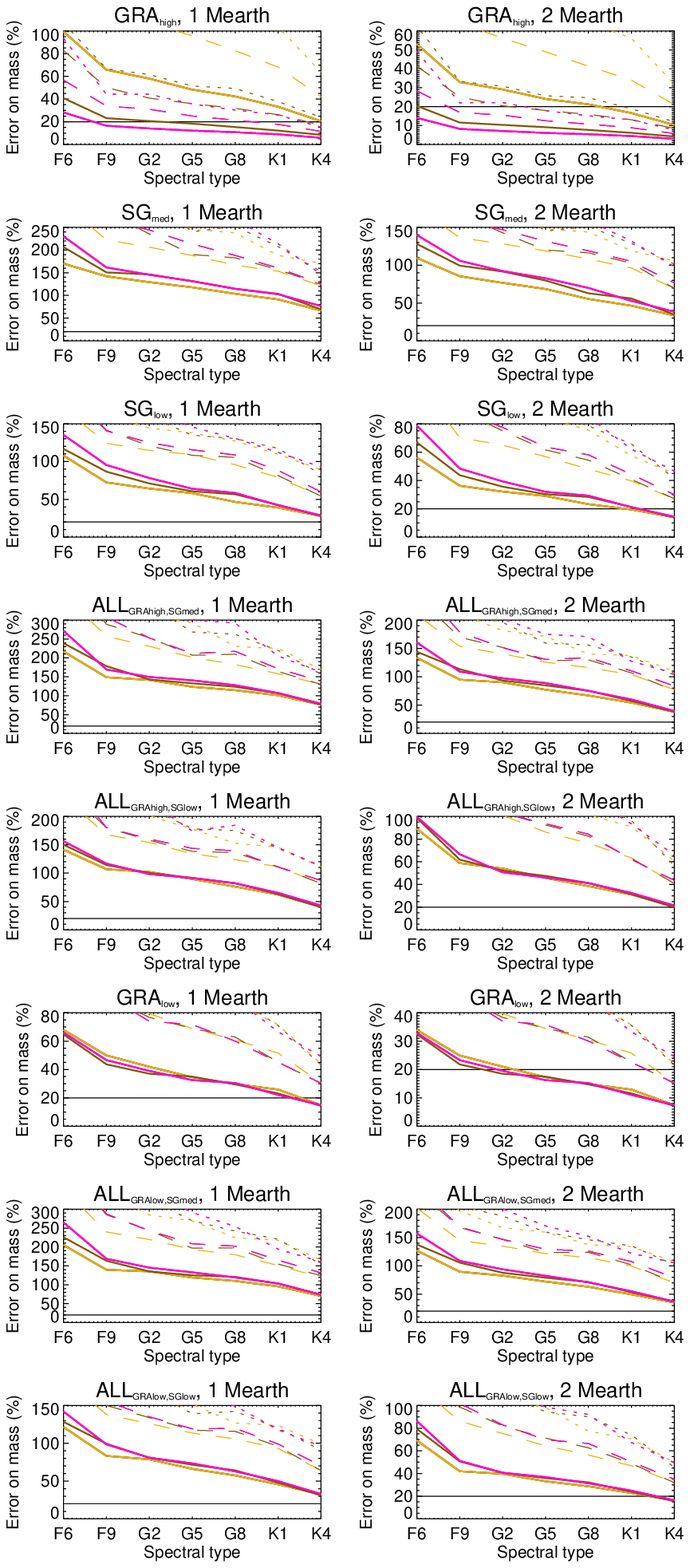}
\caption{
Same as Fig.~\ref{errm} but for a 30-day binning. 
}
\label{carac_bin30}
\end{figure}

\begin{figure}
\includegraphics{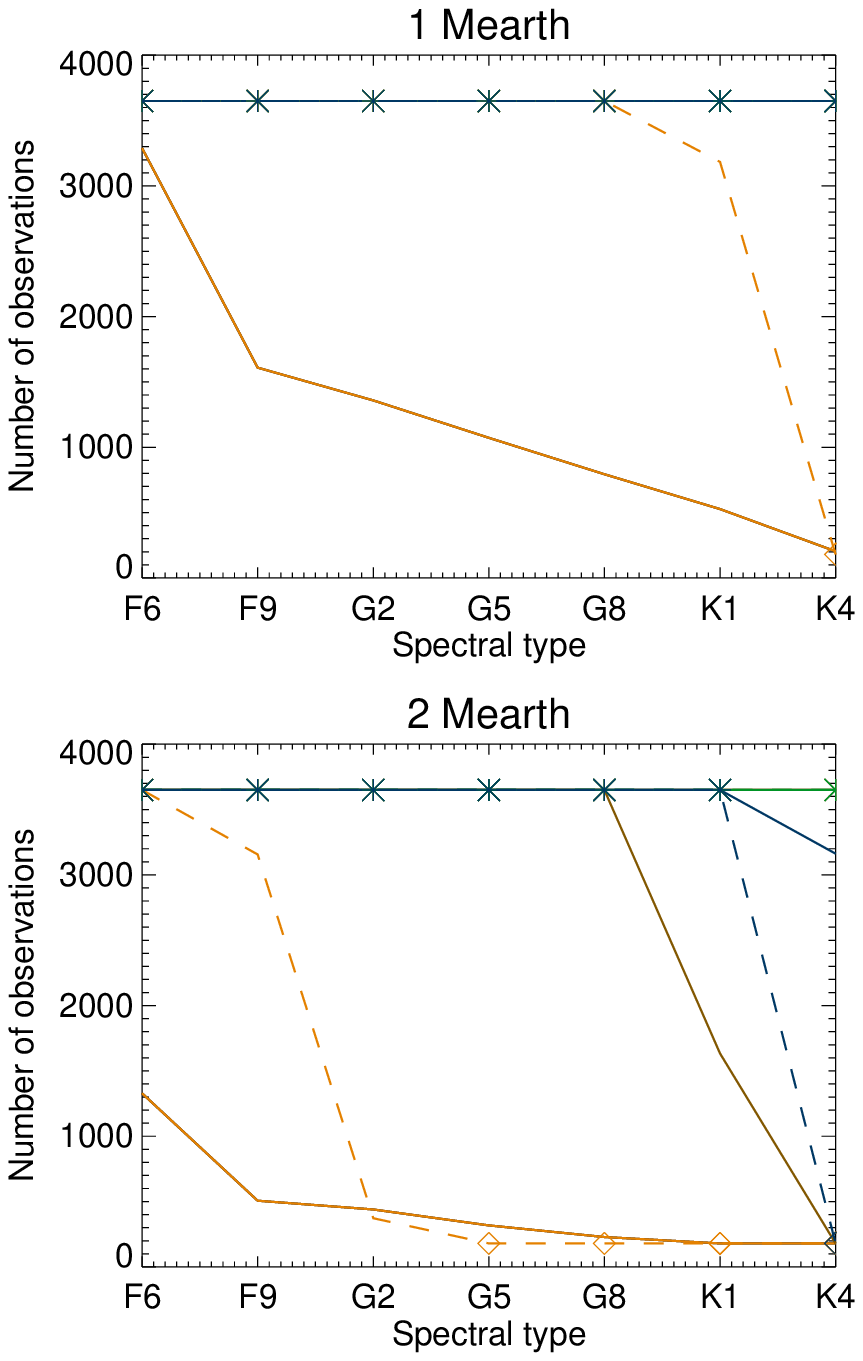}
\caption{
Same as Fig.~\ref{carac20} but for a 30-day binning. 
}
\label{carac20_bin30}
\end{figure}

Figure~\ref{carac_bin30} shows the uncertainty on the mass for a binning over 30 days (see Sect.~5.4) and Fig.~\ref{carac20_bin30} shows the number of observations necessary to reach an uncertainty on the mass of 20\%. The results are discussed in Sect.~5.4.

\section{Additional blind test results}

Additional blind tests have been performed to compare different conditions. We explore the sampling, the planet mass and effects of inclination.

\subsection{Case 1 M$_{\rm Earth}$, 180 points}

\begin{figure}
\includegraphics{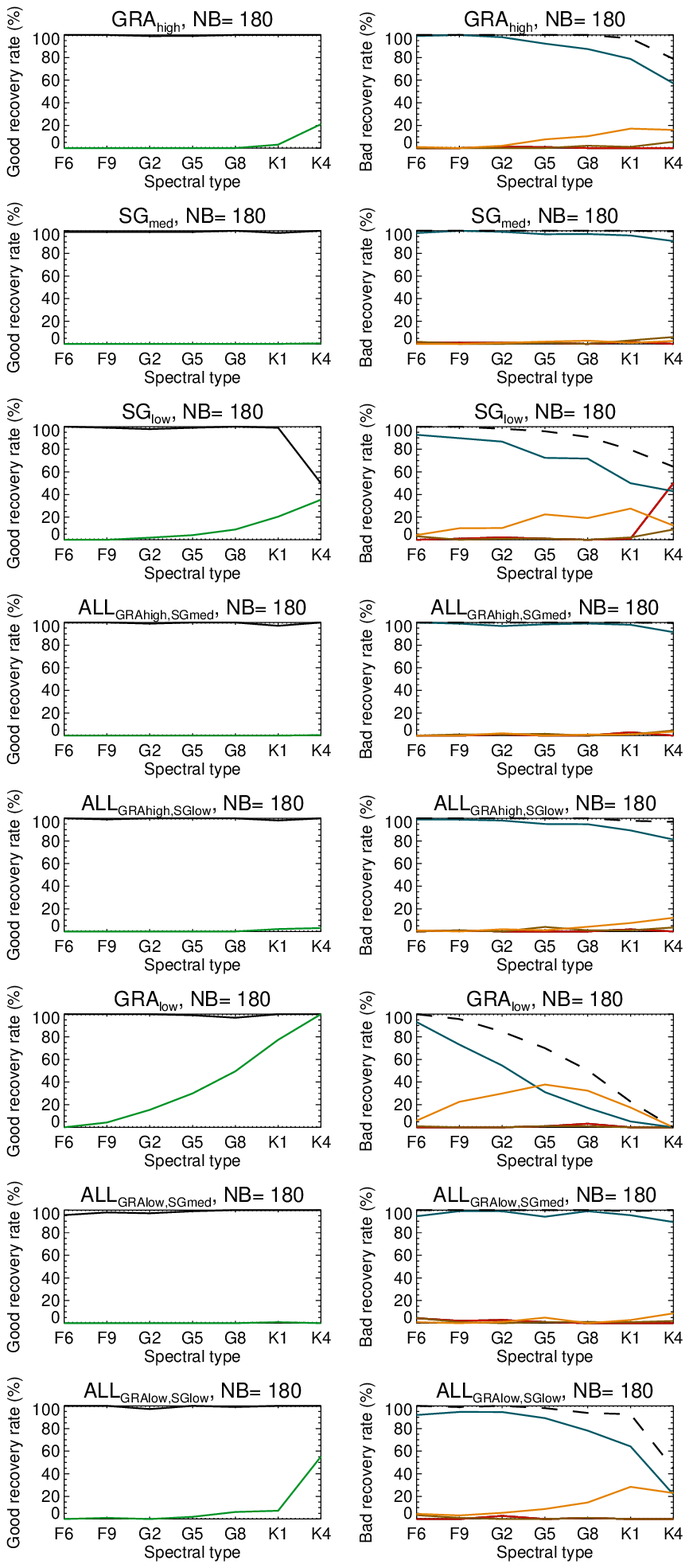}
\caption{
Same as Fig.~\ref{blind_pc} but for 180 points. 
}
\label{blind_pc_N180}
\end{figure}

\begin{figure}
\includegraphics{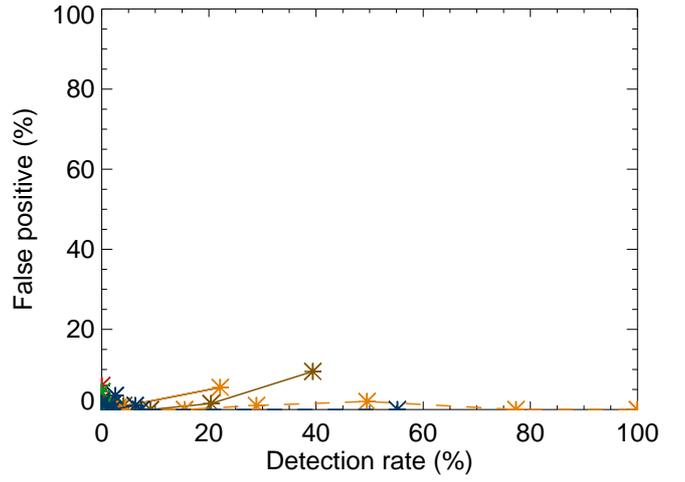}
\caption{
Same as Fig.~\ref{blind_roc} but for 180 points. 
}
\label{blind_roc_N180}
\end{figure}

\begin{figure}
\includegraphics{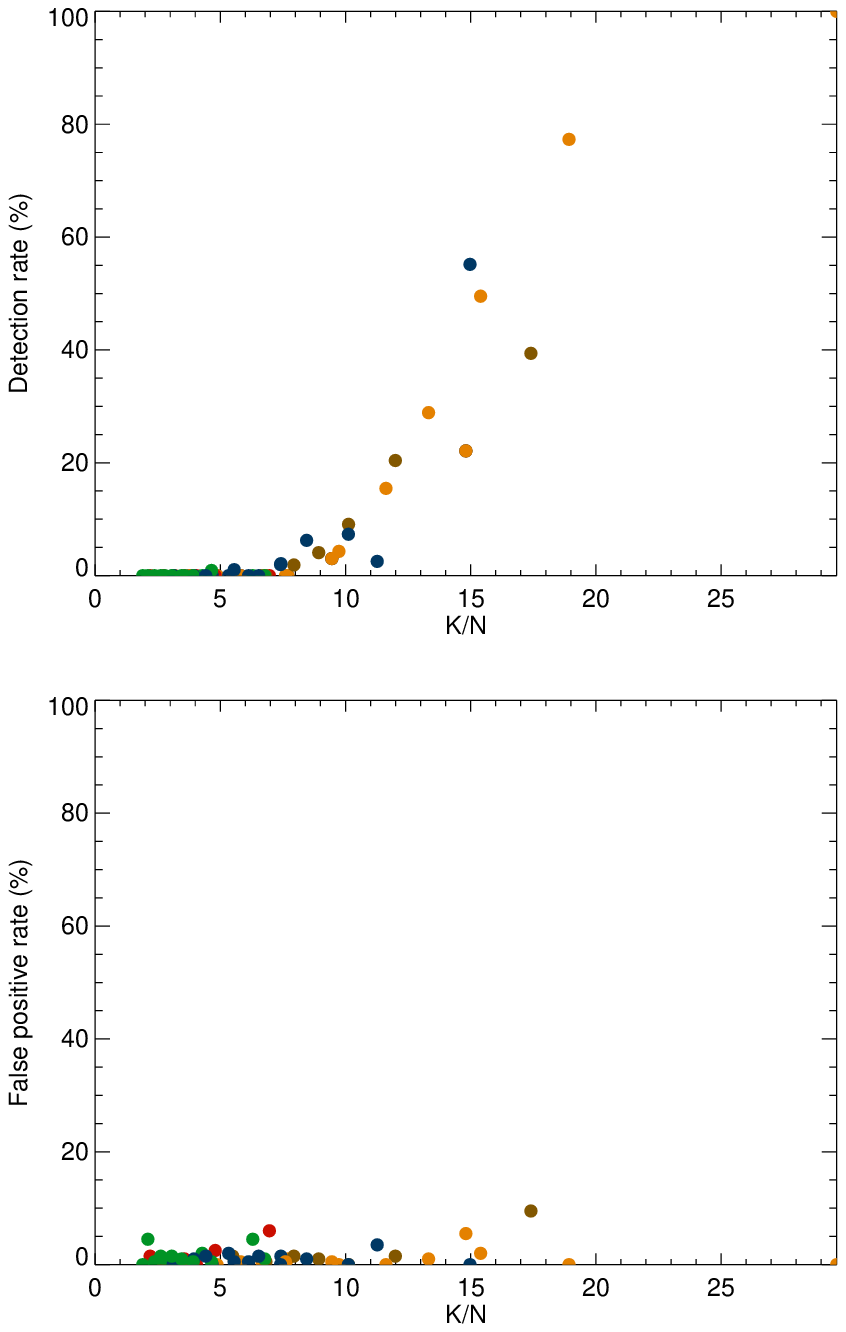}
\caption{
Same as Fig.~\ref{kn} but for 180 points. 
}
\label{kn_N180}
\end{figure}

Figures~\ref{blind_pc_N180}, \ref{blind_roc_N180}, and ~\ref{kn_N180} show the results for a blind test performed with 180 days only (1 M$_{\rm Earth}$ as in Sect 4.4). They are discussed in Sect.~4.4.6 and 4.4.8. 

\subsection{Case 2 M$_{\rm Earth}$, 1266 points}

\begin{figure}
\includegraphics{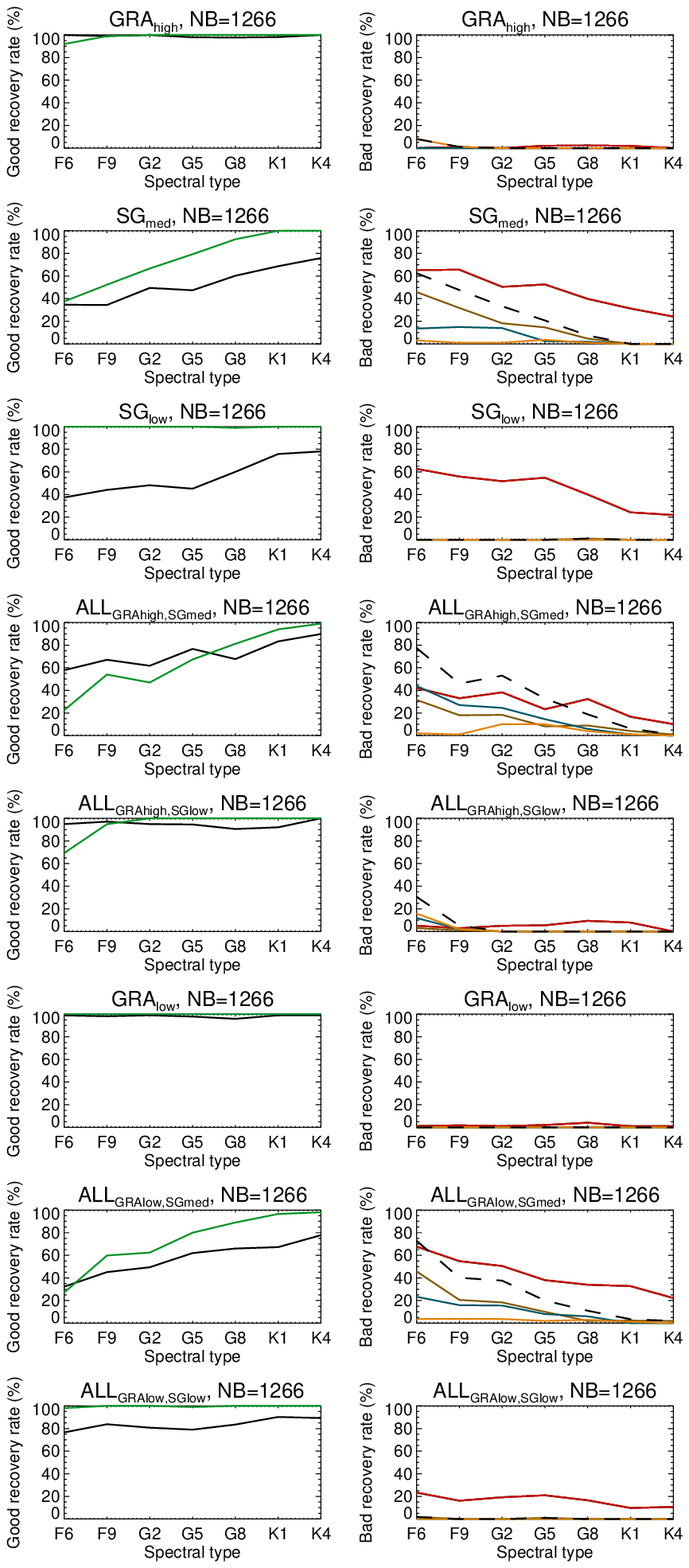}
\caption{
Same as Fig.~\ref{blind_pc} but for a 2 M$_{\rm Earth}$ planet. 
}
\label{blind_pc_2Mt}
\end{figure}

\begin{figure}
\includegraphics{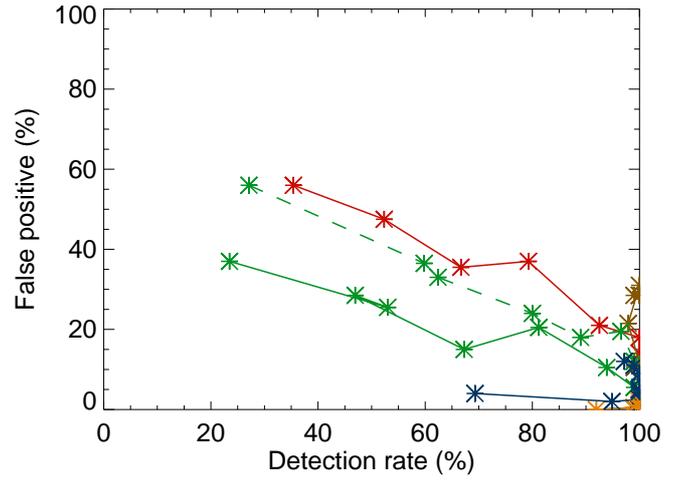}
\caption{
Same as Fig.~\ref{blind_roc} but for a 2 M$_{\rm Earth}$ planet. 
}
\label{blind_roc_2Mt}
\end{figure}

\begin{figure}
\includegraphics{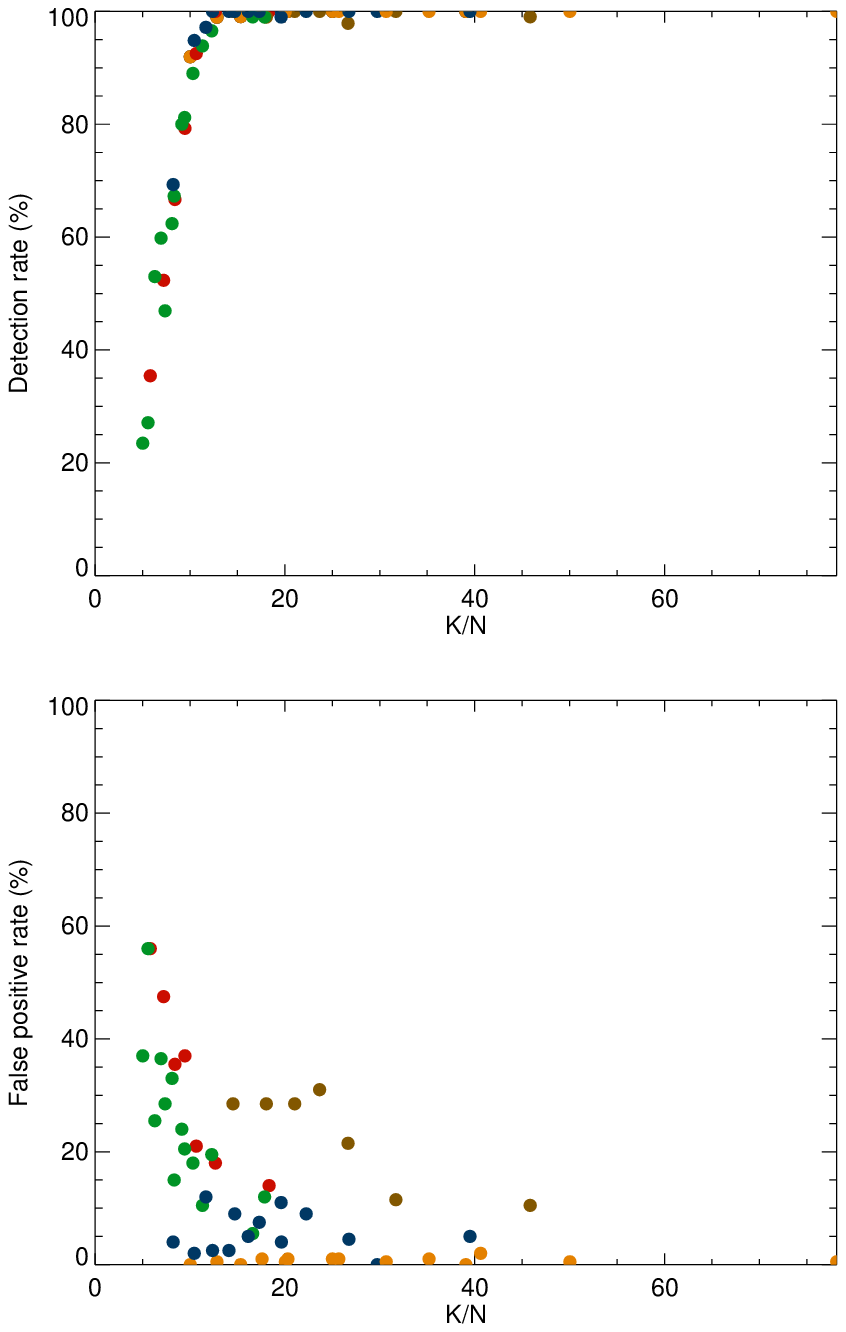}
\caption{
Same as Fig.~\ref{kn} but for a 2 M$_{\rm Earth}$ planet. 
}
\label{kn_2Mt}
\end{figure}

Figures~\ref{blind_pc_2Mt}, ~\ref{blind_roc_2Mt}, and ~\ref{kn_2Mt} show the results for a blind test performed for 2 M$_{\rm Earth}$ (1266 points as in Sect 4.4). They are discussed in Sect.~4.4.6 and 4.4.8.

\subsection{Case 1 M$_{\rm Earth}$ with inclination distribution, 1266 points}

\begin{figure}
\includegraphics{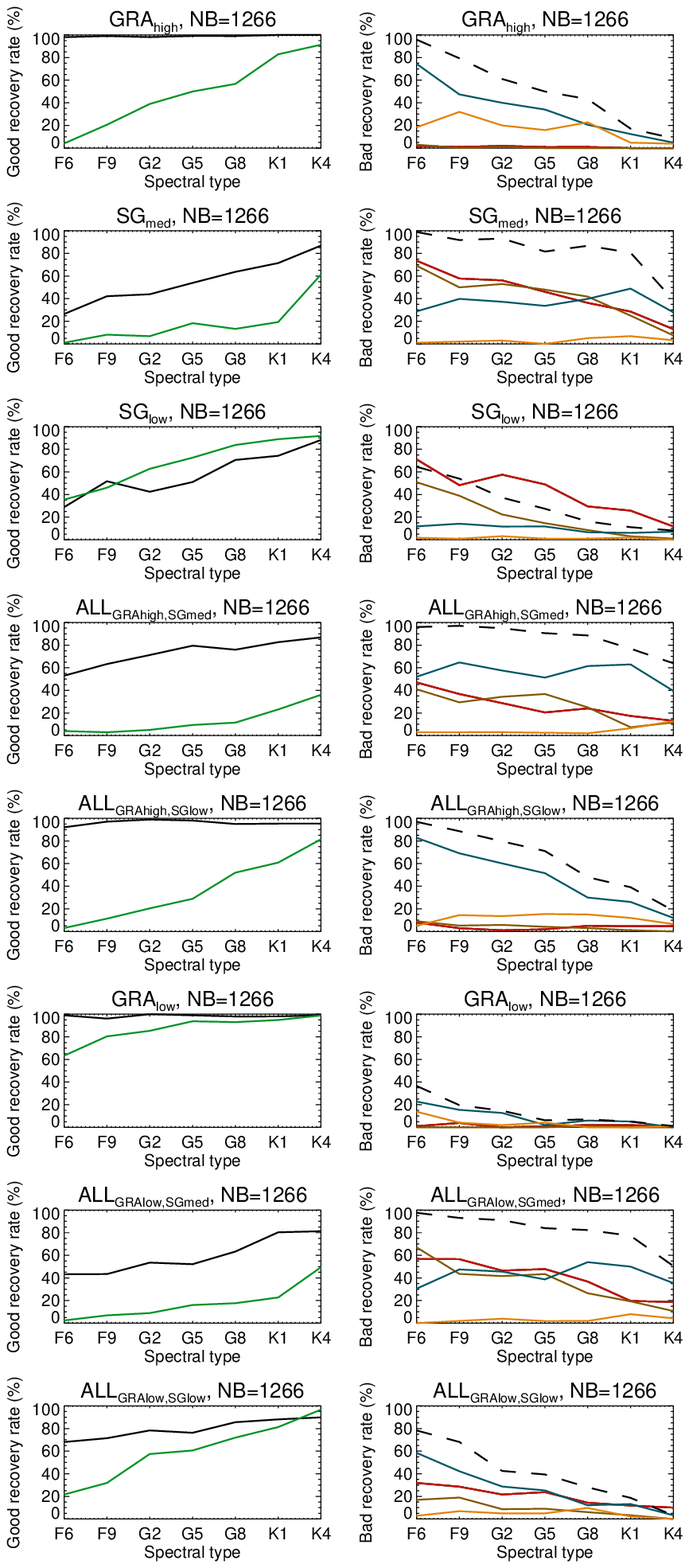}
\caption{
Same as Fig.~\ref{blind_pc} but with inclination distribution. 
}
\label{blind_pc_incl}
\end{figure}

\begin{figure}
\includegraphics{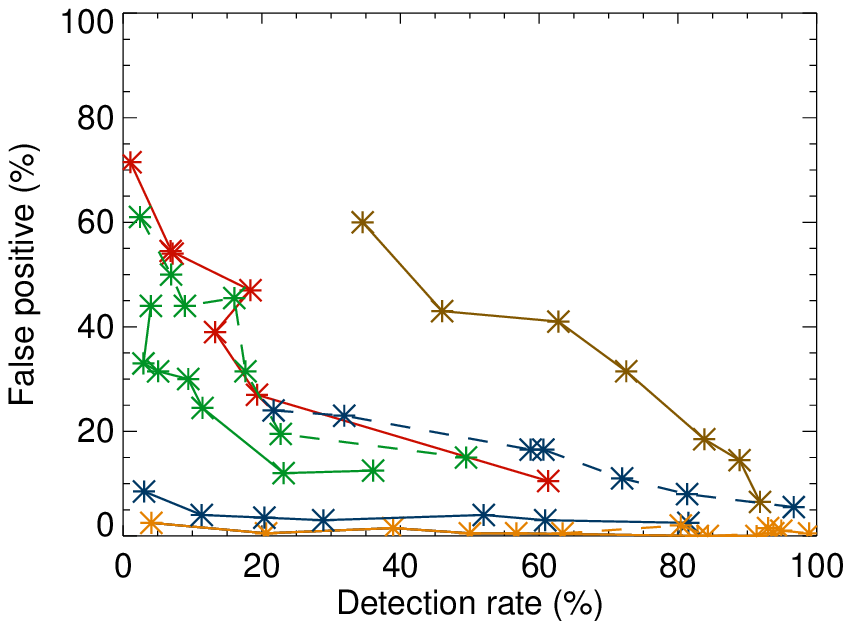}
\caption{
Same as Fig.~\ref{blind_roc} but with inclination distribution. 
}
\label{blind_roc_incl}
\end{figure}

Figures~\ref{blind_pc_incl},  and ~\ref{blind_roc_incl}, show the results for a blind test performed with a realistic distribution of the inclination angle. 
They are discussed in Sect.~4.4.6.

\subsection{Case 2 M$_{\rm Earth}$ with inclination distribution, 1266 points}

\begin{figure}
\includegraphics{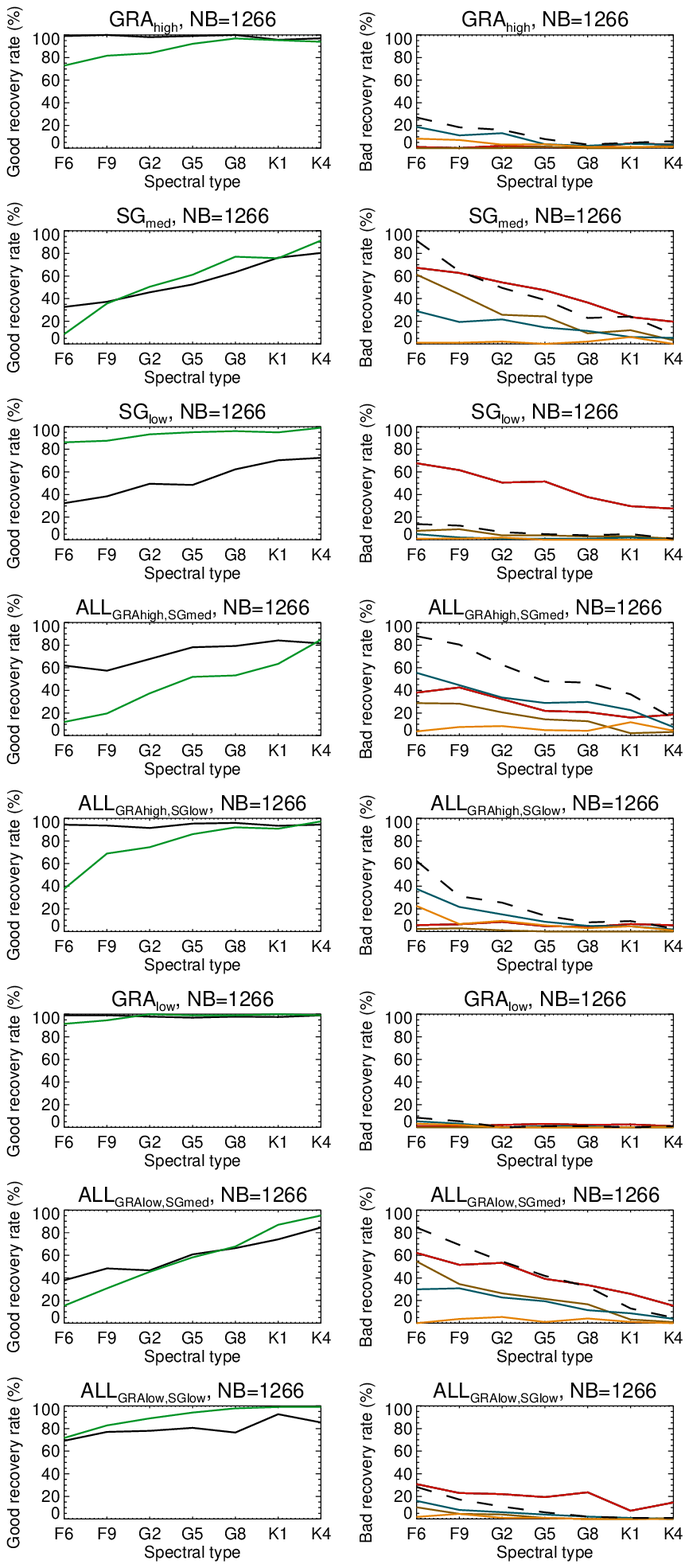}
\caption{
Same as Fig.~\ref{blind_pc} but with inclination distribution and 2~M$_{\rm Earth}$ planet. 
}
\label{blind_pc_incl2}
\end{figure}

\begin{figure}
\includegraphics{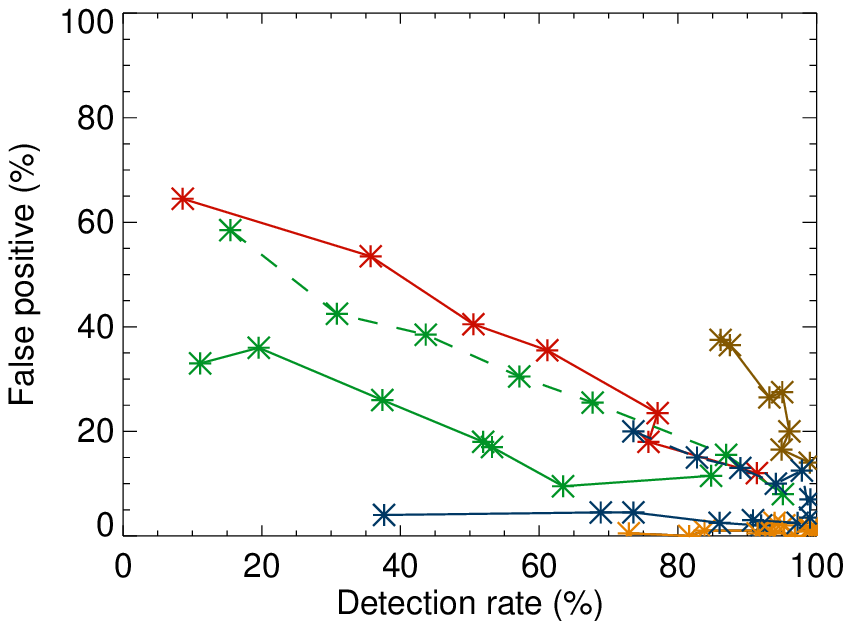}
\caption{
Same as Fig.~\ref{blind_roc} but with inclination distribution and 2~M$_{\rm Earth}$ planet. 
}
\label{blind_roc_incl2}
\end{figure}

Figures~\ref{blind_pc_incl2},  and ~\ref{blind_roc_incl2}, show the results for a blind test performed with a distribution of the inclination angle and a 2~M$_{\rm Earth}$ planet. 
They are discussed in Sect.~4.4.6.

\subsection{Case 1 M$_{\rm Earth}$, 3-year coverage}

\begin{figure}
\includegraphics{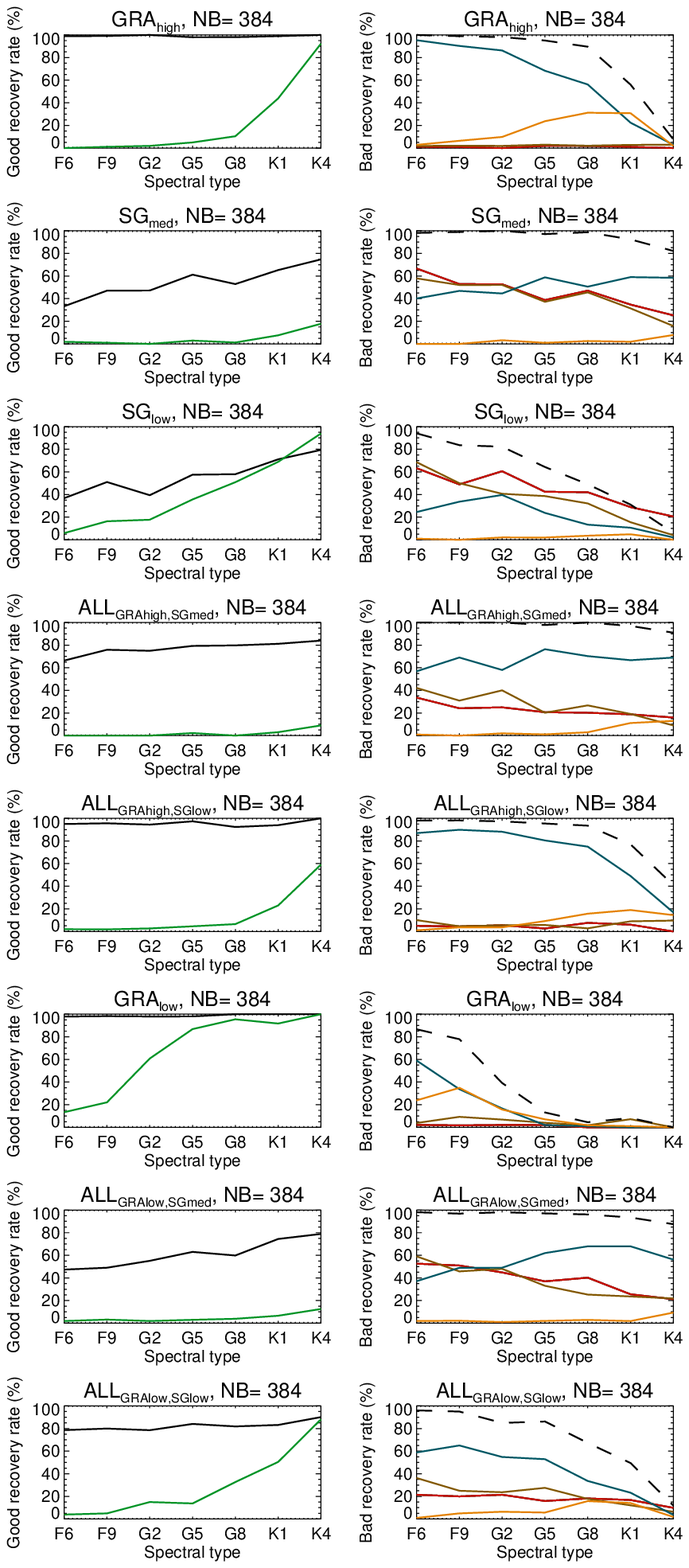}
\caption{
Same as Fig.~\ref{blind_pc} but for three-year coverage, 384 points. 
}
\label{blind_pc_3y}
\end{figure}

\begin{figure}
\includegraphics{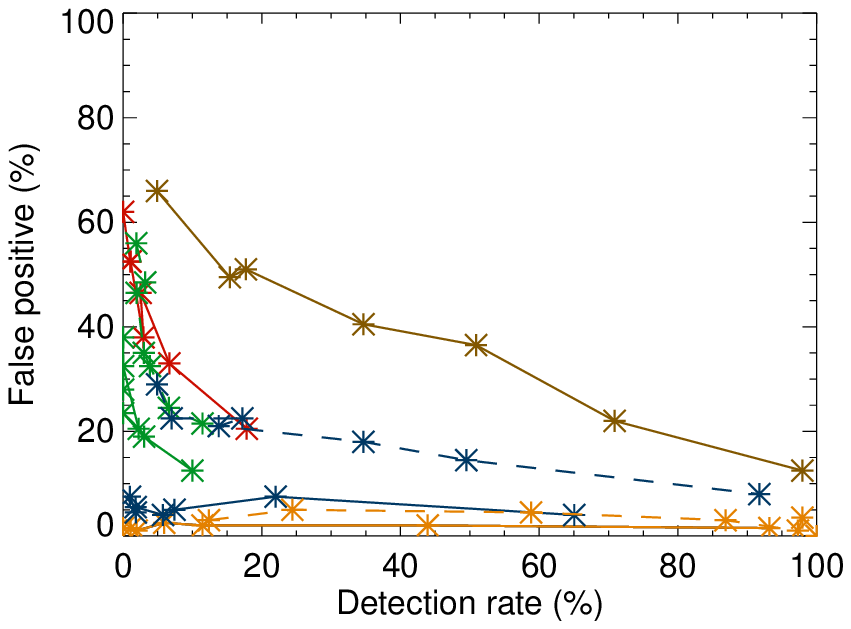}
\caption{
Same as Fig.~\ref{blind_roc} but for three-year coverage, 384 points. 
}
\label{blind_roc_3y}
\end{figure}

\begin{figure}
\includegraphics{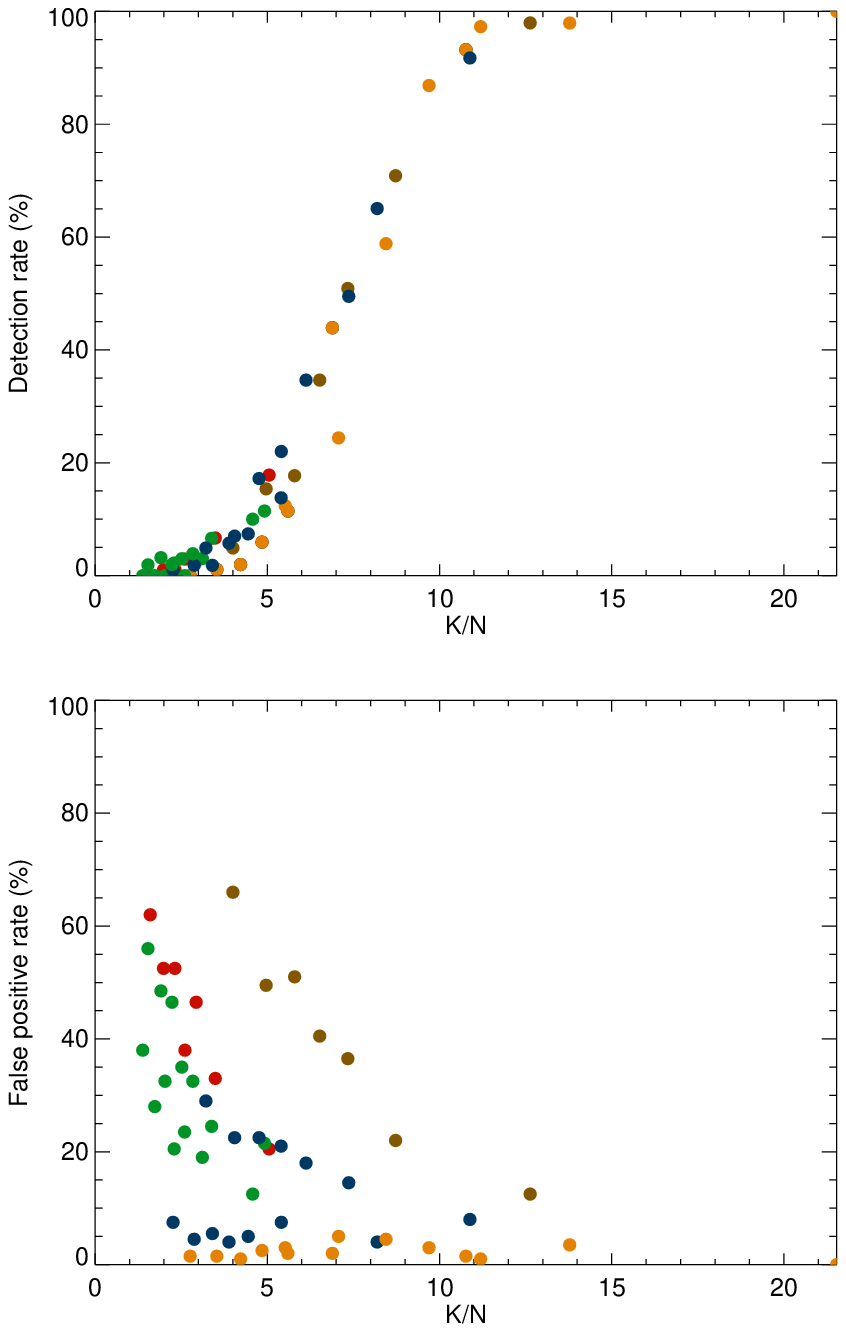}
\caption{
Same as Fig.~\ref{kn} but for three-year coverage, 384 points. 
}
\label{kn_3y}
\end{figure}

Figures~\ref{blind_pc_3y}, ~\ref{blind_roc_3y}, and ~\ref{kn_3y} show the results for a blind test performed with 384 points, over 3 years instead of 10 years (1 M$_{\rm Earth}$ as in Sect 4.4). They are discussed in Sect.~5.3. 

\subsection{Case 1 M$_{\rm Earth}$, 1266 points, 30 day binning}

\begin{figure}
\includegraphics{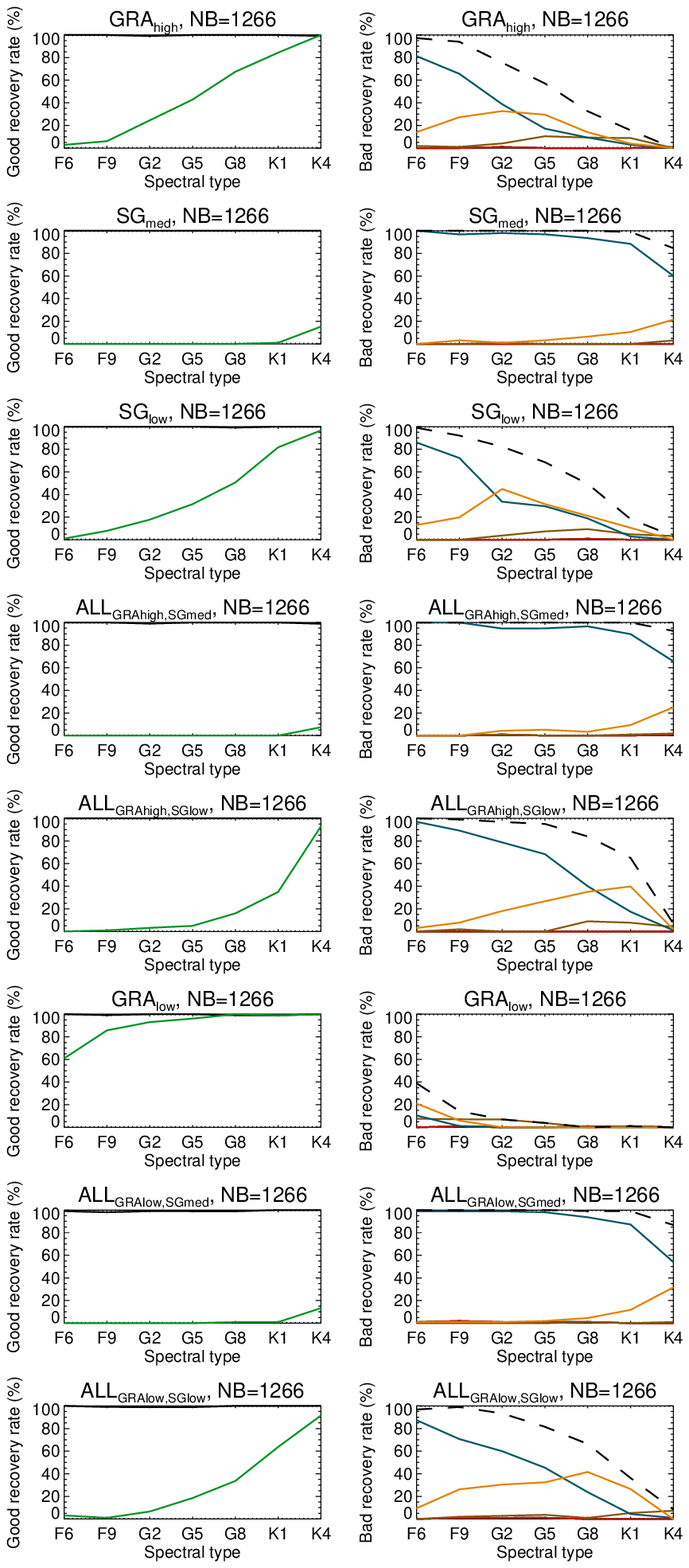}
\caption{
Same as Fig.~16 but for a 30-day binning. 
}
\label{blind_pc_bin30}
\end{figure}

\begin{figure}
\includegraphics{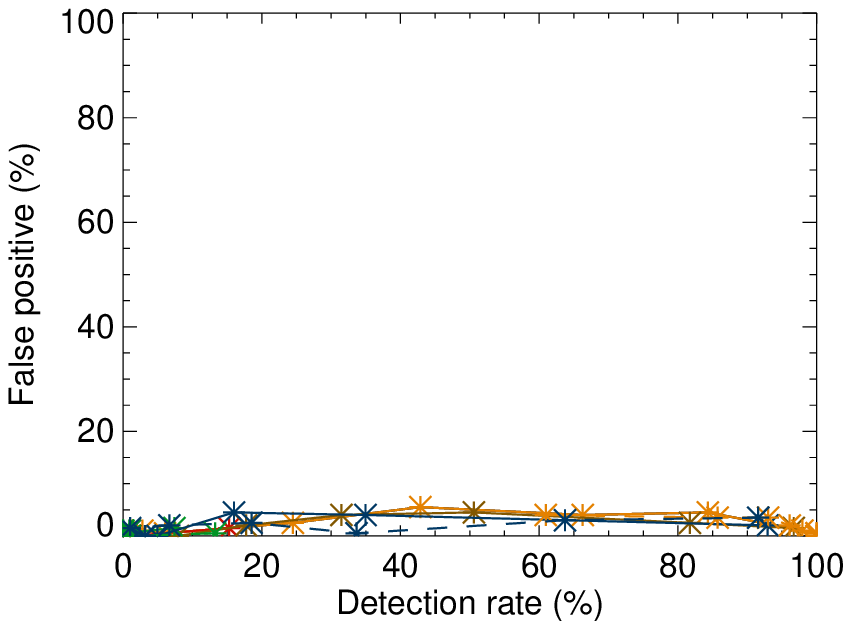}
\caption{
Same as Fig.~\ref{blind_roc} but for a 30-day binning. 
}
\label{blind_roc_bin30}
\end{figure}

\begin{figure}
\includegraphics{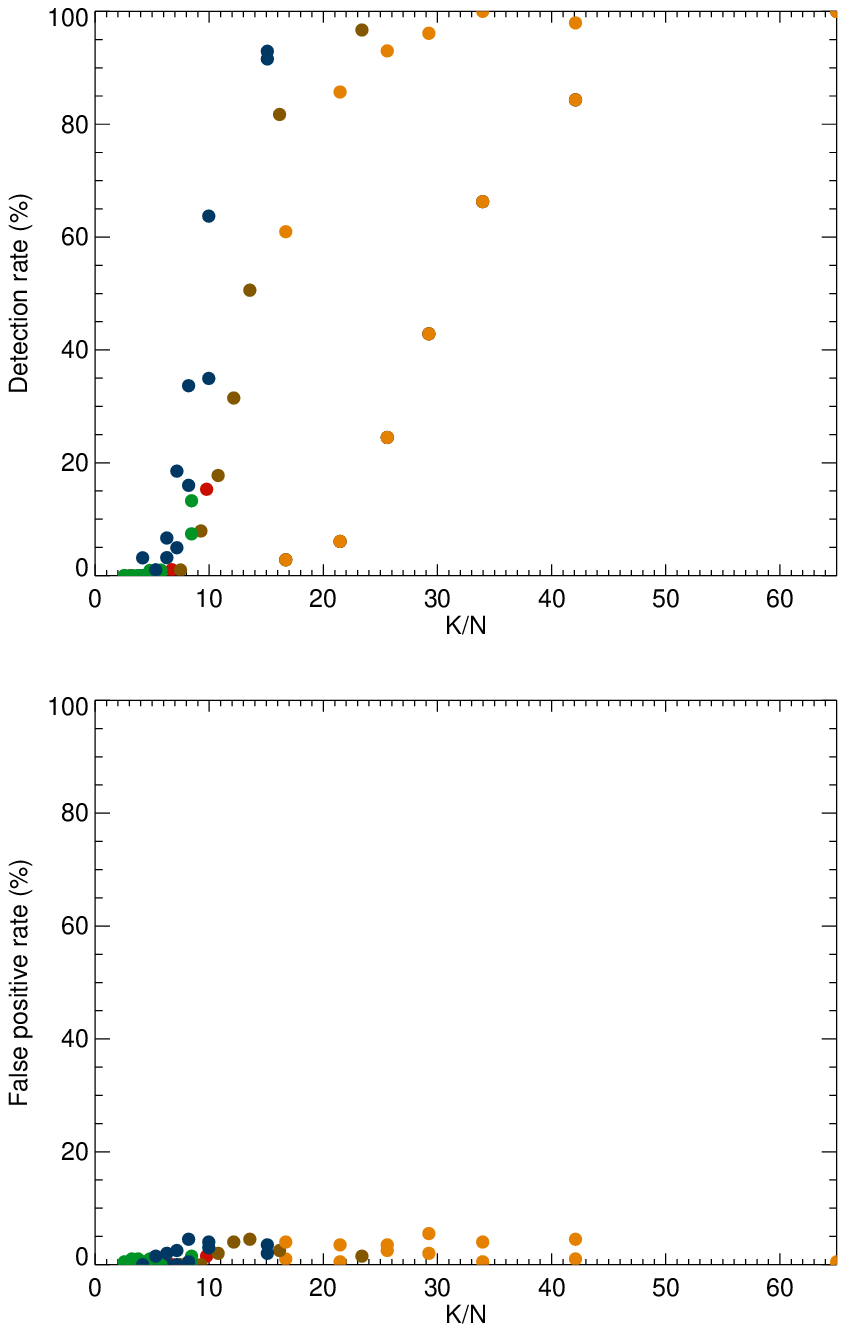}
\caption{
Same as Fig.~\ref{kn} but for a 30-day binning. 
}
\label{kn_bin30}
\end{figure}

Figures~\ref{blind_pc_bin30}, ~\ref{blind_roc_bin30}, and ~\ref{kn_bin30} show the results for a blind test performed with a 30 day binning, for 1 M$_{\rm Earth}$ (1266 points as in Sect 4.4). They are discussed in Sect.~5.4.

\end{appendix}

\end{document}